\documentclass{conm-p-l}%
\usepackage{amsmath}
\usepackage{graphicx}%
\usepackage{amsfonts}%
\usepackage{amssymb}
\theoremstyle{plain}

\newtheorem{conjecture}{Conjecture}
\newtheorem{corollary}{Corollary}

\newtheorem{definition}{Definition}

\newtheorem{lemma}{Lemma}

\newtheorem{problem}{Problem}
\newtheorem{proposition}{Proposition}
\newtheorem{remark}{Remark}

\newtheorem{theorem}{Theorem}
\numberwithin{equation}{section}

\begin{document}
\title[Quantum Hidden Subgroup Algorithms]{Quantum Hidden Subgroup Algorithms: \\A Mathematical Perspective}
\author{Samuel J. Lomonaco, Jr.}
\address[Samuel J. Lomonaco, Jr.]{Department of Computer Science and Electrical Engineering\\
University of Maryland Baltimore County\\
1000 Hilltop Circle\\
Baltimore, MD 21250}
\email[Samuel J. Lomonaco, Jr.]{Lomonaco@UMBC.EDU}
\urladdr{WebPage: http://www.csee.umbc.edu/\symbol{126}lomonaco}
\author{Louis H. Kauffman}
\address[Louis H. Kauffman]{Department of Mathematics, Statistics, and Computer Science\\
University of Illinois at Chicago, Chicago, IL 60607-7045}
\email[Louis H. Kauffman]{kauffman@uic.edu}
\urladdr{WebPage: http://math.uic.edu/\symbol{126}kauffman}
\thanks{This effort partially supported by the Defense Advanced Research Projects
Agency (DARPA) and Air Force Research Laboratory, Air Force Materiel Command,
USAF, under agreement number F30602-01-2-0522, the National Institute for
Standards and Technology (NIST), and by L-O-O-P Fund Grant WADC2000. \ The
U.S. Government is authorized to reproduce and distribute reprints for
Government purposes notwithstanding any copyright annotations thereon. \ The
views and conclusions contained herein are those of the authors and should not
be interpreted as necessarily representing the official policies or
endorsements, either expressed or implied, of the Defense Advanced Research
Projects Agency, the Air Force Research Laboratory, or the U.S. Government.
\ (Copyright 2002 by authors. Reproduction of this article, in its entirety,
is permitted for non-commercial purposes.) \ }
\date{January 21, 2002}
\copyrightinfo{2002}
{by authors. Reproduction of this article, in its entirety, is permitted for non-commertial purposes.}

\begin{abstract}
The ultimate objective of this paper is to create a stepping stone to the
development of new quantum algorithms. \ The strategy chosen is to begin by
focusing on the class of abelian quantum hidden subgroup algorithms, i.e., the
class of abelian algorithms of the Shor/Simon genre. \ Our strategy is to make
this class of algorithms as mathematically transparent as possible. \ By the
phrase ``mathematically transparent'' we mean to expose, to bring to the
surface, and to make explicit the concealed mathematical structures that are
inherently and fundamentally a part of such algorithms. \ In so doing, we
create symbolic abelian quantum hidden subgroup algorithms that are analogous
to the those symbolic algorithms found within such software packages as Axiom,
Cayley, Maple, Mathematica, and Magma. \ 

\bigskip

As a spin-off of this effort, we create three different generalizations of
Shor's quantum factoring algorithm to free abelian groups of finite rank. \ We
refer to these algorithms as wandering (or vintage $\mathbb{Z}_{Q})$ Shor
algorithms. \ They are essentially quantum algorithms on free abelian groups
$A$ of finite rank $n$ which, with each iteration, first select a random
cyclic direct summand $\mathbb{Z}$ of the group $A$ and then apply one
iteration of the standard Shor algorithm to produce a random character of the
``approximating'' finite group $\widetilde{A}=\mathbb{Z}_{Q}$, called the
group probe. \ These characters are then in turn used to find either the order
$P$ of a maximal cyclic subgroup $\mathbb{Z}_{P}$ of the hidden quotient group
$H_{\varphi}$, or the entire hidden quotient group $H_{\varphi}$. An integral
part of these wandering quantum algorithms is the selection of a very special
random transversal $\iota_{\mu}:\widetilde{A}\longrightarrow A$, which we
refer to as a Shor transversal. The algorithmic time complexity of the first
of these wandering Shor algorithms is found to be $O\left(  n^{2}\left(  \lg
Q\right)  ^{3}\left(  \lg\lg Q\right)  ^{n+1}\right)  $.

\end{abstract}
\maketitle
\tableofcontents

\bigskip

\part{\textbf{Preamble\bigskip}}

\section{Introduction}

\bigskip

The ultimate objective of this paper is to create a stepping stone to the
development of new quantum algorithms. \ The strategy chosen is to begin by
focusing on the class of abelian quantum hidden subgroup algorithms (QHSAs),
i.e., the class of abelian algorithms of the Shor/Simon genre. \ Our strategy
is to make this class of algorithms as mathematically transparent as possible.
\ By the phrase ``mathematically transparent,'' we mean to expose, to bring to
the surface, and to make explicit the concealed mathematical structures that
are inherently and fundamentally a part of such algorithms. \ In so doing, we
create a class of symbolic abelian QHSAs that are analogous to those symbolic
algorithms found within such software packages as Axiom, Cayley, Magma, Maple,
and Mathematica.

\bigskip

During this mathematical analysis, the differences between the Simon and Shor
quantum algorithms become dramatically apparent. \ This is in spite of the
fact that these two share a common ancestor, namely, the quantum random group
character generator \textsc{QRand}, described herein. \ While the Simon
algorithm is a QHSA on finite abelian groups which produces random characters
of the hidden quotient group, \ the Shor algorithm is a QHSA on free abelian
finite rank groups which produces random characters of a group which
``approximate'' the hidden quotient group. \ It is misleading, and a frequent
cause of much confusion in the open literature, to call them both essentially
the same QHSA. \ 

\bigskip

Surprisingly, these two very different algorithms touch an amazing array of
different mathematical disciplines, from the obvious to the not-so-obvious,
requiring the integration of many diverse fields of mathematics. \ Shor's
quantum factoring algorithm, for example, depends heavily on the interplay of
two metrics on the unit circle $\mathbb{S}^{1}$, namely the arclength metric
\textsc{Arc}$_{2\pi}$ and the chordal metric \textsc{Chord}$_{2\pi}$. \ This
observation greatly simplifies the analysis of the Shor factoring algorithm,
while at the same time revealing more of the structure concealed within the algorithm.

\bigskip

As a spin-off of this effort, we create three different generalizations of
Shor's quantum factoring algorithm to free abelian groups of finite rank,
found in sections \ref{VintageSummary} and \ref{Alternate1}. We refer to these
algorithms as wandering (or vintage $\mathbb{Z}_{Q}$) Shor algorithms. \ They
are essentially QHSAs on free abelian finite rank $n$ groups $A$ which, with
each iteration, first select a random cyclic direct summand $\mathbb{Z}$ of
the group $A$ and then apply one iteration of the standard Shor algorithm to
produce a random character of the ``approximating'' finite group
$\widetilde{A}$, called a group probe. These algorithms find either the order
$P$ of a maximal cyclic subgroup $\mathbb{Z}_{P}$ of the hidden quotient group
$H_{\varphi}$, or the entire hidden quotient group $H_{\varphi}$. \ An
integral part of these wandering algorithms is the selection of a very special
random transversal $\iota_{\mu}:\widetilde{A}\longrightarrow A$, which we
refer to as a Shor transversal. \ The algorithmic time complexity of the first
of these wandering (or vintage $\mathbb{Z}_{Q}$) algorithms is found in
theorem \ref{ZComplexity} of section \ref{ComplexitySection} to be $O\left(
n^{2}\left(  \lg Q\right)  ^{3}\left(  \lg\lg Q\right)  ^{n+1}\right)  $,
where $n$ denotes the fixed finite rank of the free abelian group $A$.
\ Theorem \ref{ZComplexity} is based on the assumptions also found in section
\ref{ComplexitySection}. This asymptotic bound is by no means the tightest possible.

\bigskip

Throughout this paper, it is assumed that the reader is familiar with the
class of quantum hidden subgroup algorithms. \ For an introductions to this
subject, please refer, for example, to any one of the references
\cite{Cheung1}, \cite{Jozsa2}, \cite{Jozsa3}, \cite{Kitaev1}, \cite{Lomonaco2}%
, \cite{Nielsen1}, \cite{Shor1}, \cite{Shor2}. \ This paper focuses, in
particular, on the abelian hidden subgroup problem (HSP), with eye toward
future work by the authors on the non-abelian HSP. \ There is a great deal of
literature on the abelian HSP, for example, \cite{Brassard1}, \cite{Ekert1},
\cite{Jozsa2}, \cite{Jozsa3}, \cite{Jozsa4}, \cite{Kitaev1}, \cite{Mosca1},
\cite{Nielsen1}, \cite{Shor1}, \cite{Shor2}, \cite{Simon1}. \ For literature
on the non-abelian hidden subgroup problem, see for example, \cite{Ettinger1},
\cite{Ivanyos1}, \cite{Ekert1}, \cite{Jozsa3}, \cite{Lomonaco2},
\cite{Nielsen1}, \cite{Pueschel1}, \cite{Roetteler1}, \cite{Russell1},
\cite{vanDam1}.

\bigskip

\section{An example of Shor's quantum factoring algorithm}

\bigskip

As an example of what we would like to make mathematically transparent,
consider the following instance of Peter Shor's quantum factoring algorithm.
\ A great part of this paper is devoted to exposing and bringing to the
surface the many concealed mathematical structures that are inherently and
fundamentally part of this example. \ 

Perhaps you see them? \ Perhaps you find them to be self evident? \ If you do,
then you need read no more of this paper, although you are most certainly
welcome to read on. \ If, on the other hand, the following example leaves you
with a restless, uneasy feeling of not fully understanding what is really
going on (i.e., of not fully understanding what concealed mathematical
structures are lurking underneath these calculations), then you are invited to
read the remainder of this paper. \ 

\bigskip

Peter Shor's quantum factoring algorithm reduces the task of factoring a
positive integer $N$ to first finding a random integer $a$ relatively prime to
$N$, and then next to determining the period $P$ of the following function
\[%
\begin{array}
[c]{ccl}%
\mathbb{Z} & \overset{\varphi}{\longrightarrow} & \mathbb{Z}\operatorname{mod}%
N\\
x & \longmapsto & a^{x}\operatorname{mod}N\text{ ,}%
\end{array}
\]
where $\mathbb{Z}$ denotes the additive group of integers, and where
$\mathbb{Z}\operatorname{mod}N$ denotes the integers $\operatorname{mod}N$
under multiplication\footnote{A random integer $a$ with $\gcd\left(
a,N\right)  =1$ is found by selecting a random integer, and then applying the
Euclidean algorithm to determine whether or not it is relatively prime to $N$.
\ If not, then the $\gcd$ is a non-trivial factor of $N$, and there is no need
to proceed futher. \ However, this possibility is highly unlikely if $N$ is
large.}.

\bigskip

Since $\mathbb{Z}$ is an infinite group, Shor chooses to work instead with the
finite additive cyclic group $\mathbb{Z}_{Q}$ of order $Q=2^{m}$, where
$N^{2}\leq Q<2N^{2},$ and with the ``approximating'' map
\[%
\begin{array}
[c]{ccll}%
\mathbb{Z}_{Q} & \overset{\widetilde{\varphi}}{\longrightarrow} &
\mathbb{Z}\operatorname{mod}N & \\
x & \longmapsto & a^{x}\operatorname{mod}N\text{ ,} & 0\leq x<Q
\end{array}
\]

Shor begins by constructing a quantum system with two quantum registers
\[
\left|  \text{\textsc{Left}\_\textsc{Register}}\right\rangle \left|
\text{\textsc{Right}\_\textsc{Register}}\right\rangle \text{ ,}%
\]
the left intended to hold the arguments $x$ of $\widetilde{\varphi}$, the
right to hold the corresponding values of $\widetilde{\varphi}$. \ This
quantum system has been constructed with a unitary transformation
\[
U_{\widetilde{\varphi}}:\left|  x\right\rangle \left|  1\right\rangle
\longmapsto\left|  x\right\rangle \left|  \widetilde{\varphi}\left(  x\right)
\right\rangle
\]
implementing the ``approximating'' map $\widetilde{\varphi}$.

\bigskip

As an example, let us use Shor's algorithm to factor the enormous
\raisebox{-0.1003in}{\includegraphics[
trim=0.000000in -0.010344in 0.000000in 0.010420in,
height=0.2681in,
width=0.2491in
]%
{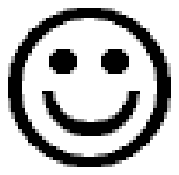}%
}%
\ integer $N=21$, assuming that $a=2$ has been randomly chosen. \ Thus,
$Q=2^{9}=512$.

\bigskip

Unknown to Peter Shor, the period is $P=6$, and hence, $Q=6\cdot85+2$. \ 

\bigskip

Shor proceeds by executing the following steps:

\bigskip

\begin{itemize}
\item[\fbox{$\mathbb{STEP}$\textbf{ 0}}] Initialize%
\[
\left|  \psi_{0}\right\rangle =\left|  0\right\rangle \left|  1\right\rangle
\]
\bigskip

\item[\fbox{$\mathbb{STEP}$\textbf{ 1}}] Apply the Fourier transform
\[
\mathcal{F}:\left|  u\right\rangle \longmapsto\frac{1}{\sqrt{512}}\sum
_{x=0}^{511}\omega^{ux}\left|  x\right\rangle
\]
to the left register, where $\omega=\exp(2\pi i/512)$ is a primitive $512$-th
root of unity, to obtain
\[
\left|  \psi_{1}\right\rangle =\frac{1}{\sqrt{512}}\sum_{x=0}^{511}\left|
x\right\rangle \left|  1\right\rangle
\]
\bigskip

\item[\fbox{$\mathbb{STEP}$\textbf{ 2}}] Apply the unitary transformation
\[
U_{\widetilde{\varphi}}:\left|  x\right\rangle \left|  1\right\rangle
\longmapsto\left|  x\right\rangle \left|  2^{x}\operatorname{mod}%
21\right\rangle
\]
to obtain
\[
\left|  \psi_{2}\right\rangle =\frac{1}{\sqrt{512}}\sum_{x=0}^{511}\left|
x\right\rangle \left|  2^{x}\operatorname{mod}21\right\rangle
\]
\bigskip

\item[\fbox{$\mathbb{STEP}$\textbf{ 3}}] Once again apply the Fourier
transform
\[
\mathcal{F}:\left|  x\right\rangle \longmapsto\frac{1}{\sqrt{512}}\sum
_{y=0}^{511}\omega^{xy}\left|  y\right\rangle
\]
to the left register to obtain
\begin{align*}
\left|  \psi_{3}\right\rangle  &  =\frac{1}{512}\sum_{x=0}^{511}\sum
_{y=0}^{511}\omega^{xy}\left|  y\right\rangle \left|  2^{x}\operatorname{mod}%
21\right\rangle =\frac{1}{512}\sum_{y=0}^{511}\left|  y\right\rangle \left(
\sum_{x=0}^{511}\omega^{xy}\left|  2^{x}\operatorname{mod}21\right\rangle
\right) \\
& \\
&  =\frac{1}{512}\sum_{y=0}^{511}\left|  y\right\rangle \left|  \Upsilon
\left(  y\right)  \right\rangle
\end{align*}
where
\[
\left|  \Upsilon\left(  y\right)  \right\rangle =\sum_{x=0}^{511}\omega
^{xy}\left|  2^{x}\operatorname{mod}21\right\rangle
\]
\bigskip

\item[\fbox{$\mathbb{STEP}$\textbf{ 4}}] Measure the left register. \ Then
with Probability
\[
Prob_{\widetilde{\varphi}}\left(  y\right)  =\frac{\left\langle \ \Upsilon
\left(  y\right)  \mid\Upsilon\left(  y\right)  \ \right\rangle }{\left(
512\right)  ^{2}}%
\]
the state will ``collapse'' to $\left|  y\right\rangle $ with the value
measured being the integer $y$, where $0\leq y<Q$.
\end{itemize}

\bigskip

Let us digress for a moment to find a more usable expression for the
probability distribution $Prob_{\widetilde{\varphi}}\left(  y\right)  $.

\bigskip%
\begin{align*}
\left|  \Upsilon\left(  y\right)  \right\rangle  &  =\sum_{x=0}^{511}%
\omega^{xy}\left|  2^{x}\operatorname{mod}21\right\rangle \\
& \\
&  =\sum_{x_{1}=0}^{85-1}\sum_{x_{0}=0}^{6-1}\omega^{\left(  6x_{1}%
+x_{0}\right)  y}\left|  2^{6x_{1}+x_{0}}\operatorname{mod}21\right\rangle
+\sum_{x_{0}=0}^{2-1}\omega^{\left(  6\cdot85+x_{0}\right)  y}\left|
2^{6\cdot85+x_{0}}\operatorname{mod}21\right\rangle
\end{align*}
But the order of $a=2$ modulo $21$ is $P=6$, i.e., $P=6$ is the smallest
positive integer such that $2^{6}=1\operatorname{mod}21$. \ Hence, the above
expression becomes
\begin{align*}
\left|  \Upsilon\left(  y\right)  \right\rangle  &  =\left(  \sum_{x_{1}%
=0}^{84}\omega^{6x_{1}y}\right)  \sum_{x_{0}=0}^{5}\omega^{x_{0}y}\left|
2^{x_{0}}\operatorname{mod}21\right\rangle +\omega^{6\cdot85y}\sum_{x_{0}%
=0}^{1}\omega^{x_{0}y}\left|  2^{x_{0}}\operatorname{mod}21\right\rangle \\
& \\
&  =\left(  \sum_{x_{1}=0}^{85}\omega^{6x_{1}y}\right)  \sum_{x_{0}=0}%
^{1}\omega^{x_{0}y}\left|  2^{x_{0}}\operatorname{mod}21\right\rangle +\left(
\sum_{x_{1}=0}^{84}\omega^{6x_{1}y}\right)  \sum_{x_{0}=2}^{5}\omega^{x_{0}%
y}\left|  2^{x_{0}}\operatorname{mod}21\right\rangle
\end{align*}
Since the kets $\left\{  \ \left|  2^{x_{0}}\operatorname{mod}21\right\rangle
\mid0\leq x_{0}<6\ \right\}  $ are all distinct, we have
\[
\left\langle \ \Upsilon\left(  y\right)  \mid\Upsilon\left(  y\right)
\ \right\rangle =2\left|  \sum_{x_{1}=0}^{85}\omega^{6x_{1}y}\right|
^{2}+4\left|  \sum_{x_{1}=0}^{84}\omega^{6x_{1}y}\right|  ^{2}\text{ .}%
\]
After a little algebraic manipulation, we finally have the following
expression for $Prob_{\widetilde{\varphi}}\left(  y\right)  $:
\[
Prob_{\varphi}\left(  y\right)  =\frac{\left\langle \ \Upsilon\left(
y\right)  \mid\Upsilon\left(  y\right)  \ \right\rangle }{\left(  512\right)
^{2}}=\left\{
\begin{array}
[c]{cc}%
\frac{\sin^{2}\left(  \frac{\pi y}{128}\right)  +2\sin^{2}\left(  \frac{\pi
y}{256}\right)  }{\left(  131072\right)  \sin^{2}\left(  \frac{3\pi y}%
{256}\right)  } & \text{if }y\neq0\text{ or }256\\
& \\
\frac{10923}{65536} & \text{if }y=0\text{ or }256
\end{array}
\right.
\]
A plot of $Prob_{\widetilde{\varphi}}\left(  y\right)  $ is shown in Figure
1.
\begin{center}
\includegraphics[
height=3.4091in,
width=4.3708in
]%
{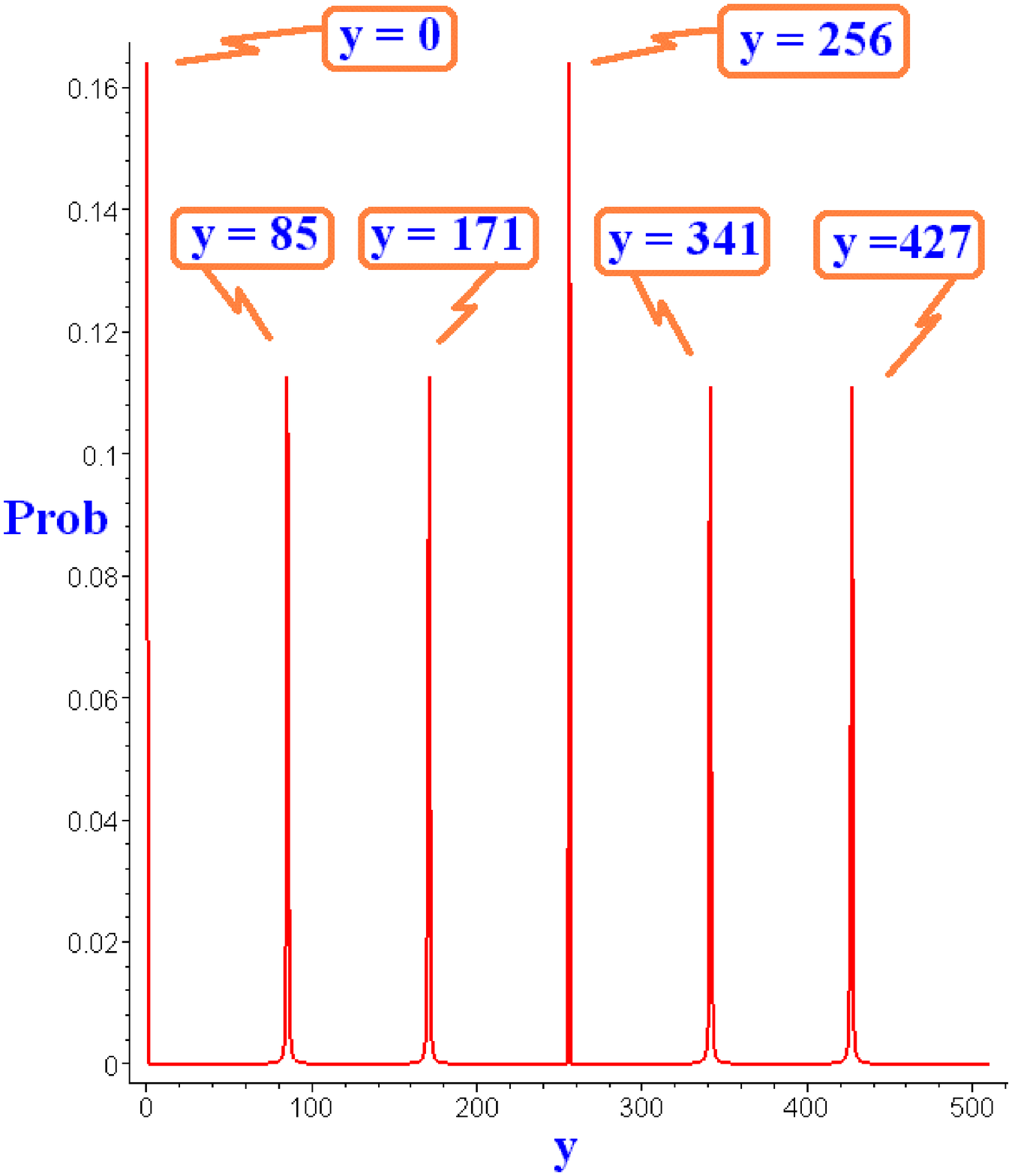}%
\\
\textbf{Figure 1. \ A plot of }$\mathbf{Prob}_{\widetilde{\mathbf{\varphi}}%
}\mathbf{(y)}$\textbf{.}%
\end{center}

\bigskip

The peaks in the above plot of $Prob_{\widetilde{\varphi}}\left(  y\right)  $
occur at the integers
\[
y=0,\ 85,\ 171,\ 256,\ 341,\ 427\text{.}%
\]
The probability that at least one of these six integers will occur is quite
high. \ It is actually $0.78^{+}$. \ Indeed, the probability distribution has
been intentionally engineered to make the probability of these particular
integers as high as possible. \ And there is a good reason for doing so.

\bigskip

The above six integers are those for which the corresponding rational $y/Q$ is
``closest'' to a rational of the form $d/P$. \ By ``closest'' we mean that
\[
\left|  \frac{y}{Q}-\frac{d}{P}\right|  <\frac{1}{2Q}<\frac{1}{2P^{2}}\text{
.}%
\]
In particular,
\[
\frac{0}{512},\ \frac{85}{512},\ \frac{171}{512},\frac{256}{512}%
,\ \frac{341}{512},\ \frac{427}{512}%
\]
are rationals respectively ``closest'' to the rationals
\[
\frac{0}{6},\ \frac{1}{6},\ \frac{2}{6},\ \frac{3}{6},\ \frac{4}{6}%
,\ \frac{5}{6}\text{ .}%
\]
So by theorem \ref{CFConvergent} of Appendix A, the six rational numbers
$0/6,\ 1/6,\ \ldots\ ,\ 5/6$ are convergents of the continued fraction
expansions of $0/512,\ 85/512,\ \ldots\ ,\ 427/512$, respectively. \ Hence,
each of the six rationals $0/6,\ 1/6,\ \ldots\ ,\ 5/6$ can be found with the
recursion given in Appendix A.

\bigskip

But ... , we are not searching for rationals of the form $d/P$. \ Instead, we
seek only the denominator $P=6$. \ 

\bigskip

Unfortunately, the denominator $P=6$ can only be gotten from the continued
fraction recursion when the numerator and denominator of $d/P$ are relatively
prime. \ Given that the algorithm has selected one of the random integers
$0,\ 85,\ \ldots\ ,\ 427$, the probability that the corresponding rational
$d/P$ has relatively prime numerator and denominator is $\phi\left(  6\right)
/6=1/3$, where $\phi\left(  -\right)  $ denotes the Euler totient function.
\ So the probability of finding $P=6$ is actually not $0.78^{+}$, but is
instead $0.23^{-}$.

\bigskip

From Peter Shor's perspective, the expression for the probability distribution
is not known, since the period $P$ is not known. \ All that Peter sees is a
random integer $y$ produced by the probability distribution $Prob_{\widetilde
{\varphi}}$. \ However, he does know an approximate lower bound for the
probability that the random $y$ produced by $Prob_{\widetilde{\varphi}}$ is a
``closest'' one, namely the approximate lower bound $4/\pi^{2}=0.41^{-}$.
\ Also, because\footnote{Please refer to reference \cite[Theorem 328, Section
18.4]{Hardy1}.}
\[
\lim\inf\frac{\phi(N)\ln\ln N}{N}=e^{-\gamma}\text{ ,}%
\]
where $\gamma=0.5772\cdots$ denotes Euler's constant, he knows that
\[
\frac{\phi\left(  P\right)  }{P}=\Omega\left(  \frac{1}{\lg\lg N}\right)
\text{ .}%
\]
Hence, if he repeats the algorithm $O\left(  \lg\lg N\right)  $
times\footnote{For even tighter asymptotic bounds, please refer to
\cite{Cleve1} and \cite{Preskill1}.}, he will obtain one of the desired
integers $y$ with probability bounded below by approximately $4/\pi^{2}$.

\bigskip

However, once he has in his possession a candidate $P^{\prime}$ for the actual
period $P=6$, the only way he can be sure he has the correct period $P$ is to
test $P^{\prime}$ by computing $2^{P^{\prime}}\operatorname{mod}21$. \ If the
result is $1$, he is certain he has found the correct period $P$. \ This last
part of the computation is done by the repeated squaring algorithm\footnote{By
the repeated squaring algorithm, we mean the algorithm which computes
$a^{P^{\prime}}\operatorname{mod}N$ via the expression
\[
a^{P^{\prime}}=\prod_{j}\left(  a^{2^{j}}\right)  ^{P_{j}^{\prime}}\text{,}%
\]
where $P^{\prime}=\sum_{j}P_{j}^{\prime}2^{j}$ is the radix 2 expansion of
$P^{\prime}$.}.

\bigskip

\bigskip

\bigskip

\section{Definition of the hidden subgroup problem (HSP) and hidden subgroup
algorithms (HSAs)}

\qquad\bigskip

We now proceed by defining what is meant by a hidden subgroup problem (HSP)
and a corresponding hidden subgroup algorithm. \ For other perspectives on
HSPs, please refer to \cite{Kitaev1}, \cite{Jozsa4}, \cite{Mosca1}.

\bigskip

\begin{definition}
A map $\varphi:A\longrightarrow S$ from a group $A$ into a set $S$ is said to
have \textbf{hidden subgroup structure} if there exists a subgroup
$K_{\varphi}$ of $A$, called a \textbf{hidden subgroup}, and an injection
$\iota_{\varphi}:A/K_{\varphi}\longrightarrow S$, called a \textbf{hidden
injection}, such that the diagram
\[%
\begin{array}
[c]{ccc}%
A\  & \overset{\varphi}{\longrightarrow} & S\\
\nu\searrow &  & \nearrow\iota_{\varphi}\\
& A/K_{\varphi} &
\end{array}
\]
is commutative, where $A/K_{\varphi}$ denotes the collection of right cosets
of $K_{\varphi}$ in $A$, and where $\nu:A\longrightarrow A/K_{\varphi}$ is the
natural map of $A$ onto $A/K_{\varphi}$. \ We refer to the group $A$ as the
\textbf{ambient group} and to the set $S$ as the \textbf{target set}. \ If
$K_{\varphi}$ is a normal subgroup of $A$, then $H_{\varphi}=A/K_{\varphi}$ is
a group, called the \textbf{hidden quotient group}, and $\nu:A\longrightarrow
A/K_{\varphi}$ is an epimorphism, called the \textbf{hidden epimorphism}.
\end{definition}

\bigskip

The hidden subgroup problem can be expressed as follows:

\bigskip

\begin{problem}
[\textbf{Hidden Subgroup Problem (HSP)}]Given a map with hidden subgroup
structure
\[
\varphi:A\longrightarrow S\text{ ,}%
\]
determine a hidden subgroup $K_{\varphi}$ of $A$. \ \ An algorithm solving
this problem is called a \textbf{hidden subgroup algorithm (HSA)}.
\end{problem}

\bigskip

The corresponding quantum form of this HSP is stated as follows:

\bigskip

\begin{problem}
[\textbf{Hidden Subgroup Problem: Quantum Version}]Let
\[
\varphi:A\longrightarrow S
\]
be a map with hidden subgroup structure. \ Construct a quantum implementation
of the map $\varphi$ as follows: \ \medskip

Let $\mathcal{H}_{A}$ and $\mathcal{H}_{S}$ be Hilbert spaces defined
respectively by the orthonormal bases
\[
\left\{  \ \left|  a\right\rangle \mid a\in A\ \right\}  \text{ and }\left\{
\ \left|  s\right\rangle \mid s\in S\ \right\}  \text{ ,}%
\]
and let $s_{0}=\varphi\left(  0\right)  $, where $0$ denotes the identity of
the ambient group $A$. \ Finally, let $U_{\varphi}$ be the unitary
transformation
\[%
\begin{array}
[c]{ccc}%
U_{\varphi}:\mathcal{H}_{A}\otimes\mathcal{H}_{S} & \longrightarrow &
\mathcal{H}_{A}\otimes\mathcal{H}_{S}\\
&  & \\
\left|  a\right\rangle \left|  s_{0}\right\rangle  & \longmapsto & \left|
a\right\rangle \left|  \varphi\left(  a\right)  \right\rangle
\end{array}
,
\]

Determine the hidden subgroup \ $K_{\varphi}$ with bounded probability of
error by making as few queries as possible of the blackbox $U_{\varphi}$. \ A
quantum algorithm solving this problem is called a \textbf{quantum hidden
subgroup algorithm (QHSA).}
\end{problem}

\bigskip

In this paper, we focus on the \textbf{abelian hidden subgroup problem
(AHSP)}, i.e., the HSP with the ambient group $A$ assumed to be a finitely
generated abelian group, and where the image of the hidden morphism $\varphi$
is a \textbf{finite subset} of $S$. \ (We will also on occasion assume that
the entire set $S$ is finite.)

\bigskip

In this paper we focus on the following two classes of abelian hidden subgroup
problems:\footnote{For the general abelian HSP, please refer to \cite{Cheung1}
and \cite{Kitaev1}.}

\bigskip

\begin{itemize}
\item \textbf{Vintage Simon AHSP.} \ The ambient group $A$ is finite and abelian.\medskip

\item \textbf{Vintage Shor AHSP.} \ The ambient group $A$ is free abelian of
finite rank.
\end{itemize}

\bigskip

\noindent\fbox{\textsc{Notation Convention}} For notational simplicity,
\textit{throughout this paper we will use additive notation for both the
ambient group }$A$\textit{ and the hidden subgroup }$K_{\varphi}$\textit{, and
multiplicative notation for the hidden quotient group }$H_{\varphi
}=A/K_{\varphi}$.\footnote{This follows the notational convention found in
\cite{Shor1}.}

\bigskip

\part{Algebraic Preliminaries\bigskip}

\bigskip

\section{The Character Group}

\bigskip

Let $G$ be an abelian group\label{CharacterGroupSection}. Then the
\textbf{character group} (or, \textbf{dual group)} $\widehat{G}$ of $G$ is
defined as the group of all morphisms of $G$ into the group $\mathbb{S}^{1}$,
i.e.,
\[
\widehat{G}=Hom\left(  G,\mathbb{S}^{1}\right)
\]
where $\mathbb{S}^{1}$ denotes the group of orientation preserving symmetries
of the standard circle, and where multiplication on $\widehat{G}$ is defined
as:
\[
\left(  f_{1}f_{2}\right)  \left(  g\right)  =f_{1}\left(  g\right)
f_{2}\left(  g\right)  \quad\text{for all }f_{1},f_{2}\in\widehat{G}%
\]
The elements of $\widehat{G}$ are called \textbf{characters}.\footnote{More
generally, for non-abelian groups, a character is defined as the trace of a
representation of the group.} \bigskip

\begin{remark}
The group $\mathbb{S}^{1}$ can be identified with

\begin{itemize}
\item[\textbf{1)}] The multiplicative group $U(1)=\left\{  \quad e^{2\pi
ix}\quad\mid\quad x\in\mathbb{R}\quad\right\}  $, i.e., with multiplication
defined by $e^{2\pi i\alpha}\cdot e^{2\pi i\beta}=e^{2\pi i\left(
\alpha+\beta\right)  }\bigskip$

\item[\textbf{2)}] The additive group $2\pi\mathbb{R}/2\pi\mathbb{Z}$, i.e.,
the reals modulo $2\pi$ under addition, i.e., with addition defined as
\[
2\pi\alpha+2\pi\beta\ \operatorname{mod}2\pi=2\pi\left(  \alpha+\beta
\ \operatorname{mod}1\right)
\]
\end{itemize}
\end{remark}

\bigskip

\begin{remark}
Please note that the 1-sphere $\mathbb{S}^{1}$ can be thought of as a
$\mathbb{Z}$-module under the action
\[
\left(  n,2\pi\alpha\right)  \longmapsto2\pi\left(  n\alpha\operatorname{mod}%
1\right)
\]
\end{remark}

\bigskip

\begin{theorem}
Every finite abelian group $G$ is isomorphic to the direct product of cyclic
groups, i.e.,\label{AbelianGroupTheorem}
\[
G\cong\mathbb{Z}_{m_{1}}\mathbb{\times Z}_{m_{2}}\mathbb{\times\ \ldots
\ \times Z}_{m_{\ell}}\text{ ,}%
\]
where $\mathbb{Z}_{m_{j}}$ denotes the cyclic group of order $m_{j}$.
\end{theorem}

\bigskip

\begin{theorem}
Let $G$ be a finite abelian group. If $G=G_{1}\times G_{2}$, then $\widehat
{G}=\widehat{G}_{1}\times\widehat{G}_{2}$.
\end{theorem}

\bigskip

\begin{theorem}
$\widehat{\mathbb{Z}}_{m}\cong\mathbb{Z}_{m}$
\end{theorem}

\bigskip

\begin{corollary}
If $G$ is a finite abelian group, then $G\cong\widehat{G}$.
\end{corollary}

\bigskip

\begin{remark}
The isomorphism $G\cong\widehat{G}$ can be expressed more explicitly as follows:

Let $G\cong\mathbb{Z}_{m_{1}}\mathbb{\times Z}_{m_{2}}\mathbb{\times
\ \ldots\ \times Z}_{m_{\ell}}$, and let $g_{1},g_{2},\ \ldots\ ,g_{\ell}$
denote generators of $\mathbb{Z}_{m_{1}}$, $\mathbb{Z}_{m_{2}}$%
,$\mathbb{\ \ldots\ }$, $\mathbb{Z}_{m_{\ell}}$ respectively. Moreover, let
$\omega_{1}$, $\omega_{2}$, $\ldots\ $, $\omega_{\ell}$ be $m_{1}$-th, $m_{2}%
$-th, ... , $m_{\ell}$-th primitive roots of unity, respectively. Then the
character $\widetilde{\chi}_{j}$ of $\mathbb{Z}_{m_{j}}$ defined by
\[
\widetilde{\chi}_{j}(g_{j})=\omega_{j}%
\]
generates $\widehat{\mathbb{Z}}_{m_{j}}$ as a cyclic group, i.e., the powers
$\left(  \widetilde{\chi}_{j}\right)  ^{k}$ generate $\widehat{\mathbb{Z}%
}_{m_{j}}$. Moreover, the characters $\chi_{j}$ of $G$ defined by
\[
\chi_{j}=\left(  \prod\limits_{i=0}^{j-1}\widetilde{\chi}_{i}^{0}\right)
\widetilde{\chi}_{j}\left(  \prod\limits_{i=j+1}^{\ell}\widetilde{\chi}%
_{i}^{0}\right)
\]
generate $\widehat{G}$. It follows that an isomorphism $G\cong\widehat{G}$ is
given by
\[
g_{j}\longleftrightarrow\chi_{j}%
\]
\end{remark}

\bigskip

\noindent\fbox{\textsc{Notation Convention}} \ \textit{In general, we will not
need to represent the isomorphism }$G\cong\widehat{G}$\textit{ as explicitly
as stated above. We will use the following convention. Let }$\left\{
g_{1},g_{2},\ \ldots\ ,g_{\ell}\right\}  $\textit{ and }$\left\{  \chi
_{1},\chi_{2},\ \ldots\ ,\chi_{\ell},\right\}  $\textit{ denote respectively
the set of elements of }$G$\textit{ and }$\widehat{G}$\textit{ indexed in such
a way that }%
\[
g_{j}\longleftrightarrow\chi_{j}%
\]
\textit{is the chosen isomorphism of }$G$\textit{ and }$\widehat{G}$\textit{.
\ We will at times use the notation }%
\[
\left\{
\begin{array}
[c]{c}%
g\longleftrightarrow\chi_{g}\\
\\
\chi\longleftrightarrow g_{\chi}%
\end{array}
\right.
\]

\bigskip

\section{Fourier analysis on a finite abelian group}

\bigskip\label{FourierSection}

As in the previous section, let $G$ be a \textbf{finite abelian group}%
\footnote{If $G$ is infinite, then ring multiplication `$\bullet$' is not
always well defined. \ So $\mathbb{C}G$ is not a ring, but a $\mathbb{Z}%
$-module, with a group of operators. \ One way of of making $\mathbb{C}G$ into
a ring, is to restrict the maps on $G$, e.g., to maps with compact support, to
maps with $L^{2}$ norm, etc.} and let $\widehat{G}$ denote is character group.
\ Let $g$ and $\chi$ denote respectively elements of the groups $G$ and
$\widehat{G}$.

\bigskip

Let $\mathbb{C}G$ and $\mathbb{C}\widehat{G}$ denote the corresponding group
algebras of $G$ and $\widehat{G}$ over the complex numbers $\mathbb{C}$.
\ Hence, $\mathbb{C}G$ consists of all maps $f:G\longrightarrow\mathbb{C}$.
\ Addition `$+$', multiplication `$\bullet$', and scalar multiplication are
defined as:
\[
\left\{
\begin{array}
[c]{lllll}%
\left(  f_{1}+f_{2}\right)  (g) & = & f_{1}(g)+f_{2}(g) & \forall g\in G & \\
&  &  &  & \\
\left(  f_{1}\bullet f_{2}\right)  (g) & = &
{\displaystyle\sum\limits_{h\in G}}
f_{1}(h)f_{2}(h^{-1}g) & \forall g\in G & \text{(\textbf{Convolution})}\\
&  &  &  & \\
\left(  \lambda f\right)  (g) & = & \lambda f(g) & \forall\lambda\in
\mathbb{C}\text{ and }\forall g\in G &
\end{array}
\right.
\]

\bigskip

\noindent\textbf{Caveat.} \ \textit{Please note that the symbol }$g$\textit{
has at least three different meanings:}

\begin{itemize}
\item[$\blacklozenge$] \textsc{Interpretation 1.} \ The symbol $g$\textit{
denotes an element of the group }$G$\medskip

\item[$\blacklozenge$] \textsc{Interpretation 2.} \ The symbol $g$\textit{
denotes a pointwise map}
\[
g:G\longrightarrow\mathbb{C}%
\]
\textit{defined by}
\[
g(g^{\prime})=\left\{
\begin{array}
[c]{cl}%
1 & \text{if }g=g^{\prime}\\
& \\
0 & \text{otherwise}%
\end{array}
\right.
\]
Thus,
\[
f=%
{\displaystyle\sum\limits_{g\in G}}
f(g)g\text{ denotes }g\longmapsto f(g)
\]
\textit{Hence, }$g\in\mathbb{C}G$\textit{. \ Since }$G$\textit{ is isomorphic
as a group to the set of pointwise maps }$\left\{  g:G\longrightarrow
\mathbb{C}\mid g\in G\right\}  $\textit{ under convolution, we can and do
identify the group elements of }$G$\textit{ with the pointwise maps }%
$g\in\mathbb{C}G$\textit{. \ Thus, }\textsc{interpretations 1}\textit{ and
}\textsc{2}\textit{ lead to no ambiguity at the algebraic level.}\medskip

\item[$\blacklozenge$] \textsc{Interpretation 3.} \ The symbol $g$\textit{
denotes a character of }$\widehat{G}$\textit{ defined by}
\[
g\left(  \chi\right)  =\chi\left(  g\right)
\]
\textit{Thus, with this interpretation, }$g\in\widehat{\widehat{G}}%
\subset\mathbb{C}\widehat{\widehat{G}}$\textit{. \ This third interpretation
can, in some instances, lead to some unnecessary confusion. \ When this
intended interpretation is possibly not clear from context, we will resort to
the notation} \
\[
g^{\bullet}%
\]
\textit{for }\textsc{interpretation 3}\textit{ of the symbol }$g$\textit{.
Thus, for example, }%
\[
f^{\bullet}=%
{\displaystyle\sum\limits_{g\in G}}
f(g)g^{\bullet}\text{ denotes the map }\chi\longmapsto%
{\displaystyle\sum\limits_{g\in G}}
f(g)\chi\left(  g\right)
\]
\end{itemize}

\bigskip

In like manner, \textit{the symbol }$\chi$\textit{ has at least three
different meanings:}

\begin{itemize}
\item \textsc{Interpretation }$\widehat{\text{\textsc{1}}}$\textsc{.} \ The
symbol $\chi$\textit{ denotes an element of the group }$\widehat{G}$\medskip

\item[$\blacklozenge$] \textsc{Interpretation }$\widehat{\text{\textsc{2}}}%
$\textsc{.} \ The symbol $\chi$\textit{ denotes a pointwise map}
\[
\chi:\widehat{G}\longrightarrow\mathbb{C}%
\]
\textit{defined by}
\[
\chi(\chi^{\prime})=\left\{
\begin{array}
[c]{cl}%
1 & \text{if }\chi=\chi^{\prime}\\
& \\
0 & \text{otherwise}%
\end{array}
\right.
\]
Thus,
\[
\widehat{f}=%
{\displaystyle\sum\limits_{\chi\in\widehat{G}}}
\widehat{f}(\chi)\chi\text{ denotes }\chi\longmapsto\widehat{f}(\chi)
\]
\textit{Hence, }$\chi\in\mathbb{C}\widehat{G}$\textit{. \ Since }$\widehat{G}%
$\textit{ is isomorphic as a group to the set of pointwise maps }$\left\{
\chi:\widehat{G}\longrightarrow\mathbb{C}\mid\chi\in\widehat{G}\right\}
$\textit{ under convolution, we can and do identify the group elements of
}$\widehat{G}$\textit{ with the pointwise maps }$\chi\in\mathbb{C}\widehat{G}%
$\textit{. \ Thus, }\textsc{interpretations }$\widehat{\text{\textsc{1}}}%
$\textit{ and }$\widehat{\text{\textsc{2}}}$\textit{ lead to no ambiguity at
the algebraic level.}\medskip

\item[$\blacklozenge$] \textsc{Interpretation }$\widehat{\text{\textsc{3}}}%
$\textsc{.} \ The symbol $\chi$\textit{ denotes a character map of }%
$G$\textit{ onto }$\mathbb{C}$ \textit{defined by}
\[
g\longmapsto\chi\left(  g\right)
\]
\textit{Thus, with this interpretation, }$\chi\in\mathbb{C}G$\textit{. \ This
third interpretation can, in some instances, \ also lead to some unnecessary
confusion. \ When this intended interpretation is possibly not clear from
context, we will resort to the notation} \
\[
\chi^{\bullet}%
\]
\textit{for }\textsc{interpretation }$\widehat{\text{\textsc{3}}}$\textit{ of
the symbol }$\chi$\textit{. Thus, for example, }%
\[
\widehat{f}^{\bullet}=%
{\displaystyle\sum\limits_{\chi\in\widehat{G}}}
\widehat{f}(\chi)\chi^{\bullet}\text{ denotes the map }g\longmapsto%
{\displaystyle\sum\limits_{\chi\in\widehat{G}}}
\widehat{f}(\chi)\chi\left(  g\right)
\]
\end{itemize}

\bigskip

We define complex inner products on the group algebras $\mathbb{C}G$ and
$\mathbb{C}\widehat{G}$ as follows:
\[
\left\{
\begin{array}
[c]{llll}%
\left(  f_{1},f_{2}\right)  & = & \frac{1}{\left|  G\right|  }%
{\displaystyle\sum\limits_{g\in G}}
f_{1}(g)\overline{f_{2}(g)} & \forall f_{1},f_{2}\in\mathbb{C}G\\
&  &  & \\
\left(  \widehat{f}_{1},\widehat{f}_{2}\right)  & = & \frac{1}{\left|
G\right|  }%
{\displaystyle\sum\limits_{\chi\in\widehat{G}}}
\widehat{f}_{1}(\chi)\overline{\widehat{f}_{2}(\chi)} & \forall\widehat{f}%
_{1},\widehat{f}_{2}\in\mathbb{C}\widehat{G}%
\end{array}
\right.
\]
where $\overline{f_{2}(g)}$ and $\overline{\widehat{f}_{2}(\chi)}$ denote
respectively the complex conjugates of $f_{2}(g)$ and $\widehat{f}_{2}(\chi)$.

\bigskip

The corresponding norms are defined as
\[
\left\{
\begin{array}
[c]{llll}%
\left\|  f\right\|  & = & \sqrt{\left(  f,f\right)  } & \forall f\in
\mathbb{C}G\\
&  &  & \\
\left\|  \widehat{f}\right\|  & = & \sqrt{\left(  \widehat{f},\widehat
{f}\right)  } & \forall\widehat{f}\in\mathbb{C}\widehat{G}%
\end{array}
\right.
\]

\bigskip

As an immediate consequence of the above definitions, we have:
\[%
\begin{array}
[c]{lll}%
\left(  g_{1},g_{2}\right)  & = & \left\{
\begin{array}
[c]{ll}%
1 & \text{if }g_{1}=g_{2}\\
& \\
0 & \text{otherwise}%
\end{array}
\right.
\end{array}
\text{ \quad\quad and \quad\quad}%
\begin{array}
[c]{lll}%
\left(  \chi_{1},\chi_{2}\right)  & = & \left\{
\begin{array}
[c]{ll}%
1 & \text{if }\chi_{1}=\chi_{2}\\
& \\
0 & \text{otherwise}%
\end{array}
\right.
\end{array}
\]
It also follows from the standard character identities that
\[%
\begin{array}
[c]{lll}%
\left(  g_{1}^{\bullet},g_{2}^{\bullet}\right)  & = & \left\{
\begin{array}
[c]{ll}%
1 & \text{if }g_{1}^{\bullet}=g_{2}^{\bullet}\\
& \\
0 & \text{otherwise}%
\end{array}
\right.
\end{array}
\text{ \quad\quad and \quad\quad}%
\begin{array}
[c]{lll}%
\left(  \chi_{1}^{\bullet},\chi_{2}^{\bullet}\right)  & = & \left\{
\begin{array}
[c]{ll}%
1 & \text{if }\chi_{1}^{\bullet}=\chi_{2}^{\bullet}\\
& \\
0 & \text{otherwise}%
\end{array}
\right.
\end{array}
\]

\bigskip

We are now in a position to define the Fourier transform on a finite abelian
group $G$.

\bigskip

\begin{definition}
The \textbf{Fourier transform} $\mathcal{F}$ for a finite abelian group $G$ is
defined as
\[%
\begin{array}
[c]{ccl}%
\mathbb{C}G & \overset{\mathcal{F}}{\longrightarrow} & \qquad\qquad
\mathbb{C}\widehat{G}\\
f & \longmapsto & \widehat{f}=\frac{1}{\sqrt{\left|  G\right|  }}%
{\displaystyle\sum\limits_{g\in G}}
f(g)\overline{g^{\bullet}}=\frac{1}{\sqrt{\left|  G\right|  }}%
{\displaystyle\sum\limits_{\chi\in\widehat{G}}}
\left(
{\displaystyle\sum\limits_{g\in G}}
f(g)\overline{\chi}\left(  g\right)  \right)  \chi
\end{array}
\]
Hence,
\[
\widehat{f}\left(  \chi\right)  =\sqrt{\left|  G\right|  }\left(
f,\chi^{\bullet}\right)  =\frac{1}{\sqrt{\left|  G\right|  }}%
{\displaystyle\sum\limits_{g\in G}}
f(g)\overline{\chi\left(  g\right)  }%
\]
\end{definition}

\bigskip

\begin{proposition}%
\[
f=\frac{1}{\sqrt{\left|  G\right|  }}%
{\displaystyle\sum\limits_{\chi\in\widehat{G}}}
\widehat{f}(\chi)\chi^{\bullet}%
\]
\end{proposition}

\begin{proof}%
\begin{align*}
\frac{1}{\sqrt{\left|  G\right|  }}%
{\displaystyle\sum\limits_{\chi\in\widehat{G}}}
\widehat{f}(\chi)\chi(g_{0})  &  =\frac{1}{\sqrt{\left|  G\right|  }}%
{\displaystyle\sum\limits_{\chi\in\widehat{G}}}
\frac{1}{\sqrt{\left|  G\right|  }}%
{\displaystyle\sum\limits_{g\in G}}
f(g)\overline{\chi\left(  g\right)  }\chi(g_{0})\\
& \\
&  =\frac{1}{\left|  G\right|  }%
{\displaystyle\sum\limits_{g\in G}}
f(g)%
{\displaystyle\sum\limits_{\chi\in\widehat{G}}}
\overline{\chi\left(  g\right)  }\chi(g_{0})=f(g_{0})
\end{align*}
\end{proof}

We define the inverse Fourier transform as follows:

\bigskip

\begin{definition}
The inverse Fourier transform $\mathcal{F}^{-1}$ is defined as
\[%
\begin{array}
[c]{ccl}%
\mathbb{C}\widehat{G} & \overset{\mathcal{F}^{-1}}{\longrightarrow} &
\qquad\qquad\mathbb{C}G\\
\widehat{f} & \longmapsto & f=\frac{1}{\sqrt{\left|  G\right|  }}%
{\displaystyle\sum\limits_{\chi\in\widehat{G}}}
\widehat{f}(\chi)\chi^{\bullet}%
\end{array}
\]
Hence,
\[
f\left(  g\right)  =\sqrt{\left|  G\right|  }\left(  \widehat{f}%
,\overline{g^{\bullet}}\right)  =\frac{1}{\sqrt{\left|  G\right|  }}%
{\displaystyle\sum\limits_{\chi\in\widehat{G}}}
\widehat{f}(\chi)\chi\left(  g\right)
\]
\end{definition}

\bigskip

\begin{theorem}
[Plancherel identity]%
\[
\left\|  f\right\|  =\left\|  \widehat{f}\right\|
\]
\end{theorem}

\begin{proof}%
\begin{align*}
\left\|  f\right\|  ^{2}  &  =\left(  f,f\right)  =\frac{1}{\left|  G\right|
}%
{\displaystyle\sum\limits_{g\in G}}
\left|  f(g)\right|  ^{2}\\
&  =\frac{1}{\left|  G\right|  }%
{\displaystyle\sum\limits_{g\in G}}
\frac{1}{\sqrt{\left|  G\right|  }}%
{\displaystyle\sum\limits_{\chi\in\widehat{G}}}
\widehat{f}(\chi)\chi\left(  g\right)  \overline{\left(  \frac{1}%
{\sqrt{\left|  G\right|  }}%
{\displaystyle\sum\limits_{\chi^{\prime}\in\widehat{G}}}
\widehat{f}(\chi^{\prime})\chi^{\prime}\left(  g\right)  \right)  }\\
& \\
&  =\frac{1}{\left|  G\right|  ^{2}}%
{\displaystyle\sum\limits_{g\in G}}
{\displaystyle\sum\limits_{\chi\in\widehat{G}}}
{\displaystyle\sum\limits_{\chi^{\prime}\in\widehat{G}}}
\widehat{f}(\chi)\overline{\widehat{f}(\chi^{\prime})}\chi\left(  g\right)
\overline{\chi^{\prime}}\left(  g\right) \\
& \\
&  =\frac{1}{\left|  G\right|  ^{2}}%
{\displaystyle\sum\limits_{\chi\in\widehat{G}}}
{\displaystyle\sum\limits_{\chi^{\prime}\in\widehat{G}}}
\widehat{f}(\chi)\overline{\widehat{f}(\chi^{\prime})}\left(
{\displaystyle\sum\limits_{g\in G}}
\chi\left(  g\right)  \overline{\chi^{\prime}}\left(  g\right)  \right) \\
& \\
&  =\frac{1}{\left|  G\right|  }%
{\displaystyle\sum\limits_{\chi\in\widehat{G}}}
\left|  \widehat{f}(\chi)\right|  ^{2}=\left\|  \widehat{f}\right\|  ^{2}%
\end{align*}
\end{proof}

\bigskip

\section{Implementation issues: Group algebras as Hilbert spaces}

\qquad\bigskip

For implementation purposes, we will need to view group algebras also as
Hilbert spaces.\footnote{Category theorists will recognize this as a forgetful functor.}

\bigskip

In particular, $\mathbb{C}G$ and $\mathbb{C}\widehat{G}$ can be respectively
viewed as the Hilbert spaces $\mathcal{H}_{G}$ and $\mathcal{H}_{\widehat{G}}$
defined by the respective orthonormal bases
\[
\left\{  \left|  g\right\rangle \mid g\in G\right\}  \text{ and }\left\{
\left|  \chi\right\rangle \mid\chi\in\widehat{G}\right\}  \text{.}%
\]

\bigskip

In this context, the \textbf{Fourier transform} $\mathcal{F}$ becomes
\[%
\begin{array}
[c]{ccl}%
\mathcal{H}_{G} & \overset{\mathcal{F}}{\longrightarrow} & \qquad
\qquad\mathcal{H}_{\widehat{G}}\\
\left|  f\right\rangle =%
{\displaystyle\sum\limits_{g\in G}}
f(g)\left|  g\right\rangle  & \longmapsto & \left|  \widehat{f}\right\rangle
=\frac{1}{\sqrt{\left|  G\right|  }}%
{\displaystyle\sum\limits_{\chi\in\widehat{G}}}
\left(
{\displaystyle\sum\limits_{g\in G}}
f(g)\overline{\chi}\left(  g\right)  \right)  \left|  \chi\right\rangle
\end{array}
\]
and the \textbf{inverse Fourier transform} $\mathcal{F}^{-1}$ becomes
\[%
\begin{array}
[c]{ccl}%
\mathcal{H}_{\widehat{G}} & \overset{\mathcal{F}^{-1}}{\longrightarrow} &
\qquad\qquad\mathcal{H}_{G}\\
\left|  \widehat{f}\right\rangle  & \longmapsto & \left|  f\right\rangle
=\frac{1}{\sqrt{\left|  G\right|  }}%
{\displaystyle\sum\limits_{g\in G}}
\left(
{\displaystyle\sum\limits_{\chi\in\widehat{G}}}
\widehat{f}(\chi)\chi\left(  g\right)  \right)  \left|  g\right\rangle
\end{array}
\]

\vspace{0.4in}

One important and useful identification is to use the Hilbert space
isomorphism
\[%
\begin{array}
[c]{ccc}%
\mathcal{H}_{G} & \longleftrightarrow & \mathcal{H}_{\widehat{G}}\\
\left|  g\right\rangle  & \longleftrightarrow & \left|  \chi_{g}\right\rangle
\\
\left|  g_{\chi}\right\rangle  & \longleftrightarrow & \left|  \chi
\right\rangle
\end{array}
\]
to identify the two Hilbert spaces $\mathcal{H}_{G}$\ and $\mathcal{H}%
_{\widehat{G}}$. \ As a result, the Fourier transform $\mathcal{F}$ and it's
inverse $\mathcal{F}^{-1}$ can both be viewed as transforms taking the Hilbert
space $\mathcal{H}_{G}$ to itself, i.e.,
\[
\mathcal{H}_{G}%
\begin{array}
[c]{c}%
\overset{\mathcal{F}}{\longrightarrow}\\
\underset{\mathcal{F}^{-1}}{\longleftarrow}%
\end{array}
\mathcal{H}_{G}%
\]

\begin{remark}
This last identification is crucial for the implementation of hidden subgroup algorithms.
\end{remark}

\bigskip

\part{\textsc{QRand}$_{\varphi}()$: The Progenitor of All QHSAs\bigskip}

\bigskip

\section{Implementing $Prob_{\varphi}\left(  \chi\right)  $ with quantum
subroutine \textsc{QRand}$_{\varphi}()$}

\bigskip\label{Qrand}

Let
\[
\varphi:A\longrightarrow S
\]
be a map from a finite abelian group $A$ into a finite set $S$.

\bigskip

We use additive notation for the group $A$; and let $s_{0}=\varphi\left(
0\right)  $ denote the image of the identity $0$ of $A$ under the map
$\varphi$.

\bigskip

Let $\mathcal{H}_{A}$, $\mathcal{H}_{\widehat{A}}$, and $\mathcal{H}_{S}$
denote the Hilbert spaces respectively defined by the orthonormal bases
\[
\left\{  \ \overset{}{\left|  a\right\rangle }\mid a\in A\ \right\}  \text{,
}\left\{  \ \overset{}{\left|  \chi\right\rangle }\mid\overset{}{\chi}%
\in\widehat{A}\ \right\}  \text{, and }\left\{  \ \overset{}{\left|
s\right\rangle }\mid s\in S\ \right\}  \text{ .}%
\]

\bigskip

We assume that we are given a quantum system which implements the unitary
transformation $U_{\varphi}$ defined by
\[%
\begin{array}
[c]{ccc}%
\mathcal{H}_{A}\otimes\mathcal{H}_{S} & \overset{U_{\varphi}}{\longrightarrow}
& \mathcal{H}_{A}\otimes\mathcal{H}_{S}\\
&  & \\
\left|  a\right\rangle \left|  s_{0}\right\rangle  & \longmapsto & \left|
a\right\rangle \left|  \varphi\left(  a\right)  \right\rangle
\end{array}
\]

\bigskip

We will use the above implementation to construct a quantum subroutine
\textsc{QRand}$_{\varphi}\left(  {}\right)  $ which produces a probability
distribution
\[
Prob_{\varphi}:\widehat{A}\longrightarrow\left[  0,1\right]
\]
on the character group $\widehat{A}$ of the group $A$.

\bigskip

Before doing so, we will, as explained in the previous section, make use of
various identifications, such as respectively identifying the Fourier and
inverse Fourier transforms $\mathcal{F}_{A}$ and $\mathcal{F}_{A}^{-1}$ on the
group $A$
\[
\mathbb{C}A=\mathcal{H}_{A}%
\begin{array}
[c]{c}%
\overset{\mathcal{F}_{A}}{\longrightarrow}\\
\underset{\mathcal{F}_{A}^{-1}}{\longleftarrow}%
\end{array}
\mathcal{H}_{\widehat{A}}=\mathbb{C}\widehat{A}%
\]
with
\[
\mathbb{C}A=\mathcal{H}_{A}%
\begin{array}
[c]{c}%
\overset{\mathcal{F}_{A}}{\longrightarrow}\\
\underset{\mathcal{F}_{A}^{-1}}{\longleftarrow}%
\end{array}
\mathcal{H}_{A}=\mathbb{C}A
\]

\vspace{0.4in}

\begin{center}
\fbox{\textsc{Quantum Subroutine QRand}$_{\varphi}()$}
\end{center}

\bigskip

\begin{itemize}
\item[\fbox{\textbf{Step 0.}}] Initialization
\[
\left|  \psi_{0}\right\rangle =\left|  0\right\rangle \left|  s_{0}%
\right\rangle
\]
\bigskip

\item[\fbox{\textbf{Step 1.}}] Application of the inverse Fourier transform
$\mathcal{F}_{A}^{-1}$ of $A$
\[
\left|  \psi_{1}\right\rangle =\left(  \mathcal{F}_{A}^{-1}\otimes
1_{S}\right)  \left|  \psi_{0}\right\rangle =\frac{1}{\sqrt{\left|  A\right|
}}%
{\displaystyle\sum\limits_{a\in A}}
\left|  a\right\rangle \left|  s_{0}\right\rangle
\]
where $\left|  A\right|  $ denotes the cardinality of the group $A$.\bigskip

\item[\fbox{\textbf{Step 2.}}] Application of the unitary transformation
$U_{\varphi}$
\[
\left|  \psi_{2}\right\rangle =U_{\varphi}\left|  \psi_{1}\right\rangle
=\frac{1}{\sqrt{\left|  A\right|  }}%
{\displaystyle\sum\limits_{a\in A}}
\left|  a\right\rangle \left|  \varphi\left(  a\right)  \right\rangle
\]
\bigskip

\item[\fbox{\textbf{Step 3.}}] Application of the Fourier transform
$\mathcal{F}_{A}$ of $A$
\begin{align*}
\left|  \psi_{3}\right\rangle  &  =\left(  \mathcal{F}_{A}^{-1}\otimes
1_{S}\right)  \left|  \psi_{2}\right\rangle =\frac{1}{\left|  A\right|  }%
{\displaystyle\sum\limits_{a\in A}}
{\displaystyle\sum\limits_{\chi\in\widehat{A}}}
\chi\left(  a\right)  \left|  \chi\right\rangle \left|  \varphi\left(
a\right)  \right\rangle \\
& \\
&  =%
{\displaystyle\sum\limits_{\chi\in\widehat{A}}}
\frac{\left\|  \left|  \varphi\left(  \chi^{\bullet}\right)  \right\rangle
\right\|  }{\left|  A\right|  }\left|  \chi\right\rangle \frac{\left|
\varphi\left(  \chi^{\bullet}\right)  \right\rangle }{\left\|  \left|
\varphi\left(  \chi^{\bullet}\right)  \right\rangle \right\|  }%
\end{align*}
where
\[
\left|  \varphi\left(  \chi^{\bullet}\right)  \right\rangle =%
{\displaystyle\sum\limits_{a\in A}}
\chi\left(  a\right)  \left|  \varphi(a)\right\rangle
\]
\bigskip
\end{itemize}

\begin{remark}
This notation is meant to be suggestive, since under the identification
$\mathcal{H}_{A}=\mathbb{C}A$ we have
\[
\left|  \varphi\left(  \chi^{\bullet}\right)  \right\rangle =%
{\displaystyle\sum\limits_{a\in A}}
\chi\left(  a\right)  \varphi\left(  a\right)  =\varphi\left(
{\displaystyle\sum\limits_{a\in A}}
\chi\left(  a\right)  a\right)  =\varphi\left(  \chi^{\bullet}\right)
\]
\bigskip
\end{remark}

\begin{itemize}
\item[\fbox{\textbf{Step 4.}}] Measurement of the left quantum register.
\ Thus, with probability
\[
Prob_{\varphi}\left(  \chi\right)  =\frac{\left\|  \varphi\left(
\chi^{\bullet}\right)  \right\|  ^{2}}{\left|  A\right|  ^{2}}%
\]
the character $\chi$ is the resulting measured value, and the quantum system
``collapses'' to the state
\[
\left|  \psi_{4}\right\rangle =\left|  \chi\right\rangle \frac{\left|
\varphi\left(  \chi^{\bullet}\right)  \right\rangle }{\left\|  \varphi\left(
\chi^{\bullet}\right)  \right\|  }%
\]
\bigskip

\item[\fbox{\textbf{Step 5.}}] \textsc{Output} the character $\chi$, and
\textsc{stop}.
\end{itemize}

\vspace{0.5in}

\begin{remark}
The quantum subroutine \textsc{QRand}$_{\varphi}()$ can also be viewed as a
subroutine with the state $\left|  \chi\right\rangle \left|  \varphi\left(
\chi^{\bullet}\right)  \right\rangle $ as a side effect.
\end{remark}

\bigskip

As a result of the above description of \textsc{QRand}$_{\varphi}()$, we have
the following theorem:

\bigskip

\begin{theorem}
Let
\[
\varphi:A\longrightarrow S
\]
be a map from a finite abelian group $A$ into a finite set $S$. \ Then the
quantum subroutine \textsc{QRand}$_{\varphi}()$ is an implementation of the
probability distribution $Prob_{\varphi}\left(  \chi\right)  $ on the group
$\widehat{A}$ of characters of $A$ given by
\[
Prob_{\varphi}\left(  \chi\right)  =\frac{\left\|  \varphi\left(
\chi^{\bullet}\right)  \right\|  ^{2}}{\left|  A\right|  ^{2}}\text{ ,}%
\]
for all $\chi\in\widehat{A}$, where $\chi^{\bullet}$ denotes
\[
\chi^{\bullet}=%
{\displaystyle\sum\limits_{a\in A}}
\chi\left(  a\right)  a\in\mathbb{C}A
\]
\end{theorem}

\bigskip

\begin{remark}
Please note that the above theorem is true whether or not the map
$\varphi:A\longrightarrow S$ has a hidden subgroup.
\end{remark}

\bigskip

We will, on occasion, refer to the probability distribution
\[
Prob_{\varphi}:\widehat{A}\longrightarrow\left[  0,1\right]
\]
on the character group $\widehat{A}$ as the \textbf{stochastic source
}$\mathcal{S}_{\varphi}\left(  \chi\right)  $ which produces a symbol $\chi
\in\widehat{A}$ with probability $Prob_{\varphi}\left(  \chi\right)  $. \ (See
\cite{Lomonaco1}.) \ Thus, \textsc{QRand}$_{\varphi}\left(  \chi\right)  $ is
an algorithmic implementation of the stochastic source $\mathcal{S}_{\varphi
}\left(  \chi\right)  $.

\bigskip

\part{Vintage Simon Algorithms\bigskip}

We now begin the development of the class of vintage Simon QHSAs. \ These are
QHSAs for which the ambient group $A$ is finite abelian. \ \bigskip

\section{Properties of the probability distribution $Prob_{\varphi}\left(
\chi\right)  $ when $\varphi$ has a hidden subgroup}

\bigskip

Let
\[
\varphi:A\longrightarrow S
\]
be a map from a finite abelian group $A$ to a set $S$. \ We now assume that
$\varphi$ has a hidden subgroup $K_{\varphi}$, and hence, a hidden quotient
group $H_{\varphi}=A/K_{\varphi}$. \ 

\bigskip

Let
\[
\nu:A\longrightarrow H_{\varphi}=A/K_{\varphi}%
\]
denote the corresponding natural epimorphism respectively. \ Then since
$Hom_{\mathbb{Z}}\left(  -,2\pi\mathbb{R}/2\pi\mathbb{Z}\right)  $ is a left
exact contravariant functor, the map
\[%
\begin{array}
[c]{rcl}%
\widehat{\nu}:\widehat{H_{\varphi}} & \longrightarrow & \widehat{A}\\
\eta & \longmapsto & \eta\circ\nu
\end{array}
\]
is a monomorphism\footnote{See \cite{Cartan1}.}.

\bigskip

Since $\widehat{\nu}$ is a monomorphism, each character $\eta$ of the hidden
quotient group $H_{\varphi}$ can be identified with a character $\chi$ of $A$
for which $\chi\left(  k\right)  =1$ for every element of $K_{\varphi}$. \ In
other words, $\widehat{H_{\varphi}}$ can be identified with all characters of
$A$ which are trivial on $K_{\varphi}$.

\bigskip

\begin{theorem}
Let\label{ProbTheorem}
\[
\varphi:A\longrightarrow S
\]
be a map from a finite abelian group $A$ into a finite set $S$. \ If there
exists a hidden subgroup $K_{\varphi}$ of $\varphi$, and hence a hidden
quotient group $H_{\varphi}=A/K_{\varphi}$ of $\varphi$, then the probability
distribution $Prob_{\varphi}\left(  \chi\right)  $ on $\widehat{A}$
implemented by the quantum subroutine \textsc{QRand}$_{\varphi}()$ is given
by
\[
Prob_{\varphi}\left(  \chi\right)  =\left\{
\begin{array}
[c]{cl}%
\frac{1}{\left|  H_{\varphi}\right|  } & \text{if }\chi\text{ }\in\text{
}\widehat{H_{\varphi}}\\
& \\
0 & \text{otherwise}%
\end{array}
\right.
\]
In other words, in this particular case, $Prob_{\varphi}\left(  \chi\right)  $
is nothing more than the uniform probability distribution on the character
group $\widehat{H_{\varphi}}$ of the hidden quotient group $H_{\varphi}$.
\end{theorem}

\begin{proof}
Since $\varphi$ has a hidden subgroup $K_{\varphi}$, there exists a hidden
injection
\[
\iota_{\varphi}:H_{\varphi}\longrightarrow S
\]
from the hidden quotient group $H_{\varphi}=A/K_{\varphi}$ to the set $S$ such
that the diagram
\[%
\begin{array}
[c]{ccc}%
\underset{\mathstrut}{A} & \overset{\varphi}{\longrightarrow} & S\\
\nu\downarrow\  & \underset{}{\overset{\mathstrut}{\nearrow}}\overset
{\mathstrut}{\iota_{\varphi}} & \\
\overset{\mathstrut}{H_{\varphi}} &  &
\end{array}
\]
is commutative, where $\nu:A\longrightarrow H_{\varphi}$ denotes the hidden
natural epimorphism of $A$ onto the quotient group $H_{\varphi}=A/K_{\varphi}$.

Next let
\[
\iota_{\nu}:H_{\varphi}\longrightarrow A
\]
be a transversal map of the subgroup $K_{\varphi}$ in $A$, i.e., a map such
that
\[
\nu\circ\iota_{\nu}=id_{H_{\varphi}}\text{ .}%
\]
In other words, $\iota_{\nu}$ sends each element $h$ of $H_{\varphi}$ to a
unique element of the coset $\varphi^{-1}\left(  h\right)  $.

Recalling that
\[
\varphi\left(  \chi^{\bullet}\right)  =%
{\displaystyle\sum\limits_{a\in A}}
\chi\left(  a\right)  \varphi\left(  a\right)  \text{ ,}%
\]
we have
\begin{align*}
\varphi\left(  \chi^{\bullet}\right)   &  =%
{\displaystyle\sum\limits_{a\in A}}
\chi\left(  a\right)  \iota_{\varphi}\nu a=%
{\displaystyle\sum\limits_{h\in H_{\varphi}}}
\left(
{\displaystyle\sum\limits_{k\in K_{\varphi}}}
\chi\left(  \iota_{v}h+k\right)  \right)  \iota_{\varphi}h\\
& \\
&  =%
{\displaystyle\sum\limits_{h\in H_{\varphi}}}
\chi\left(  \iota_{v}h\right)  \left(
{\displaystyle\sum\limits_{k\in K_{\varphi}}}
\chi\left(  k\right)  \right)  \iota_{\varphi}h=\left(
{\displaystyle\sum\limits_{k\in K_{\varphi}}}
\chi\left(  k\right)  \right)  \left(
{\displaystyle\sum\limits_{h\in H_{\varphi}}}
\chi\left(  \iota_{v}h\right)  \iota_{\varphi}h\right)
\end{align*}

Thus,
\begin{align*}
\left\|  \varphi\left(  \chi^{\bullet}\right)  \right\|  ^{2}  &  =\left|
{\displaystyle\sum\limits_{k\in K_{\varphi}}}
\chi\left(  k\right)  \right|  ^{2}\left\|
{\displaystyle\sum\limits_{h\in H_{\varphi}}}
\chi\left(  \iota_{v}h\right)  \iota_{\varphi}h\right\|  ^{2}\\
& \\
&  =\left|
{\displaystyle\sum\limits_{k\in K_{\varphi}}}
\chi\left(  k\right)  \right|  ^{2}%
{\displaystyle\sum\limits_{h\in H_{\varphi}}}
\left|  \chi\left(  \iota_{v}h\right)  \right|  ^{2}=\left|
{\displaystyle\sum\limits_{k\in K_{\varphi}}}
\chi\left(  k\right)  \right|  ^{2}\left|  H_{\varphi}\right|
\end{align*}

But by a standard character identity\footnote{See \cite{Fulton1}.}, we have
\[%
{\displaystyle\sum\limits_{k\in K_{\varphi}}}
\chi\left(  k\right)  =\left\{
\begin{array}
[c]{ll}%
\left|  K_{\varphi}\right|  =\left|  A\right|  /\left|  H_{\varphi}\right|  &
\text{if }\chi\in\widehat{H_{\varphi}}\\
& \\
0 & \text{otherwise}%
\end{array}
\right.
\]
Hence, it follows that
\[
Prob_{\varphi}\left(  \chi\right)  =\frac{\left\|  \varphi\left(
\chi^{\bullet}\right)  \right\|  ^{2}}{\left|  A\right|  ^{2}}=\left\{
\begin{array}
[c]{ll}%
\frac{1}{\left|  H_{\varphi}\right|  } & \text{if }\chi\in\widehat{H_{\varphi
}}\\
& \\
0 & \text{otherwise}%
\end{array}
\right.
\]
\end{proof}

\bigskip

\section{A Markov process $\mathcal{M}_{\varphi}$ induced by $Prob_{\varphi}$}

\bigskip

Before we can discuss the class of vintage Simon quantum hidden subgroup
algorithms, we need to develop the mathematical machinery to deal with the
following question:

\bigskip

\noindent\textbf{Question.} \ Let $\varphi:A\longrightarrow S$ be a map from a
finite abelian group $A$ to a finite set $S$. \ Assume that the map $\varphi$
has a hidden group $K_{\varphi}$, and hence a hidden quotient group
$H_{\varphi}$. \ From theorem \ref{ProbTheorem} of the previous section, we
know that the probability distribution
\[
Prob_{\varphi}:\widehat{A}\longrightarrow\left[  0,1\right]
\]
is effectively the uniform probability distribution on the character group
$\widehat{H_{\varphi}}$ of the hidden quotient group $H_{\varphi}$. \ How many
times do we need to query the probability distribution $Prob_{\varphi}$ to
obtain enough characters of $H_{\varphi}$ to generate the entire character
group $\widehat{H_{\varphi}}$?

\bigskip

We begin with a definition:

\bigskip

\begin{definition}
Let
\[
Prob_{G}:G\longrightarrow\lbrack0,1]
\]
be a probability distribution on a finite abelian group $G$, and let $G_{+}$
denote the subgroup of $G$ generated by all elements $g$ of $G$ such that
$Prob_{G}\left(  g\right)  >0$. \ The \textbf{Markov process} $\mathcal{M}%
_{G}$ associated with a probability distribution $Prob_{G}$ is the Markov
process with the subgroups $G_{\alpha}$ of $G_{+}$ as states, and with
transition probabilities given by
\[
Prob\left(  G_{a}\rightsquigarrow G_{\beta}\right)  =Prob_{G}\left\{  g\in
G_{+}\mid G_{\beta}\text{ is generated by }g\text{ and the elements of
}G_{\alpha}\right\}  \text{ ,}%
\]
where $G_{a}\rightsquigarrow G_{\beta}$ denotes the transition from state
$G_{\alpha}$ to state $G_{\beta}$. \ The \textbf{initial state} of the Markov
process $\mathcal{M}_{G}$ is the trivial subgroup $G_{0}$. \ The subgroup
$G_{+}$ is called the \textbf{absorbing subgroup} of $G$. \ The
\textbf{transition matrix} $T$ of the Markov process is the matrix indexed on
the states according to some chosen fixed linear ordering with $\left(
G_{\alpha},G_{\beta}\right)  $-th entry $T_{\alpha\beta}$ given by
$Prob\left(  G_{a}\rightsquigarrow G_{\beta}\right)  $.
\end{definition}

\bigskip

The following two propositions are immediate consequences of the above definition:

\bigskip

\begin{proposition}
Let
\[
Prob_{G}:G\longrightarrow\lbrack0,1]
\]
be a probability distribution on a finite abelian group $G$. \ Then the Markov
process $\mathcal{M}_{G}$ is an absorbing Markov process with sole absorbing
state $G_{+}$, a state which once entered can never be left. \ The remaining
states are transient states, i.e., states once left can never again be
entered. \ Hence,%
\[
\lim\limits_{n\rightarrow\infty}Prob_{G}\left(  G_{0}\underset{n}%
{\rightsquigarrow}G_{\alpha}\right)  =\left\{
\begin{array}
[c]{ccc}%
1 & \text{if} & G_{\alpha}=G_{+}\\
&  & \\
0 & \text{if} & G_{\alpha}\neq G_{+}%
\end{array}
\right.
\]
\end{proposition}

In other words, if the Markov process $\mathcal{M}_{G}$ starts in state
$G_{0}$, it will eventually end up permanently in the absorbing state $G_{+}$. \ 

\bigskip

\begin{proposition}
Let $T$ be the transition matrix of the Markov process associated with the
probability distribution
\[
Prob_{G}:G\longrightarrow\lbrack0,1]
\]
Then the probability $Prob\left(  G_{\alpha}\underset{n}{\rightsquigarrow
}G_{\beta}\right)  $ that the Markov process $\mathcal{M}_{G}$ starting in
state $G_{\alpha}$ is in state $G_{\beta}$ after $n$ transitions is equal to
the $\left(  G_{\alpha},G_{\beta}\right)  $-th entry of the matrix $T^{n}$,
i.e.,
\[
Prob\left(  G_{\alpha}\underset{n}{\rightsquigarrow}G_{\beta}\right)  =\left(
T^{n}\right)  _{\alpha\beta}%
\]
\ 
\end{proposition}

\bigskip

Under certain circumstances, we can work with a much simpler Markov process.

\bigskip

\begin{proposition}
Let $G$ be a finite abelian group with probability distribution
\[
Prob_{G}:G\longrightarrow\lbrack0,1]
\]
such that $Prob_{G}$ is the uniform probability distribution on the absorbing
group $G_{+}$. \ Partition the states of the associated Markov process
$\mathcal{M}_{G}$ into the collection of sets
\[
\left\{  \mathcal{G}_{j}\mid j\text{ divides }\left|  G_{+}\right|  \right\}
\text{ ,}%
\]
where $\mathcal{G}_{j}$ is the set of all states $G_{\alpha}$ of
$\mathcal{M}_{G}$ of group order $j$. \ 

If
\[
Prob\left(  G_{i}\rightsquigarrow\mathcal{G}_{j}\right)  =%
{\displaystyle\sum\limits_{G_{j}\in\mathcal{G}_{j}}}
Prob\left(  G_{i}\rightsquigarrow G_{j}\right)
\]
has the same value for all $G_{i}\in\mathcal{G}_{i}$, then the states of
$\mathcal{M}_{G}$ can be \textbf{combined} (\textbf{lumped}) to form a Markov
process $\mathcal{M}_{G}^{Lumped}$ with states $\left\{  \mathcal{G}_{j}\mid
j\text{ divides }\left|  G_{+}\right|  \right\}  $, and with transition
probabilities given by
\[
Prob^{Lumped}\left(  \mathcal{G}_{i}\rightsquigarrow\mathcal{G}_{j}\right)
=Prob\left(  G_{i}\rightsquigarrow\mathcal{G}_{j}\right)  \text{ , }%
\]
where $G_{i}$ is an arbitrarily chosen element of $\mathcal{G}_{i}$, and with
initial state $\mathcal{G}_{1}=\left\{  G_{0}\right\}  $. \ 

Moreover, the resulting $\mathcal{M}_{G}^{Lumped}$ is also an absorbing Markov
process with sole absorbing state $\mathcal{G}_{\left|  G_{+}\right|
}=\left\{  G_{+}\right\}  $, with all other states transient, and such that
\[
Prob\left(  G_{0}\underset{k}{\rightsquigarrow}G_{+}\right)  =Prob\left(
\mathcal{G}_{1}\underset{k}{\rightsquigarrow}\mathcal{G}_{\left|
G_{+}\right|  }\right)
\]
\end{proposition}

\bigskip

As a consequence of the above proposition and theorem \ref{ProbTheorem}, we have:

\bigskip

\begin{corollary}
Let $\varphi:A\longrightarrow S$ be a map from a finite abelian group $A$ to a
finite set $S$, which has a hidden subgroup $K_{\varphi}$, and hence a hidden
quotient group $H_{\varphi}$. \ Moreover, let the ambient group $A$ be the
direct sum of cyclic groups of the same prime order $p$, i.e., let
\[
A=\underset{1}{\overset{n}{\oplus}}\mathbb{Z}_{p}\text{ .}%
\]

Then the combined (lumped) process $\mathcal{M}_{\widehat{A}}^{Lumped}$ is a
Markov process such that
\[
Prob\left(  \widehat{A}_{0}\underset{k}{\rightsquigarrow}\widehat{H_{\varphi}%
}\right)  =Prob\left(  \mathcal{G}_{1}\underset{k}{\rightsquigarrow
}\mathcal{G}_{\left|  \widehat{H_{\varphi}}\right|  }\right)
\]
Moreover, if the states of $\mathcal{M}_{\widehat{A}}^{Lumped}$ are linearly
ordered as
\[
\mathcal{G}_{i}<\mathcal{G}_{j}\text{ if and only if }i\text{ divides }j\text{
,}%
\]
then the transition matrix $T$ of $\mathcal{M}_{\widehat{A}}^{Lumped}$ is
given by
\[
T=\left(
\begin{array}
[c]{ccccccc}%
1 & 0 & 0 & 0 & \cdots & 0 & 0\\
1-\frac{1}{p} & \frac{1}{p} & 0 & 0 & \cdots & 0 & 0\\
0 & 1-\frac{1}{p^{2}} & \frac{1}{p^{2}} & 0 & \cdots & 0 & 0\\
0 & 0 & 1-\frac{1}{p^{3}} & \frac{1}{p^{3}} & \cdots & 0 & 0\\
\vdots & \vdots & \vdots & \vdots & \ddots & \vdots & \vdots\\
0 & 0 & 0 & 0 & \cdots & \frac{1}{p^{n-1}} & 0\\
0 & 0 & 0 & 0 & \cdots & 1-\frac{1}{p^{n}} & \frac{1}{p^{n}}%
\end{array}
\right)
\]
Hence,
\[
Prob\left(  \widehat{A}_{0}\underset{k}{\rightsquigarrow}\widehat{H_{\varphi}%
}\right)  =\left(  T^{k}\right)  _{n1}\text{ ,}%
\]
from which it easily follows that
\[
Prob\left(  \widehat{A}_{0}\underset{k}{\rightsquigarrow}\widehat{H_{\varphi}%
}\right)  >1-\frac{1}{p-1}\left(  \frac{1}{p}\right)  ^{k-n}S\geq
1-\frac{1}{\left(  p-1\right)  p^{2}}%
\]
for $k\geq n+2$.
\end{corollary}

\bigskip

\bigskip

\section{Vintage Simon quantum hidden subgroup algorithms (QHSAs)}

\qquad\bigskip

We are now prepared to extend Simon's quantum algorithm to an entire class of
QHSAs on finite abelian groups.

\bigskip

Let
\[
\varphi:A\longrightarrow S
\]
be a map from a finite abelian group $A$ to a finite set $S$ for which there
exists a hidden subgroup $K_{\varphi}$, and hence, a hidden quotient group
$H_{\varphi}=A/K_{\varphi}$.

\bigskip

Following our usual convention, we use additive notation for the ambient group
$A$ and multiplicative notation for the hidden quotient group $H_{\varphi}$.

\bigskip

As mentioned in section 2 of this paper, it follows from the standard theory
of abelian groups (i.e., Theorem \ref{AbelianGroupTheorem}) that the ambient
group $A$ can be decomposed into the finite direct sum of cyclic groups
$\mathbb{Z}_{m_{0}}$, $\mathbb{Z}_{m_{1}}$, $\ldots$ , $\mathbb{Z}_{m_{\ell
-1}}$, i.e.,
\[
A=\mathbb{Z}_{m_{0}}\oplus\mathbb{Z}_{m_{1}}\oplus\ldots\oplus\mathbb{Z}%
_{m_{\ell-1}}\text{ ,}%
\]
We denote respective generators of the above cyclic groups by
\[
a_{0},a_{1},\ldots,a_{\ell-1}\text{ .}%
\]

\bigskip

Consequently, each character $\chi$ of the ambient group $A$ can be uniquely
expressed as
\[
\chi:%
{\displaystyle\sum\limits_{j=0}^{\ell-1}}
\alpha_{j}a_{j}\longmapsto\exp\left(  2\pi i%
{\displaystyle\sum\limits_{j=0}^{\ell-1}}
\alpha_{j}\frac{y_{j}}{m_{j}}\right)  \text{ ,}%
\]
where $0\leq y_{j}<m_{j}$ for $j=0,1,\ldots,\ell-1$. Thus, we have a
one-to-one correspondence between the characters $\chi$ of $A$ and $\ell
$-tuples of rationals (modulo 1) of the form
\[
\left(  \frac{y_{0}}{m_{0}},\frac{y_{1}}{m_{1}},\ldots,\frac{y_{\ell-1}%
}{m_{\ell-1}}\right)  \text{ ,}%
\]
where
\[
0\leq y_{j}<m_{j}\text{, \ }j=0,1,\ldots,\ell-1\text{ .}%
\]
As a result, we can and do use the following notation to refer uniquely to
each and every character $\chi$ of $A$
\[
\chi=\chi_{(\frac{y_{0}}{m_{0}},\frac{y_{1}}{m_{1}},\ldots,\frac{y_{\ell-1}%
}{m_{\ell-1}})}\text{ .}%
\]

\bigskip

\begin{definition}
Let
\[
A=\mathbb{Z}_{m_{0}}\oplus\mathbb{Z}_{m_{1}}\oplus\cdots\oplus\mathbb{Z}%
_{m_{\ell-1}}%
\]
be a direct sum decomposition of a finite abelian group $A$ into finite cyclic
groups. \ Let
\[
a_{0},a_{1},\ldots,a_{n-1}%
\]
denote respective generators of the cyclic groups in this direct sum decomposition.

Then an integer matrix
\[
\mathfrak{G}=\left[  \alpha_{ij}\right]  _{k\times n}\text{ \ \ \ \ \ }%
\operatorname{mod}\left(  m_{0},m_{1},\ldots,m_{n-1}\right)
\]
is said to be a \textbf{generator matrix} of a subgroup $K$ of $A$ provided
\[
\left\{
{\displaystyle\sum\limits_{j=0}^{n-1}}
\alpha_{ij}a_{j}\mid0\leq i<k\right\}
\]
is a complete set of generators of the subgroup $K$.

A matrix of rationals $\operatorname{mod}1$
\[
\mathfrak{H}=\left[  \frac{y_{ij}}{m_{j}}\right]  _{\ell\times n}\text{
\ \ \ \ \ }\operatorname{mod}1
\]
is said to be a \textbf{dual generator matrix} of a subgroup $K$ of $A$
provided
\[
\left\{  \chi_{(\frac{y_{i0}}{m_{0}},\frac{y_{i1}}{m_{1}},\ldots
\frac{y_{i(n-1)}}{m_{n-1}})}\mid0\leq i<\ell\right\}
\]
is a complete set of generators of the character group $\widehat{H}$ of the
quotient group $H=A/K$. \ 
\end{definition}

\bigskip

Let $\mathcal{M}_{\varphi}$ be the Markov process associated with the
probability distribution
\[
Prob_{\varphi}:\widehat{A}\longrightarrow\lbrack0,1]
\]
on the character group $\widehat{A}$ of the ambient group $A$.

\bigskip

Let $0\leq\epsilon\ll1$ be a chosen threshold.

\bigskip

Then a vintage Simon algorithm is given below:

\vspace{0.4in}

\begin{center}
\fbox{\textsc{Vintage Simon}$(\varphi,\epsilon)$}
\end{center}

\begin{itemize}
\item[\fbox{\textbf{Step 1.}}] Select a positive integer $\ell$ such that
\[
Prob_{\widehat{A}}\left(  \widehat{A}_{0}\underset{\ell}{\rightsquigarrow
}\widehat{H_{\varphi}}\right)  <1-\epsilon
\]
\bigskip

\item[\fbox{\textbf{Step 2.}}] Initialize running dual generator matrix
\[
\mathfrak{H}=\left[  \quad\right]
\]
\bigskip

\item[\fbox{\textbf{Step 3.}}] Query the probability distribution
$Prob_{\varphi}$ $\ell$ times to obtain $\ell$ characters (not necessarily
distinct) of the hidden quotient group $H_{\varphi},$ while incrementing the
running dual generator matrix $\mathfrak{H}$. \ \medskip

\item[\quad] \textsc{Loop} $i$ \textsc{From} $0$ \textsc{To} $\ell-1$
\textsc{Do}%
\begin{align*}
\chi_{(\frac{y_{i0}}{m_{0}},\frac{y_{i1}}{m_{1}},\ldots\frac{y_{i(n-1)}%
}{m_{n-1}})}  &  =\text{\textsc{QRand}}_{\varphi}()\\
& \\
\mathfrak{H\qquad\qquad\qquad}  &  =\left[
\begin{array}
[c]{c}%
\begin{array}
[c]{cccc}%
\frac{y_{i0}}{m_{0}} & \frac{y_{i1}}{m_{1}} & \ldots & \frac{y_{i(n-1)}%
}{m_{n-1}}%
\end{array}
\\
------------\\
\mathfrak{H}%
\end{array}
\right]
\end{align*}
\medskip\ 

\item[\quad] \textsc{Loop Lower Boundary;}\bigskip

\item[\fbox{\textbf{Step 4.}}] Compute the generator matrix $\mathfrak{G}$
from the dual generator matrix $\mathfrak{H}$ by using Gaussian elimination to
solve the system of equations
\[%
{\displaystyle\sum\limits_{j=0}^{n-1}}
\frac{y_{ij}}{m_{j}}x_{j}=0\operatorname{mod}1\text{ \ \ \ \ \ \ \ \ \ \ }%
0\leq i<N_{0}%
\]
for unknown $x_{j}\operatorname{mod}m_{j}$.\bigskip

\item[\fbox{\textbf{Step 5.}}] \textsc{Output} $\mathfrak{G}$ and
\textsc{Stop}.
\end{itemize}

\bigskip

\part{Vintage Shor Algorithms\bigskip}

\vspace{0.3in}

\section{Vintage Shor quantum hidden subgroup algorithms(QHSAs)}

\bigskip

Let $\varphi:A\longrightarrow S$ be a map with hidden subgroup structure. \ We
now consider QHSPs for which the ambient group $A$ is \textbf{free abelian of
finite rank} $n$.

\bigskip

Since the ambient group $A$ is infinite, at least two difficulties naturally
arise. \ One is that the associated complex vector space $\mathcal{H}_{A}$ is
now infinite dimensional, thereby causing some implementation problems. \ The
other is that the Fourier transform of a periodic function on $A$ does not
exist as a function\footnote{As a clarifying note, let $f:\mathbb{Z}%
\longrightarrow\mathbb{C}$ be a period $P$ function on $\mathbb{Z}$. \ Then
$f$ on $\mathbb{Z}$ is neither of compact support, nor of bounded $L^{2}$ or
$L^{1}$ norm. \ So the Fourier transform of $f$ does not exist as a function,
but as a generalized function, i.e., as a distribution. \ However, the
function $f$ does induce a function $\widetilde{f}:\mathbb{Z}_{P}%
\longrightarrow\mathbb{C}$ which does have a Fourier transform on
$\mathbb{Z}_{P}$ which exists as a function. The problem is that we do not
know the period of $\varphi$, and as a consequence, cannot do Fourier analysis
on the corresponding unknown finite cyclic group.}, but as a generalized function!

\bigskip

Following Shor's lead, we side-step these annoying obstacles by choosing not
to work with the ambient group $A$ and the map $\varphi$ at all. Instead, we
work with a group $\widetilde{A}$ and a map $\widetilde{\varphi}:\widetilde
{A}\longrightarrow S$ which are ``approximations'' of $A$ and $\varphi
:A\longrightarrow S$, respectively.

\bigskip

The group $\widetilde{A}$ and the \textbf{approximating map} $\widetilde
{\varphi}$ are constructed as follows:

\bigskip

Choose an epimorphism
\[
\mu:A\longrightarrow\widetilde{A}%
\]
of the ambient group $A$ onto a chosen finite group $\widetilde{A}$, called a
\textbf{group probe}. \ Next, select a \textbf{transversal} \
\[
\iota_{\mu}:\widetilde{A}\longrightarrow A\text{ }%
\]
of $\mu$, i.e., a map such that
\[
\mu\circ\iota_{\mu}=id_{\widetilde{A}}\text{ ,}%
\]
where $id_{\widetilde{A}}$, denotes the identity map on the group probe
$\widetilde{A}$. \ [Consequently, $\iota_{\mu}$ is an injection, and in most
cases not a morphism at all.] \ 

Having chosen $\mu$ and $\iota_{\mu}$, the \textbf{approximating map}
$\widetilde{\varphi}$ is defined as
\[
\widetilde{\varphi}=\varphi\circ\iota_{\mu}:\widetilde{A}\longrightarrow S
\]

\bigskip

Although the map $\widetilde{\varphi}$ is not usually a morphism, the quantum
subroutine \textsc{QRand}$_{\widetilde{\varphi}}()$ is still a well defined
quantum procedure which produces a well defined probability distribution
$Prob_{\widetilde{\varphi}}\left(  \chi\right)  $ on the character group
$\widehat{\widetilde{A}}$ of the group probe $\widetilde{A}$. \ As we shall
see, if the the map $\widetilde{\varphi}$ is a ``reasonably good
approximation'' to the original map $\varphi$, then \textsc{QRand}%
$_{\widetilde{\varphi}}()$ will with high probability produce characters
$\chi$ of the probe group $\widetilde{A}$ which are ``sufficiently close'' to
corresponding characters $\eta$ of the hidden quotient group $H_{\varphi}$.

\bigskip

Following this basic strategy, we will now use the quantum subroutine
\textsc{QRand}$_{\widetilde{\varphi}}()$ to build three classes of vintage
Shor QHSAs, where the probe group $\widetilde{A}$ is a finite cyclic group
$\mathbb{Z}_{Q}$ of order $Q$. \ In this way, we will create three classes of
quantum algorithms which form natural extensions of Shor's original quantum
factoring algorithm.

\bigskip

\section{Direct summand structure}

We digress momentarily to discuss the direct sum structure of the ambient
group $A$ when it is free abelian of finite rank $n$.\bigskip

Since the ambient group $A$ is free abelian of finite rank $n$, the hidden
subgroup $K_{\varphi}$ is also free abelian of finite rank. \ Moreover, there
exist compatible direct sum decompositions of $A$ and $K_{\varphi}$ into free
cyclic groups
\[
\left\{
\begin{array}
[c]{ccc}%
K_{\varphi} & = & P_{1}\mathbb{Z}\oplus\cdots\oplus P_{n}\mathbb{Z}\\
&  & \\
A & = & \underset{n\text{ direct summands}}{\underbrace{\underset{\mathstrut
}{\mathbb{Z}\oplus\cdots\oplus\mathbb{Z}}}}%
\end{array}
\right.  \text{ ,}%
\]
where $P_{1}$, ..., $P_{n}$ are non-negative integers, and where the inclusion
morphism
\[
K_{\varphi}=P_{1}\mathbb{Z}\oplus\cdots\oplus P_{n}\mathbb{Z}\ \hookrightarrow
\ \underset{n}{\underbrace{\underset{\mathstrut}{\mathbb{Z}\oplus\cdots
\oplus\mathbb{Z}}}}\ =A
\]
is the direct sum of the inclusion morphisms
\[
P_{j}\mathbb{Z}\ \hookrightarrow\ \mathbb{Z}%
\]

\bigskip

It should be mentioned that, since the group $K_{\varphi}$ is hidden, the
above direct sum decompositions are also hidden. \ Moreover, the selection of
a direct sum decomposition of the ambient group $A$ is operationally
equivalent to a selection a basis of $A$. \ This leads to the following definition:

\bigskip

\begin{definition}
A basis
\[
\left\{  a_{1},a_{2},\ldots,a_{n}\right\}
\]
of the ambient group $A$ corresponding to the above hidden direct sum
decomposition of $A$ is called a \textbf{hidden basis} of $A$.
\end{definition}

\bigskip

\noindent\textbf{Question.} \textit{How is the hidden basis }$\left\{
a_{1},a_{2},\ldots,a_{n}\right\}  $ \textit{of }$A$\textit{ related to any
``visible'' basis }$\left\{  a_{1}^{\prime},a_{2}^{\prime},\ldots
,a_{n}^{\prime}\right\}  $ \textit{of }$A$\textit{ that we might choose to
work with?}

\bigskip

The group of automorphisms of the free abelian group $A$ of rank $n$ is
isomorphic to the group
\[
SL_{\pm}\left(  n,\mathbb{Z}\right)
\]
of $n\times n$ invertible integer matrices. \ This is the same as the group of
$n\times n$ integer matrices of determinant $\pm1$.

\bigskip

\begin{proposition}
Let $\left\{  a_{1},a_{2},\ldots,a_{n}\right\}  $ be a hidden basis of $A$,
and let $\left\{  a_{1}^{\prime},a_{2}^{\prime},\ldots,a_{n}^{\prime}\right\}
$ be any other basis of $A$. Then there exists a unique element $M\in$
$SL_{\pm}\left(  n,\mathbb{Z}\right)  $ which carries the basis $\left\{
a_{1}^{\prime},a_{2}^{\prime},\ldots,a_{n}^{\prime}\right\}  $ into the hidden
basis $\left\{  a_{1},a_{2},\ldots,a_{n}\right\}  $. \ 
\end{proposition}

\bigskip

Since the image of $\varphi$ is finite, we know that $P_{j}>0$, for all $j$.
\ Thus, the direct sum decomposition of the inclusion morphism becomes
\[
\overset{K_{\varphi}}{\overbrace{\overset{\mathstrut}{\left(  P_{1}%
\mathbb{Z}\oplus\cdots\oplus P_{\overline{n}}\mathbb{Z}\right)  \oplus
\underset{n-\overline{n}}{\underbrace{\underset{\mathstrut}{\left(
\mathbb{Z}\oplus\cdots\oplus\mathbb{Z}\right)  }}}}}}\ \hookrightarrow
\ \overset{A}{\overbrace{\overset{\mathstrut}{\underset{\overline{n}%
}{\underbrace{\underset{\mathstrut}{\left(  \mathbb{Z}\oplus\cdots
\oplus\mathbb{Z}\right)  }}}\oplus\underset{n-\overline{n}}{\underbrace
{\underset{\mathstrut}{\left(  \mathbb{Z}\oplus\cdots\oplus\mathbb{Z}\right)
}}}}}}%
\]

\bigskip

As a consequence, the hidden quotient group $H_{\varphi}$ is the corresponding
direct sum of finite cyclic groups
\[
H_{\varphi}=\left(  \mathbb{Z}_{P_{1}}\oplus\cdots\oplus\mathbb{Z}%
_{P_{\overline{n}}}\right)  \oplus\underset{n-\overline{n}}{\underbrace
{\underset{\mathstrut}{\left(  0\oplus\cdots\oplus0\right)  }}}\text{ ,}%
\]
and the hidden epimorphism
\[
\nu:\left(  P_{1}\mathbb{Z}\oplus\cdots\oplus P_{\overline{n}}\mathbb{Z}%
\right)  \oplus\underset{n-\overline{n}}{\underbrace{\underset{\mathstrut
}{\left(  \mathbb{Z}\oplus\cdots\oplus\mathbb{Z}\right)  }}}\longrightarrow
\mathbb{Z}_{P_{1}}\oplus\cdots\oplus\mathbb{Z}_{P_{\overline{n}}}%
\]
is the direct sum of of the epimorphisms
\[
\left\{
\begin{array}
[c]{c}%
\mathbb{Z}\longrightarrow\mathbb{Z}_{P_{j}}\\
\\
\mathbb{Z}\longrightarrow0
\end{array}
\right.
\]

\bigskip

As a consequence of the above, we have:

\bigskip

\begin{definition}
Let
\[
\left\{  a_{1},a_{2},\ldots,a_{n}\right\}
\]
be a hidden basis of $A$. \ Then a corresponding \textbf{induced hidden basis}
of the hidden quotient group $H_{\varphi}$ is defined as
\[
\left\{  b_{1}=\nu\left(  a_{1}\right)  ,b_{2}=\nu\left(  a_{2}\right)
,\ldots,b_{\overline{n}}=\nu\left(  a_{\overline{n}}\right)  \right\}  \text{
,}%
\]
where $\nu:A\longrightarrow H_{\varphi}$ denotes the hidden
epimorphism.\footnote{Please note that the hidden basis $\left\{  a_{1}%
,a_{2},\ldots,a_{n}\right\}  $ of $A$ is free in the abelian category.
\ However, the induced basis $\left\{  b_{1},b_{2},\ldots,b_{\overline{n}%
}\right\}  $ of $H_{\varphi}$ is not because $H_{\varphi}$ is a torsion group.
\ $\left\{  b_{1},b_{2},\ldots,b_{\overline{n}}\right\}  $ is a basis in the
sense that it is a set of generators of $H_{\varphi}$ such that
\[
b_{1}^{k_{1}}b_{2}^{k_{2}}\cdots b_{\overline{n}}^{k_{\overline{n}}}=1
\]
implies that
\[
b_{j}^{k_{j}}=1
\]
for every $j$. \ (For more information, please refer to \cite{Hall1}.)}
\end{definition}

\bigskip

The above direct sum decompositions are summarized in the following diagram:
\[%
\begin{array}
[c]{ccrcc}%
\overset{K_{\varphi}}{\overbrace{\overset{\mathstrut}{\left(
{\displaystyle\bigoplus\limits_{j=1}^{\overline{n}}}
P_{j}\mathbb{Z}\right)  \oplus\left(
{\displaystyle\bigoplus\limits_{j=\overline{n}+1}^{n}}
\mathbb{Z}\right)  }}} & \hookrightarrow & \overset{A}{\overbrace
{\overset{\mathstrut}{\left(
{\displaystyle\bigoplus\limits_{j=1}^{\overline{n}}}
\mathbb{Z}\right)  \oplus\left(
{\displaystyle\bigoplus\limits_{j=\overline{n}+1}^{n}}
\mathbb{Z}\right)  }}} & \overset{\varphi}{\longrightarrow} & S\\
&  & \nu\searrow &  & \nearrow\iota_{\varphi}\\
&  &  & \underset{H_{\varphi}}{\underbrace{\underset{\mathstrut}{%
{\displaystyle\bigoplus\limits_{j=1}^{\overline{n}}}
\mathbb{Z}_{P_{j}}}}} &
\end{array}
\]

\bigskip

\begin{definition}
Let $H$ be a finite abelian group. \ Then a \textbf{maximal cyclic subgroup of
}$H$ is a cyclic subgroup of $H$ of highest possible order.
\end{definition}

\bigskip

\begin{proposition}
Let $b_{1},b_{2},\ldots,b_{\overline{n}}$ be the above defined induced hidden
basis of the hidden quotient group $H_{\overline{n}}=\mathbb{Z}_{P_{1}}%
\oplus\mathbb{Z}_{P_{2}}\oplus\cdots\oplus\mathbb{Z}_{P_{\overline{n}}}$.
\ Then a maximal cyclic subgroup of $H_{\varphi}$ is generated by
\[
b_{1}\oplus b_{2}\oplus\cdots\oplus b_{\overline{n}}\text{ ,}%
\]
and is isomorphic to the finite cyclic group $\mathbb{Z}_{P}$ of order
\[
P=\operatorname{lcm}\left(  P_{1},P_{2},\ldots,P_{\overline{n}}\right)  \text{
.}%
\]
\end{proposition}

\bigskip

\section{Vintage Shor QHSAs with group probe $\widetilde{A}=\mathbb{Z}_{Q}$.}

\bigskip

Choose a positive integer $Q$ and an epimorphism
\[
\mu:A\longrightarrow\mathbb{Z}_{Q}%
\]
of the free abelian group $A$ onto the finite cyclic group $\widetilde
{A}=\mathbb{Z}_{Q}$ of order $Q$.

\bigskip

Next we wish to select a transversal $\iota_{\mu}$ of the epimorphism $\mu$. \ 

\bigskip

However, at this juncture we must take care. For, not every choice of the
transversal $\iota_{\mu}$ will produce an efficient vintage Shor algorithm.
\ In fact, most choices probably will produce highly inefficient
algorithms\footnote{For example, consider $A=\mathbb{Z}$, $P=6$, $Q=64$, and
the transversal defined by $\iota_{\mu}:6n+k\longmapsto6n+k+64\left\lfloor
k/2\right\rfloor $ for $0\leq n\leq10$, where $\left\{
\begin{array}
[c]{ccl}%
0\leq k<6 & \text{if} & 0\leq n<10\\
0\leq k<4 & \text{if} & n=10
\end{array}
\right.  $. \ One reason this is a poor choice of transversal is that the
image of $\iota_{\mu}$ does not contain a representative of every coset of the
hidden subgroup $\mathbb{Z}_{P}$ of the ambient group $A$.}. \ We emphasize
that the efficiency of the class of algorithms we are about to define depends
heavily on the choice of the transversal $\iota_{\mu}$.

\bigskip

Following Shor's lead once again, we select a very special transversal
$\iota_{\mu}$. \ 

\bigskip

\begin{definition}
Let $\mu:A\longrightarrow\mathbb{Z}_{Q}$ be an epimorphism from a free abelian
group $A$ of finite rank $n$ onto a finite cyclic group $\mathbb{Z}_{Q}$ of
order $Q$, and let $\widetilde{a}$ be a chosen generator of the cyclic group
$\mathbb{Z}_{Q}.$

A transversal
\[
\iota_{\mu}:\mathbb{Z}_{Q}\longrightarrow A
\]
is said to be a \textbf{Shor transversal} provided\medskip

\begin{itemize}
\item[\textbf{1)}] $\iota_{\mu}\left(  k\widetilde{a}\right)  \longmapsto
k\iota_{\mu}\left(  \widetilde{a}\right)  $ , for all $0\leq k<Q$, and\bigskip

\item[\textbf{2)}] There exists a basis $\left\{  a_{1}^{\prime},a_{2}%
^{\prime},\ldots,a_{n}^{\prime}\right\}  $ of $A$ such that, when $\iota_{\mu
}\left(  \widetilde{a}\right)  $ is expressed in this basis, i.e., when
\[
\iota_{\mu}\left(  \widetilde{a}\right)  =\sum_{j=1}^{n}\lambda_{j}^{\prime
}a_{j}^{\prime}\text{ ,}%
\]
it follows that%
\[
\gcd\left(  \lambda_{1}^{\prime},\lambda_{2}^{\prime},\ldots,\lambda
_{n}^{\prime}\right)  =1
\]
\end{itemize}
\end{definition}

\bigskip

\begin{proposition}
\label{SL}Let $\lambda_{1}^{\prime},\lambda_{2}^{\prime},\ldots,\lambda
_{n}^{\prime}$ be $n$ integers, and let $M$ be a non-singular $n\times n$
integral matrix, i.e., an element of $SL_{\pm}\left(  n,\mathbb{Z}\right)  $.
\ \ If $\lambda_{1},\lambda_{2},\ldots,\lambda_{n}$ are $n$ integers defined
by
\[
\left(  \lambda_{1},\lambda_{2},\ldots,\lambda_{n}\right)  =\left(
\lambda_{1}^{\prime},\lambda_{2}^{\prime},\ldots,\lambda_{n}^{\prime}\right)
M\text{ ,}%
\]
then
\[
\gcd\left(  \lambda_{1},\lambda_{2},\ldots,\lambda_{n}\right)  =\gcd\left(
\lambda_{1}^{\prime},\lambda_{2}^{\prime},\ldots,\lambda_{n}^{\prime}\right)
\]
\end{proposition}

\bigskip

As a corollary, we have

\bigskip

\begin{proposition}
If condition \textbf{2)} is true with respect to one basis, then it is true
with respect to every basis.
\end{proposition}

\bigskip

An another immediate consequence of the definition of a Shor traversal, we
have the following lemma:\bigskip

\begin{lemma}
If a Shor transversal \label{ShorsCosetRepMap}
\[
\iota_{\mu}:\mathbb{Z}_{Q}\longrightarrow A\text{ ,}%
\]
is used to construct the the approximating map
\[
\widetilde{\varphi}=\varphi\circ\iota_{\mu}:\mathbb{Z}_{Q}\longrightarrow
S\text{ ,}%
\]
then the approximating map $\widetilde{\varphi}$ has the following property
\[
\widetilde{\varphi}\left(  k\widetilde{a}\right)  =\left[  \widetilde{\varphi
}\left(  \widetilde{a}\right)  \right]  ^{k}\text{ ,}%
\]
for all $0\leq k<Q$, where we have used the hidden injection $\iota_{\varphi
}:H_{\varphi}\longrightarrow S$ to identify the elements $\widetilde{\varphi
}\left(  k\widetilde{a}\right)  $ of the set $S$ with corresponding elements
of the hidden quotient group $H_{\varphi}$.
\end{lemma}

\bigskip

\section{Finding Shor transversals for vintage $\mathbb{Z}_{Q}$ Shor algorithms}

\bigskip\label{FindShorTransversal}

Surprisingly enough, it is algorithmically simpler to find a Shor transversal
$\iota_{\mu}:\mathbb{Z}_{Q}\longrightarrow A$ first, and then, as an after
thought, to construct a corresponding epimorphism $\mu:A\longrightarrow
\mathbb{Z}_{Q}$.

\bigskip

\begin{definition}
Let $A$ be an ambient group, and let $\mathbb{Z}_{Q}$ be a finite cyclic group
of order $Q$ with a selected generator $\widetilde{a}$. \ Then an injection
\[
\iota:\mathbb{Z}_{Q}\longrightarrow A
\]
is called a \textbf{Shor injection} provided

\begin{itemize}
\item[\textbf{1)}] $\iota\left(  k\widetilde{a}\right)  =k\iota\left(
\widetilde{a}\right)  $ , for all $0\leq k<Q$, and

\item[\textbf{2)}] There exists a basis $\left\{  a_{1}^{\prime},a_{2}%
^{\prime},\ldots,a_{n}^{\prime}\right\}  $ of the ambient group $A$ such that
\[
\gcd\left(  \lambda_{1}^{\prime},\lambda_{2}^{\prime},\ldots,\lambda
_{n}^{\prime}\right)  =1\text{ ,}%
\]
where
\[
\iota\left(  \widetilde{a}\right)  =%
{\displaystyle\sum\limits_{j=1}^{n}}
\lambda_{j}^{\prime}a_{j}^{\prime}\text{ .}%
\]
\end{itemize}
\end{definition}

\bigskip

\begin{proposition}
If condition \textbf{2)} is true with respect to one basis, it is true with
respect to all.
\end{proposition}

\bigskip

Next, we need to construct an epimorphism $\mu:A\longrightarrow\mathbb{Z}_{Q}$
for which $\iota:\mathbb{Z}_{Q}\longrightarrow A$ is a Shor transversal.

\bigskip

\begin{proposition}
Let $A$ be an ambient group, and let $\mathbb{Z}_{Q}$ be a finite cyclic group
of order $Q$ with a selected generator $\widetilde{a}$. \ Given a Shor
injection
\[
\iota:\mathbb{Z}_{Q}\longrightarrow A\text{ ,}%
\]
there exists an epimorphism
\[
\mu_{\iota}:A\longrightarrow\mathbb{Z}_{Q}%
\]
such that $\iota$ is a Shor transversal for $\mu_{\iota}$, i.e., such that
\[
\mu_{\iota}\circ\iota=id_{\mathbb{Z}_{Q}}\text{ ,}%
\]
where $id_{\mathbb{Z}_{Q}}$ denotes the identity morphism on $\mathbb{Z}_{Q}$.
\end{proposition}

\begin{proof}
Select an arbitrary basis $\left\{  a_{1}^{\prime},a_{2}^{\prime},\ldots
,a_{n}^{\prime}\right\}  $ of $A$. \ Then
\[
\iota\left(  \widetilde{a}\right)  =%
{\displaystyle\sum\limits_{j=1}^{n}}
\lambda_{j}^{\prime}a_{j}^{\prime}\text{ ,}%
\]
where
\[
\gcd\left(  \lambda_{1}^{\prime},\lambda_{2}^{\prime},\ldots,\lambda
_{n}^{\prime}\right)  =1\text{ .}%
\]

Hence, from the extended Euclidean algorithm, we can find integers
\[
\alpha_{1},\alpha_{2},\ldots,\alpha_{n}%
\]
for which
\[%
{\displaystyle\sum\limits_{j=1}^{n}}
\alpha_{j}\lambda_{j}^{\prime}=1\text{ .}%
\]

Define
\[
\mu:\left\{  a_{1}^{\prime},a_{2}^{\prime},\ldots,a_{n}^{\prime}\right\}
\longrightarrow\mathbb{Z}_{Q}%
\]
by
\[
\mu\left(  a_{j}^{\prime}\right)  =\alpha_{j}\widetilde{a}\text{ ,
\ }j=1,2,\ldots,n\text{.}%
\]

Since $a_{1}^{\prime},a_{2}^{\prime},\ldots,a_{n}^{\prime}$ is a free abelian
basis of the ambient group $A$, it uniquely extends to a morphism
\[
\mu:A\longrightarrow\mathbb{Z}_{Q}\text{ .}%
\]
It immediately follows that $\mu$ is an epimorphism because
\[
\mu\left(
{\displaystyle\sum\limits_{j=1}^{n}}
\lambda_{j}^{\prime}a_{j}^{\prime}\right)  =%
{\displaystyle\sum\limits_{j=1}^{n}}
\alpha_{j}\lambda_{j}^{\prime}\widetilde{a}=\widetilde{a}\text{ .}%
\]
\end{proof}

\bigskip

Thus the task of finding an epimorphism $\mu:A\longrightarrow\mathbb{Z}_{Q}$
and a corresponding Shor transversal reduces to the task of finding $n$
integers $\lambda_{1}^{\prime},\lambda_{2}^{\prime},\ldots,\lambda_{n}%
^{\prime}$ such that
\[
\gcd\left(  \lambda_{1}^{\prime},\lambda_{2}^{\prime},\ldots,\lambda
_{n}^{\prime}\right)  =1\text{ .}%
\]
This leads of to the following probabilistic subroutine which finds a random
Shor traversal:\bigskip

\begin{center}
\fbox{\textsc{Random}\_\textsc{Shor}\_\textsc{transversal}$\left(  \ \left\{
a_{1}^{\prime},a_{2}^{\prime},\ldots,a_{n}^{\prime}\right\}  ,\ Q,\ \widetilde
{a},\ n\ \right)  $}\bigskip
\end{center}

\begin{itemize}
\item[\quad] \quad\# \textsc{Input}: \ \ \ \ \ A basis $\left\{  a_{1}%
^{\prime},a_{2}^{\prime},\ldots,a_{n}^{\prime}\right\}  $ of $A$, a positive
integer $Q$,

\item[\quad] \quad\# \ \ \ \ \ \ \ \ \ \ \ \ \ \ \ \ a selected generator
$\widetilde{a}$ of $A$, and the rank $n$ of $A$

\item[\quad] \quad\# \textsc{Output}: \ \ Shor transversal $\iota_{\mu
}:\mathbb{Z}_{Q}\longrightarrow A$

\item[\quad] \quad\# \textsc{Side Effect}: Epimorphism $\mu:A\longrightarrow
\mathbb{Z}_{Q}$

\item[\quad] \quad\# \textsc{Side Effect}: Random integers $\lambda
_{1}^{\prime},\lambda_{2}^{\prime},\ldots,\lambda_{n}^{\prime}$

\item[\quad] 

\item[\quad] \quad\textsc{Global: }$\mu:A\longrightarrow\mathbb{Z}_{Q}$

\item[\quad] \quad\textsc{Global: }$\lambda_{1}^{\prime},\lambda_{2}^{\prime
},\ldots,\lambda_{n}^{\prime}$

\item[\quad] 

\item[\fbox{\textbf{Step 0}}] \textsc{If }$n=1$\textsc{ Then }$\left(
\text{\textsc{Set }}\lambda_{1}^{\prime}=1\text{\textsc{ And Goto}
}\fbox{\textbf{Step 4}}\right)  $ \bigskip

\item[\fbox{\textbf{Step 1}}] \quad Select with replacement $n$ random
$\lambda_{1}^{\prime},\lambda_{2}^{\prime},\ldots,\lambda_{n}^{\prime}$ from
$\left\{  1,2,\ldots,Q\right\}  $.\bigskip

\item[\fbox{\textbf{Step 2}}] \quad Use the extended Euclidean algorithm to
determine
\[
d=\gcd\left(  \lambda_{1}^{\prime},\lambda_{2}^{\prime},\ldots,\lambda
_{n}^{\prime}\right)
\]
and integers $\alpha_{1},\alpha_{2},\ldots,\alpha_{n}$ such that $\sum
_{j=1}^{n}\alpha_{j}\lambda_{j}^{\prime}=d$\bigskip

\item[\fbox{\textbf{Step 3}}] \textsc{If }$d\neq1$\textsc{ Then Goto
}\fbox{\textbf{Step 1}}\textsc{ Else Goto }\fbox{\textbf{Step 4}}\bigskip
\end{itemize}

\begin{itemize}
\item[\fbox{\textbf{Step 4}}] \quad Construct Shor transversal $\iota_{\mu
}:\mathbb{Z}_{Q}\longrightarrow A$ as $\iota_{\mu}\left(  k\widetilde
{a}\right)  =k\sum_{j=1}^{n}\lambda_{j}^{\prime}a_{j}^{\prime}$, for $0\leq
k<Q$\bigskip

\item[\fbox{\textbf{Step 5}}] \quad Construct epimorphism $\mu
:A\longrightarrow\mathbb{Z}_{Q}$ as
\[
\mu\left(  a_{j}^{\prime}\right)  =\alpha_{j}\widetilde{a}\text{ for all
}j=1,2,\ldots,n
\]
\bigskip

\item[\fbox{\textbf{Step 6}}] \quad\textsc{Output }transversal $\iota_{\mu
}:\mathbb{Z}_{Q}\longrightarrow A$ and \textsc{Stop}
\end{itemize}

\vspace{0.3in}

\begin{theorem}
\label{ShorTransversal}For $n>1$, the average case complexity of the
\textsc{Random}\_\textsc{Shor}\_\textsc{transversal} subroutine is
\[
O\left(  n\left(  \lg Q\right)  ^{3}\right)  \text{ .}%
\]
\end{theorem}

\begin{proof}
The computationally dominant part of this subroutine is the main loop Steps 1
through 3.

Each iteration of the main loop executes the extended Euclidean algorithm $n$
times to find the $\gcd$ $d$. \ Since the computational complexity of the
extended Euclidean algorithm\footnote{See \cite[Chap. 31]{Cormen1}.} is
$O\left(  \left(  \lg Q\right)  ^{3}\right)  $, it follows that the
computational cost of one iteration of steps 1 through 3 is
\[
O\left(  n\left(  \lg Q\right)  ^{3}\right)  \text{ .}%
\]

But by Corollary \ref{Omega1Corollary} of Appendix B,
\[
Prob_{Q}\left(  \gcd\left(  \lambda_{1}^{\prime},\lambda_{2}^{\prime}%
,\ldots,\lambda_{n}^{\prime}\right)  =1\right)  =\Omega\left(  1\right)
\text{ .}%
\]
Thus, the average number of iterations before a successful exit to Step 4 is
$O\left(  1\right)  $. \ 

Hence, the average case complexity of steps 1 through 4 is
\[
O\left(  n\left(  \lg Q\right)  ^{3}\right)  \text{ .}%
\]
\end{proof}

\bigskip

\begin{remark}
Our objective in this paper is to find reasonable asymptotic bounds, not the
tightest possible bounds. \ For example,\ the above bound is by no means the
tightest possible. \ For a tighter bound for the Euclidean algorithm is
$O\left(  \left(  \lg Q\right)  ^{2}\right)  $ which can be found in
\cite{Cormen1}. \ Thus, the bound found in the above theorem can be tightened
to at least $O\left(  n\left(  \lg Q\right)  ^{2}\right)  $.
\end{remark}

\bigskip

\section{Maximal Shor transversals}

\bigskip

Unfortunately, the definition of a Shor transversal is in some instances not
strong enough to extend Shor's quantum factoring algorithm to ambient groups
which are free abelian groups of finite rank. \ From necessity, we are forced
to make the following definition.

\bigskip

\begin{definition}
Let $a_{1},a_{2},\ldots,a_{n}$ be a hidden basis of the ambient group $A$, let
$\widetilde{a}$ be a chosen generator the cyclic group probe $\mathbb{Z}_{Q}$,
and let $H_{\varphi}=\mathbb{Z}_{P_{1}}\oplus\mathbb{Z}_{P_{2}}\oplus
\cdots\oplus\mathbb{Z}_{P_{n}}$ be the corresponding hidden direct sum
decomposition. \ A \textbf{maximal Shor transversal} is a Shor transversal
$\iota_{\mu}:\mathbb{Z}_{Q}\longrightarrow A$ such that
\[
\gcd(\lambda_{j},P_{j})=1\text{, \ \ for }0\leq j<n\text{ ,}%
\]
where the integers $\lambda_{1},\lambda_{2},\ldots,\lambda_{n}$ are defined
by
\[
\iota_{\mu}\left(  \widetilde{a}\right)  =\lambda_{1}a_{1}+\lambda_{2}%
a_{2}+\ldots,+\lambda_{n}a_{n}%
\]
\end{definition}

\bigskip

\begin{remark}
Thus, for maximal Shor traversals, $\iota_{\mu}\left(  \widetilde{a}\right)  $
maps via the hidden epimorphism $\nu:A\longrightarrow A/K_{\varphi}$ to a
maximum order element of the hidden quotient group $H_{\varphi}$.
\end{remark}

\bigskip

One of the difficulties of the above definition is that it does not appear to
be possible to determine whether or not a Shor transversal is maximal without
first knowing the hidden direct sum decomposition of the hidden quotient group
$H_{\varphi}$. \ We address this important issue in the following corollary,
which is an immediate consequence of corollary \ref{conditionalprob} (found in
Appendix B):

\bigskip

\begin{corollary}
\label{Corgcd}Let
\[
P_{1},P_{2},\ldots,P_{n}%
\]
be $n$ fixed positive integers, and let $Q$ be an integer such that
\[
Q\geq\operatorname{lcm}\left(  P_{1},P_{2},\ldots,P_{n}\right)  \text{ ,}%
\]
where $n>1$.

If Conjecture \ref{Conjecture1} (found in Appendix B) is true, then the
probability that the subroutine \textsc{Random}\_\textsc{Shor}%
\_\textsc{transversal} produces a maximal Shor transversal is
\[
\Omega\left(  \frac{1}{\prod_{j=1}^{n}\lg\lg P_{j}}\right)  =\Omega\left(
\frac{1}{\left(  \lg\lg Q\right)  ^{n}}\right)  \text{ .}%
\]
\end{corollary}

\bigskip

\section{Identifying characters of cyclic groups with points on the unit
circle $\mathbb{S}^{1}$ in the complex plane $\mathbb{C}$.}

\bigskip\label{Questions}

We will now begin to develop an answer to the following question:

\bigskip

\noindent\textbf{Question. 1.} \textit{Are the characters of }the group probe
$\mathbb{Z}_{Q}$\textit{ produced by the quantum subroutine }\textsc{QRand}%
$_{\widetilde{\varphi}}()$\textit{ ``close enough'' to the characters of a
maximal cyclic subgroup of the hidden quotient group }$H_{\varphi}$?

\bigskip

If \textsc{QRand}$_{\widetilde{\varphi}}()$ produces a character $\chi$ of
$\mathbb{Z}_{Q}$\textit{ which is ``close enough'' to some character }$\eta$
of a maximal cyclic subgroup $\mathbb{Z}_{P}$ of the hidden quotient group
$H_{\varphi}$, then the character $\chi$ can be used to find the corresponding
closest character $\eta$ of $\mathbb{Z}_{P}$. \ Each time such a character
$\eta$ is found, something more is known about the hidden quotient group
$H_{\varphi}$ and the hidden subgroup $K_{\varphi}$. \ In this way, we have
the conceptual genesis of a class of vintage $\mathbb{Z}_{Q}$ Shor algorithms.\ 

\bigskip

But before we can answer the above question, we need to answer a more
fundamental question, namely:

\bigskip

\noindent\textbf{Question. 2.} \textit{What do we mean by ``close
enough''}?\textit{ \ I.e, what do we mean by saying that a character }$\chi
$\textit{ of }$\widetilde{A}=\mathbb{Z}_{Q}$\textit{ is ``close enough'' to
some character }$\eta$\textit{ of }$\mathbb{Z}_{P}$?

\bigskip

To answer this last question, we need to introduce two additional concepts:

\begin{itemize}
\item[\textbf{1)}] The concept of a common domain for the characters $\chi$ of
$\mathbb{Z}_{Q}$ and the characters $\eta$ of $\mathbb{Z}_{P}$.

\item[\textbf{2)}] The concept of a group norm which is to be used to define
when two characters are ``close.''
\end{itemize}

\bigskip

In this section, we address item \textbf{1)}. \ In the next, item \textbf{2)}.

\bigskip

We begin by noting that the character group $\widehat{\mathbb{Z}}$ of the
infinite cyclic group $\mathbb{Z}$ is simply the group $\mathbb{S}^{1}$,
i.e.,
\[
\widehat{\mathbb{Z}}=\mathbb{S}^{1}=\left\{  \chi_{\theta}:n\longmapsto
e^{2\pi i\theta n}\mid0\leq\theta<1\right\}
\]
In other words, the characters of $\mathbb{Z}$ can be identified with the
points on the unit radius circle in the complex plane $\mathbb{C}$.

\bigskip

Moreover, given an arbitrary epimorphism
\[
\tau:\mathbb{Z}\longrightarrow\mathbb{Z}_{m}%
\]
of the infinite cyclic group $\mathbb{Z}$ onto a finite cyclic group
$\mathbb{Z}_{m}$, the left exact contravariant functor\footnote{For the
definition of a left exact contravariant functor, please refer to, for
example, \cite{Cartan1}.}
\[
Hom_{\mathbb{Z}}\left(  -,2\pi\mathbb{R}/2\pi\mathbb{Z}\right)
\]
transforms $\tau$ into the monomorphism
\[%
\begin{array}
[c]{rcc}%
\widehat{\tau}:\widehat{\mathbb{Z}_{m}} & \longrightarrow & \widehat
{\mathbb{Z}}\\
\eta\ \  & \longmapsto & \eta\circ\tau
\end{array}
\]
In this way the characters of $\mathbb{Z}_{m}$ can be identified with the
points of $\widehat{\mathbb{Z}}=\mathbb{S}^{1}$.

\bigskip

Thus, to find a common domain $\mathbb{S}^{1}$ for the characters of the group
probe $\mathbb{Z}_{Q}$ and the maximal cyclic group $\mathbb{Z}_{P}$, all that
need be done is to find epimorphisms $\widetilde{\mu}:\mathbb{Z}%
\longrightarrow\mathbb{Z}_{Q}$ and $\widetilde{\tau}:\mathbb{Z}\longrightarrow
\mathbb{Z}_{P}$. \ This is accomplished as follows:

\bigskip

Let $a$ be a generator of the infinite cyclic group $\mathbb{Z}$, and let
$a_{1},a_{2},\ldots,a_{n}$ be a hidden basis of the ambient group $A$. \ Then
the epimorphisms $\widetilde{\mu}$ and $\widetilde{\tau}$ are defined as
\[%
\begin{array}
[c]{rcc}%
\widetilde{\mu}:\mathbb{Z} & \longrightarrow & \mathbb{Z}_{Q}\\
ka & \longmapsto & k\widetilde{a}%
\end{array}
\text{ \ \ and \ \ }%
\begin{array}
[c]{rcc}%
\widetilde{\tau}:\mathbb{Z} & \longrightarrow & \mathbb{Z}_{P}\\
ka & \longmapsto & \nu\left[  k\left(  a_{1}+a_{2}+\ldots+a_{n}\right)
\right]
\end{array}
\text{ ,}%
\]
where $\widetilde{a}$ is the selected generator of the group probe
$\mathbb{Z}_{Q}$, and where $\nu:A\longrightarrow H_{\varphi}$ is the hidden epimorphism.

\bigskip

Thus, as a partial answer to \textbf{Question 2} of Section \ref{Questions}, a
character $\chi$ of the probe group $\widetilde{A}=\mathbb{Z}_{Q}$\textit{ is
``close'' to some character }$\eta$ of the maximal cyclic subgroup
$\mathbb{Z}_{P}$ $H_{\varphi}=%
{\textstyle\bigoplus\nolimits_{j=1}^{\overline{n}}}
\mathbb{Z}_{P_{j}}$ if the corresponding points $\widehat{\widetilde{\mu}%
}\left(  \chi\right)  $ and $\widehat{\widetilde{\tau}}\left(  \eta\right)  $
on the circle $\mathbb{S}^{1}$ are ``close.''

\bigskip

But precisely what do we mean by two points of $\mathbb{S}^{1}$ being
``close'' to one another? \ To answer this question, we need to observe that
Shor's algorithm uses, in addition to the group structure of $\mathbb{S}^{1}$,
also the metric structure of $\mathbb{S}^{1}$. \ 

\bigskip

\section{Group norms}

\bigskip

We proceed to define a metric structure on the circle group $\mathbb{S}^{1}$.
\ To do so, we need to define what is meant by a group norm.

\vspace{0.3in}

\begin{definition}
A \textbf{(group theoretic)} \textbf{norm} on a group $G$ is a map
\[
\left|  \left\|  -\right\|  \right|  :G\longrightarrow\mathbb{R}%
\]
such that
\end{definition}

\begin{itemize}
\item[\textbf{1)}] $\left|  \left\|  x\right\|  \right|  \geq0$\textit{, for
all }$x$\textit{, and }$\left|  \left\|  x\right\|  \right|  =0$\textit{ if
and only if }$x$\textit{ is the group identity (which is }$1$\textit{ if we
think of }$G$\textit{ as a multiplicative group, or }$0$\textit{ if we think
of }$G$\textit{ as an additive group).}

\item[\textbf{2)}] $\left|  \left\|  x\cdot y\right\|  \right|  \leq\left|
\left\|  x\right\|  \right|  +\left|  \left\|  y\right\|  \right|  $\textit{
or }$\left|  \left\|  x+y\right\|  \right|  \leq\left|  \left\|  x\right\|
\right|  +\left|  \left\|  y\right\|  \right|  ,$\textit{ depending
respectively on whether we think of }$G$\textit{ as a multiplicative or as an
additive group.}
\end{itemize}

\vspace{0.4in}

\noindent\textbf{Caveat.} \textit{The group norms defined in this section are
different from the group algebra norms defined in Section \ref{FourierSection}.}

\bigskip\bigskip

\begin{remark}
Such a norm induces a metric
\begin{align*}
&  G\times G\longrightarrow\mathbb{R}\\
\left(  x,y\right)   &  \longmapsto\left|  \left\|  x\cdot y^{-1}\right\|
\right|  \text{ or }\left|  \left\|  x-y\right\|  \right|
\end{align*}
depending on whether multiplicative or additive notation is used.
\end{remark}

\bigskip

As mentioned in Section \ref{CharacterGroupSection}, we think of the 1-sphere
$\mathbb{S}^{1}$ interchangeably as the multiplicative group
\[
\mathbb{S}^{1}=\left\{  e^{2\pi i\alpha}\mid0\leq\alpha<1\right\}
\subset\mathbb{C}%
\]
with multiplication defined as
\[
e^{2\pi i\alpha}\cdot e^{2\pi i\beta}=e^{2\pi i\left(  \alpha+\beta\right)  }%
\]
or as the additive group of reals $\mathbb{R}$ modulo $2\pi$, i.e., as
\[
\mathbb{S}^{1}=2\pi\mathbb{R}/2\pi\mathbb{Z}=\left\{  2\pi\alpha\mid
0\leq\alpha<1\right\}
\]
with addition defined as
\[
2\pi\alpha+2\pi\beta=\left(  2\pi\alpha+2\pi\beta\right)  \operatorname{mod}%
2\pi=2\pi\left(  \alpha+\beta\operatorname{mod}1\right)
\]
It should be clear from context which of the two representation of the group
$\mathbb{S}^{1}$ is being used.

\vspace{0.4in}%

\begin{center}
\fbox{\includegraphics[
height=2.3981in,
width=2.1352in
]%
{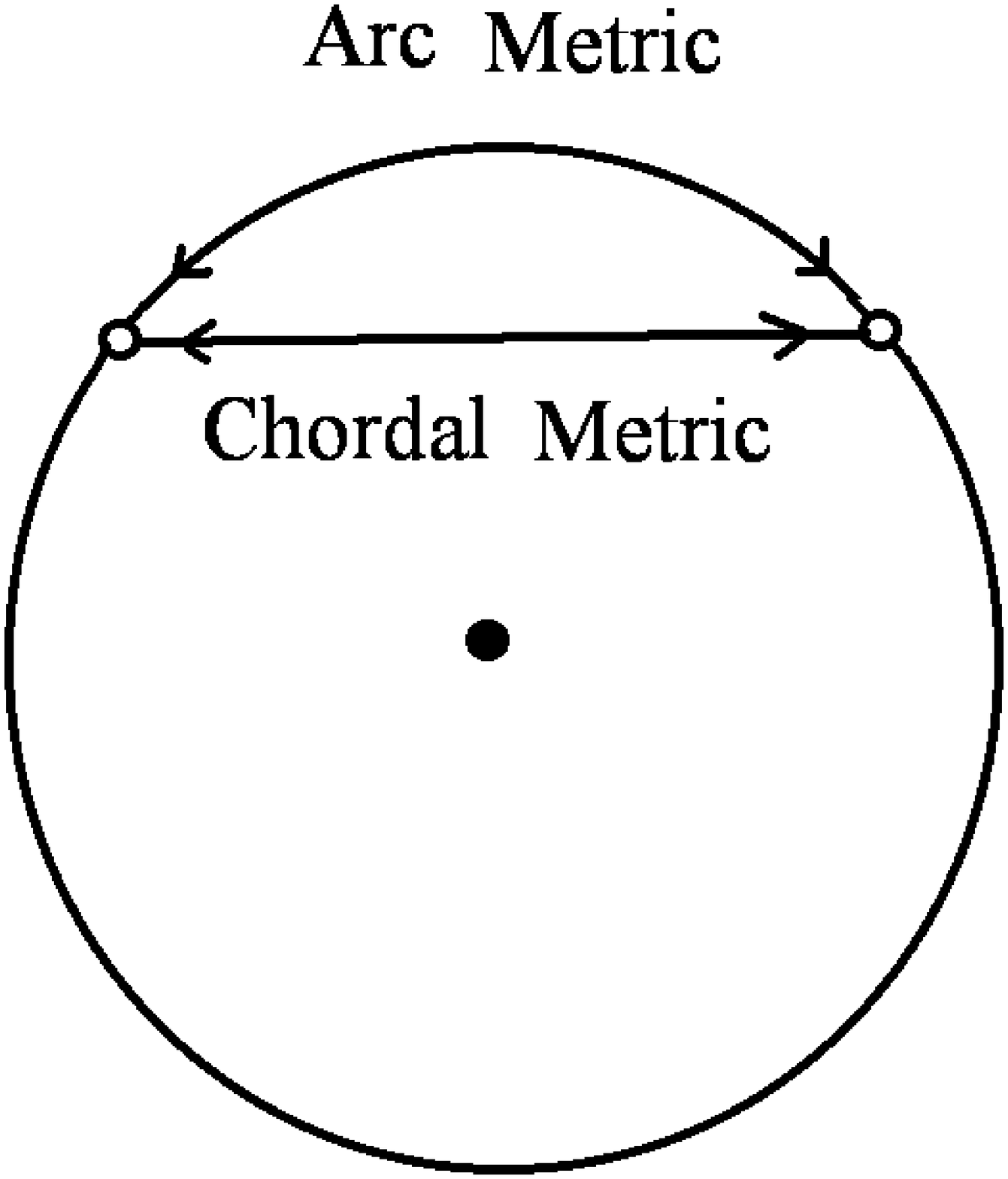}%
}\\
\textbf{Figure 2.} \ Two metrics on the unit circle $\mathbb{S}^{1}$,
\textsc{Arc}$_{2\pi}$ and \textsc{Chord}$_{2\pi}$.
\end{center}

\bigskip

There are two different norms on the 1-sphere $\mathbb{S}^{1}$ that we will be
of use to us. \ The first is the \textbf{arclength norm}, written
\textsc{Arc}$_{2\pi}$, defined by
\[
\text{\textsc{Arc}}_{2\pi}\left(  \alpha\right)  =2\pi\min\left\{  \ \left|
\alpha\right|  -\left\lfloor \left|  \alpha\right|  \right\rfloor
,\ \left\lceil \left|  \alpha\right|  \right\rceil -\left|  \alpha\right|
\ \right\}  \text{ ,}%
\]
which is simply the length of the shortest arc in the 1-sphere $\mathbb{S}%
^{1}$ connecting the point $e^{2\pi i\alpha}$ to the point $1$. \ 

\vspace{0.4in}

The second norm is the \textbf{chordal length norm}, written \textsc{Chord}%
$_{2\pi}$, defined by
\[
\text{\textsc{Chord}}_{2\pi}\left(  \alpha\right)  =2\left|  \sin\left(
\pi\alpha\right)  \right|  \text{ ,}%
\]
which is simply the length of the chord in the complex plane connecting the
point $e^{2\pi i\alpha}$ to the point $1$. \ 

\vspace{0.4in}

Shor's algorithm depends heavily on the interrelationship of these two norms.
\ We summarize these interrelationships in the following proposition:

\vspace{0.3in}

\begin{proposition}
The the norms \textsc{Arc}$_{2\pi}$ and \textsc{Chord}$_{2\pi}$ satisfy the
following conditions:\label{ChordArcProposition}

\begin{itemize}
\item[1)] \textsc{Chord}$_{2\pi}\left(  \alpha\right)  =2\sin\left(
\frac{1}{2}\text{\textsc{Arc}}_{2\pi}\left(  \alpha\right)  \right)  $\bigskip

\item[2)] $\frac{2}{\pi}$\textsc{Arc}$_{2\pi}\left(  \alpha\right)  \leq
$\textsc{Chord}$_{2\pi}\left(  \alpha\right)  \leq$\textsc{Arc}$_{2\pi}\left(
\alpha\right)  $
\end{itemize}
\end{proposition}

\vspace{0.4in}

We need the following property of the arclength norm \textsc{Arc}$_{2\pi}$:

\bigskip

\begin{proposition}
Let $n$ be a nonzero integer.\label{ArcProposition} \ If \textsc{Arc}$_{2\pi
}\left(  \alpha\right)  \leq\frac{\pi}{\left|  n\right|  }$, then
\textsc{Arc}$_{2\pi}\left(  n\alpha\right)  =\left|  n\right|  $%
\textsc{Arc}$_{2\pi}\left(  \alpha\right)  $
\end{proposition}

\bigskip

\section{Vintage $\mathbb{Z}_{Q}$ Shor QHSAs (Cont.)}

\bigskip

Our next step is to look more closely at the probability distribution
\[
Prob_{\widetilde{\varphi}}:\widehat{\widetilde{A}}\longrightarrow\left[
0,1\right]  \text{ .}%
\]

\bigskip

We seek first to use this probability distribution to determine the maximal
cyclic subgroup $\mathbb{Z}_{P}$ of the hidden quotient group $H_{\varphi}$.
\ However, as indicated by the following lemma, there are a number of
obstacles to finding the subgroup $\mathbb{Z}_{P}$.

\bigskip

\begin{lemma}
\label{Order}Let $a_{1}$, ... , $a_{n}$ denote a hidden basis of the ambient
group $A=%
{\textstyle\bigoplus\nolimits_{j=1}^{n}}
\mathbb{Z}$, and let $P_{1},P_{2},\ldots,P_{\overline{n}}$ denote the
respective orders of the corresponding cyclic direct summands of the hidden
quotient group $H_{\varphi}=%
{\textstyle\bigoplus\nolimits_{j=1}^{\overline{n}}}
\mathbb{Z}_{P_{j}}$.

Let $\widetilde{a}$ denote a chosen generator of the group probe
$\widetilde{A}=\mathbb{Z}_{Q}$, and let $\lambda_{1},\lambda_{2}%
,\ldots,\lambda_{n}$ denote the unknown integers such that
\[
\iota_{\mu}(\widetilde{a})=%
{\displaystyle\sum\limits_{j=1}^{n}}
\lambda_{j}a_{j}\in A\text{.}%
\]
Finally, use the hidden injection $\iota_{\varphi}:H_{\varphi}\longrightarrow
S$ to identify the elements of the hidden quotient group $H_{\varphi}$ with
the corresponding elements of the set $S$.\bigskip

If the approximating map $\widetilde{\varphi}$ is constructed from a Shor
transversal, then the order of $\widetilde{\varphi}\left(  \widetilde
{a}\right)  \in H_{\varphi}$ is $\overline{P}$, i.e.,
\[
order\left(  \widetilde{\varphi}\left(  \widetilde{a}\right)  \right)
=\overline{P}\text{ ,}%
\]
where $\overline{P}=\operatorname{lcm}\left(  \overline{P}_{1},\overline
{P}_{2},\ldots,\overline{P}_{n}\right)  $, and where\ $\overline{P}_{j}%
=P_{j}/\gcd\left(  \lambda_{j},P_{j}\right)  $ for $j=1,2,\ldots
,n$.\ \ Hence,
\[
\left\{  \widetilde{\varphi}\left(  k\widetilde{a}\right)  =\widetilde
{\varphi}\left(  \widetilde{a}\right)  ^{k}\mid0\leq k<\overline{P}\right\}
\]
are all distinct elements of $S$.

Moreover, if the approximating map $\widetilde{\varphi}$ is constructed from a
maximal Shor transversal, $\overline{P}=P=\operatorname{lcm}\left(
P_{1},P_{2},\ldots,P_{n}\right)  $.
\end{lemma}

\begin{proof}%
\[
\widetilde{\varphi}\left(  \widetilde{a}\right)  =\varphi\circ\iota_{\mu
}\left(  \widetilde{a}\right)  =\varphi\left(  \sum_{j}\lambda_{j}%
a_{j}\right)  =\prod_{j}\varphi\left(  a_{j}\right)  ^{\lambda_{j}}=\prod
_{j}b_{j}^{\lambda_{j}}\text{ ,}%
\]
where we have used the hidden injection $\iota_{\varphi}:H_{\varphi
}\longrightarrow S$ to identify the hidden basis element $b_{j}$ of
$H_{\varphi}$ with the element $\varphi(a_{j})$ of the set $S$.

Since the order of each $b_{j}$ is $P_{j}$, it follows from elementary group
theory that the order of $\prod_{j}b_{j}^{\lambda_{j}}$ must be $\overline{P}$.
\end{proof}

\bigskip

\begin{lemma}
\label{Chord}Let $a_{1}$, ... , $a_{n}$ be a hidden basis of the ambient group
$A=%
{\textstyle\bigoplus\nolimits_{j=1}^{n}}
\mathbb{Z}$, and let $P_{1},P_{2},\ldots,P_{n}$ denote the respective orders
of the corresponding cyclic direct summands of the hidden quotient group
$H_{\varphi}=%
{\textstyle\bigoplus\nolimits_{j=1}^{\overline{n}}}
\mathbb{Z}_{P_{j}}$.

Let $\widetilde{a}$ denote a chosen generator of the group probe
$\widetilde{A}=\mathbb{Z}_{Q}$, let $\lambda_{1},\lambda_{2},\ldots
,\lambda_{n}$ denote the unknown integers such that
\[
\iota_{\mu}(\widetilde{a})=%
{\displaystyle\sum\limits_{j=1}^{n}}
\lambda_{j}a_{j}\in A\text{ ,}%
\]
\ \ and let $\chi_{\frac{y}{Q}}$ be a character of $\mathbb{Z}_{Q}$. \ 

Finally, identify the elements of the hidden quotient group $H_{\varphi}$ with
the corresponding elements of the set $S$ via the hidden injection
$\iota_{\varphi}:H_{\varphi}\longrightarrow S$.\bigskip

If the approximating map $\widetilde{\varphi}$ is constructed from a Shor
transversal, then \bigskip

\begin{itemize}
\item When $\overline{P}y\neq0\operatorname{mod}Q$, we have
\begin{align*}
\widetilde{\varphi}\left(  \chi_{\frac{y}{Q}}^{\bullet}\right)   &  =\pm
e^{i\pi\frac{\overline{P}y}{Q}q}\frac{\text{\textsc{Chord}}_{2\pi}\left(
\frac{\overline{P}y}{Q}\left(  q+1\right)  \right)  }{\text{\textsc{Chord}%
}_{2\pi}\left(  \frac{\overline{P}y}{Q}\right)  }\sum_{k_{0}=0}^{r-1}%
\chi_{\frac{y}{Q}}\left(  k_{0}\widetilde{a}\right)  \widetilde{\varphi
}\left(  k_{0}\widetilde{a}\right) \\
&  \qquad\qquad\qquad\qquad\pm e^{i\pi\frac{\overline{P}y}{Q}(q-1)}%
\frac{\text{\textsc{Chord}}_{2\pi}\left(  \frac{\overline{P}y}{Q}q\right)
}{\text{\textsc{Chord}}_{2\pi}\left(  \frac{\overline{P}y}{Q}\right)  }%
\sum_{k_{0}=r}^{\overline{P}-1}\chi_{\frac{y}{Q}}\left(  k_{0}\widetilde
{a}\right)  \widetilde{\varphi}\left(  k_{0}\widetilde{a}\right)
\end{align*}
where $\overline{P}=\operatorname{lcm}\left(  \overline{P}_{1},\overline
{P}_{2},\ldots,\overline{P}_{n}\right)  $, where\ $\overline{P}_{j}=P_{j}%
/\gcd\left(  \lambda_{j},P_{j}\right)  $ for $j=1,2,\ldots,n$, and where\ \
\[
Q=q\overline{P}+r\text{, with }0\leq r<\overline{P}\text{.}%
\]

\item And when $\overline{P}y=0\operatorname{mod}Q$, we have
\[
\widetilde{\varphi}\left(  \chi_{\frac{y}{Q}}^{\bullet}\right)  =\left(
q+1\right)  \sum_{k_{0}=0}^{r-1}\chi_{\frac{y}{Q}}\left(  k_{0}\widetilde
{a}\right)  \widetilde{\varphi}\left(  k_{0}\widetilde{a}\right)
+q\sum_{k_{0}=r}^{\overline{P}-1}\chi_{\frac{y}{Q}}\left(  k_{0}\widetilde
{a}\right)  \widetilde{\varphi}\left(  k_{0}\widetilde{a}\right)
\]
\end{itemize}

Moreover, if the approximating map $\widetilde{\varphi}$ is constructed from a
maximal Shor transversal, then $\overline{P}=P=\operatorname{lcm}\left(
P_{1},P_{2},\ldots,P_{n}\right)  $.
\end{lemma}

\begin{proof}
We begin by identifying the elements of the hidden quotient group $H_{\varphi
}$ with the corresponding elements of the set $S$ via injection $\iota
_{\varphi}:H_{\varphi}\longrightarrow S$.\bigskip

We first consider the case when $\overline{P}y\neq0\operatorname{mod}Q$.\bigskip

Then
\begin{align*}
\widetilde{\varphi}\left(  \chi_{\frac{y}{Q}}^{\bullet}\right)   &
=\widetilde{\varphi}\left(
{\displaystyle\sum\limits_{k=0}^{Q-1}}
\chi_{\frac{y}{Q}}\left(  k\widetilde{a}\right)  k\widetilde{a}\right)  =%
{\displaystyle\sum\limits_{k=0}^{Q-1}}
\chi_{\frac{y}{Q}}\left(  k\widetilde{a}\right)  \widetilde{\varphi}\left(
k\widetilde{a}\right) \\
& \\
&  =%
{\displaystyle\sum\limits_{k=0}^{q\overline{P}-1}}
\chi_{\frac{y}{Q}}\left(  k\widetilde{a}\right)  \widetilde{\varphi}\left(
\widetilde{a}\right)  ^{k}+%
{\displaystyle\sum\limits_{k=q\overline{P}}^{Q-1}}
\chi_{\frac{y}{Q}}\left(  k\widetilde{a}\right)  \widetilde{\varphi}\left(
\widetilde{a}\right)  ^{k}\\
& \\
&  =%
{\displaystyle\sum\limits_{k_{1}=0}^{q-1}}
{\displaystyle\sum\limits_{k_{0}=0}^{\overline{P}-1}}
\chi_{\frac{y}{Q}}\left[  \left(  k_{1}\overline{P}+k_{0}\right)
\widetilde{a}\right]  \widetilde{\varphi}\left[  \widetilde{a}\right]
^{k_{1}\overline{P}+k_{0}}+%
{\displaystyle\sum\limits_{n_{0}=0}^{r-1}}
\chi_{\frac{y}{Q}}\left[  \left(  k_{1}\overline{P}+k_{0}\right)
\widetilde{a}\right]  \widetilde{\varphi}\left[  \widetilde{a}\right]
^{k_{1}\overline{P}+k_{0}}%
\end{align*}
From Lemma \ref{Order} we have
\[
\widetilde{\varphi}\left[  \widetilde{a}\right]  ^{k_{1}\overline{P}+k_{0}%
}=\widetilde{\varphi}\left[  \widetilde{a}\right]  ^{k_{0}}\text{ .}%
\]
So,%
\begin{align*}
\widetilde{\varphi}\left(  \chi_{\frac{y}{Q}}^{\bullet}\right)   &  =\left(
{\displaystyle\sum\limits_{k_{1}=0}^{q-1}}
\chi_{\frac{y}{Q}}\left[  \left(  k_{1}\overline{P}\right)  \widetilde
{a}\right]  \right)
{\displaystyle\sum\limits_{k_{0}=0}^{\overline{P}-1}}
\chi_{\frac{y}{Q}}\left(  k_{0}\widetilde{a}\right)  \widetilde{\varphi
}\left(  \widetilde{a}\right)  ^{k_{0}}\\
&  \qquad\hspace{1in}+\chi_{\frac{y}{Q}}\left(  q\overline{P}\widetilde
{a}\right)
{\displaystyle\sum\limits_{k_{0}=0}^{r-1}}
\chi_{\frac{y}{Q}}\left(  k_{0}\widetilde{a}\right)  \widetilde{\varphi
}\left(  \widetilde{a}\right)  ^{k_{0}}\\
& \\
&  =\left(
{\displaystyle\sum\limits_{k_{1}=0}^{q}}
\chi_{\frac{y}{Q}}\left[  \left(  k_{1}\overline{P}\right)  \widetilde
{a}\right]  \right)
{\displaystyle\sum\limits_{k_{0}=0}^{r-1}}
\chi_{\frac{y}{Q}}\left(  k_{0}\widetilde{a}\right)  \widetilde{\varphi
}\left(  \widetilde{a}\right)  ^{k_{0}}\\
&  \qquad\hspace{1in}+\left(
{\displaystyle\sum\limits_{k_{1}=0}^{q-1}}
\chi_{\frac{y}{Q}}\left[  \left(  k_{1}\overline{P}\right)  \widetilde
{a}\right]  \right)
{\displaystyle\sum\limits_{k_{0}=r}^{\overline{P}-1}}
\chi_{\frac{y}{Q}}\left(  k_{0}\widetilde{a}\right)  \widetilde{\varphi
}\left(  \widetilde{a}\right)  ^{k_{0}}\\
& \\
&  =\left(  \frac{e^{2\pi i\frac{\overline{P}y}{Q}(q+1)}-1}{e^{2\pi
i\frac{\overline{P}y}{Q}}-1}\right)
{\displaystyle\sum\limits_{k_{0}=0}^{r-1}}
\chi_{\frac{y}{Q}}\left(  k_{0}\widetilde{a}\right)  \widetilde{\varphi
}\left(  k_{0}\widetilde{a}\right) \\
&  \qquad\hspace{1in}+\left(  \frac{e^{2\pi i\frac{\overline{P}y}{Q}q}%
-1}{e^{2\pi i\frac{\overline{P}y}{Q}}-1}\right)
{\displaystyle\sum\limits_{k_{0}=r}^{\overline{P}-1}}
\chi_{\frac{y}{Q}}\left(  k_{0}\widetilde{a}\right)  \widetilde{\varphi
}\left(  k_{0}\widetilde{a}\right)
\end{align*}%

\begin{align*}
\qquad &  =e^{i\pi\frac{\overline{P}y}{Q}q}\left(  \frac{e^{\pi
i\frac{\overline{P}y}{Q}(q+1)}-e^{-\pi i\frac{\overline{P}y}{Q}(q+1)}}{e^{\pi
i\frac{\overline{P}y}{Q}}-e^{-\pi i\frac{\overline{P}y}{Q}}}\right)
{\displaystyle\sum\limits_{k_{0}=0}^{r-1}}
\chi_{\frac{y}{Q}}\left(  k_{0}\widetilde{a}\right)  \widetilde{\varphi
}\left(  k_{0}\widetilde{a}\right) \\
&  \qquad\hspace{1in}+e^{i\pi\frac{\overline{P}y}{Q}(q-1)}\left(  \frac{e^{\pi
i\frac{\overline{P}y}{Q}q}-e^{-\pi i\frac{\overline{P}y}{Q}q}}{e^{\pi
i\frac{\overline{P}y}{Q}}-e^{-\pi i\frac{\overline{P}y}{Q}}}\right)
{\displaystyle\sum\limits_{k_{0}=r}^{\overline{P}-1}}
\chi_{\frac{y}{Q}}\left(  k_{0}\widetilde{a}\right)  \widetilde{\varphi
}\left(  k_{0}\widetilde{a}\right) \\
& \\
&  =e^{i\pi\frac{\overline{P}y}{Q}q}\left(  \frac{\sin\left(  \pi
\frac{\overline{P}y}{Q}(q+1)\right)  }{\sin\left(  \pi\frac{\overline{P}y}%
{Q}\right)  }\right)
{\displaystyle\sum\limits_{k_{0}=0}^{r-1}}
\chi_{\frac{y}{Q}}\left(  k_{0}\widetilde{a}\right)  \widetilde{\varphi
}\left(  k_{0}\widetilde{a}\right) \\
&  \qquad\hspace{1in}+e^{i\pi\frac{\overline{P}y}{Q}(q-1)}\left(
\frac{\sin\left(  \pi\frac{\overline{P}y}{Q}q\right)  }{\sin\left(
\pi\frac{\overline{P}y}{Q}\right)  }\right)
{\displaystyle\sum\limits_{k_{0}=r}^{\overline{P}-1}}
\chi_{\frac{y}{Q}}\left(  k_{0}\widetilde{a}\right)  \widetilde{\varphi
}\left(  k_{0}\widetilde{a}\right)
\end{align*}

For the exceptional case when $\overline{P}y=0\operatorname{mod}Q$, we need
only observe that
\[%
{\displaystyle\sum\limits_{k_{1}=0}^{q}}
\chi_{\frac{y}{Q}}\left[  \left(  k_{1}\overline{P}\right)  \widetilde
{a}\right]  =q+1\text{ and }%
{\displaystyle\sum\limits_{k_{1}=0}^{q-1}}
\chi_{\frac{y}{Q}}\left(  k_{1}\overline{P}\right)  \widetilde{a}=q\text{ .}%
\]
\end{proof}

\bigskip

As an immediate consequence of above lemmas \ref{Order} and \ref{Chord}, we have:

\bigskip

\begin{corollary}
If the approximating map $\widetilde{\varphi}$ is constructed from a Shor
transversal, then%
\[
\left\|  \widetilde{\varphi}\left(  \chi_{\frac{y}{Q}}^{\bullet}\right)
\right\|  ^{2}=\left\{
\begin{array}
[c]{cc}%
\frac{r\text{\textsc{Chord}}_{2\pi}^{2}\left[  \frac{\overline{P}y}{Q}y\left(
q+1\right)  \right]  +\left(  \overline{P}-r\right)  \text{\textsc{Chord}%
}_{2\pi}^{2}\left[  \frac{\overline{P}y}{Q}yq\right]  }{\text{\textsc{Chord}%
}_{2\pi}^{2}\left(  \frac{\overline{P}y}{Q}\right)  } & \text{if }\overline
{P}y\neq0\operatorname{mod}Q\\
& \\
r\left(  q+1\right)  ^{2}+\left(  \overline{P}-r\right)  q^{2} & \text{if
}\overline{P}y=0\operatorname{mod}Q
\end{array}
\right.
\]

Moreover, if the approximating map $\widetilde{\varphi}$ is constructed from a
maximal Shor transversal, then $\overline{P}=P=\operatorname{lcm}\left(
P_{1},P_{2},\ldots,P_{n}\right)  $.
\end{corollary}

\vspace{0.4in}

As a consequence of the inequalities found in \textbf{Proposition
\ref{ChordArcProposition}}, we have:

\bigskip

\begin{corollary}
If the approximating map $\widetilde{\varphi}$ is constructed from a Shor
transversal, then when $\overline{P}y\neq0\operatorname{mod}Q$ we have
\label{InequalityCorollary}%
\[
\left\|  \widetilde{\varphi}\left(  \chi_{\frac{y}{Q}}^{\bullet}\right)
\right\|  ^{2}\geq\frac{4}{\pi^{2}}\left(  \frac{r\text{\textsc{Arc}}_{2\pi
}^{2}\left[  \frac{\overline{P}y}{Q}y\left(  q+1\right)  \right]  +\left(
\overline{P}-r\right)  \text{\textsc{Arc}}_{2\pi}^{2}\left[  \frac{\overline
{P}y}{Q}yq\right]  }{\text{\textsc{Arc}}_{2\pi}^{2}\left(  \frac{\overline
{P}y}{Q}\right)  }\right)
\]

Moreover, if the approximating map $\widetilde{\varphi}$ is constructed from a
maximal Shor transversal, then $\overline{P}=P=\operatorname{lcm}\left(
P_{1},P_{2},\ldots,P_{n}\right)  $.
\end{corollary}

\bigskip%

\begin{center}
\fbox{\includegraphics[
trim=0.000000in 0.000000in 0.002153in 0.000000in,
height=3.3831in,
width=3.4126in
]%
{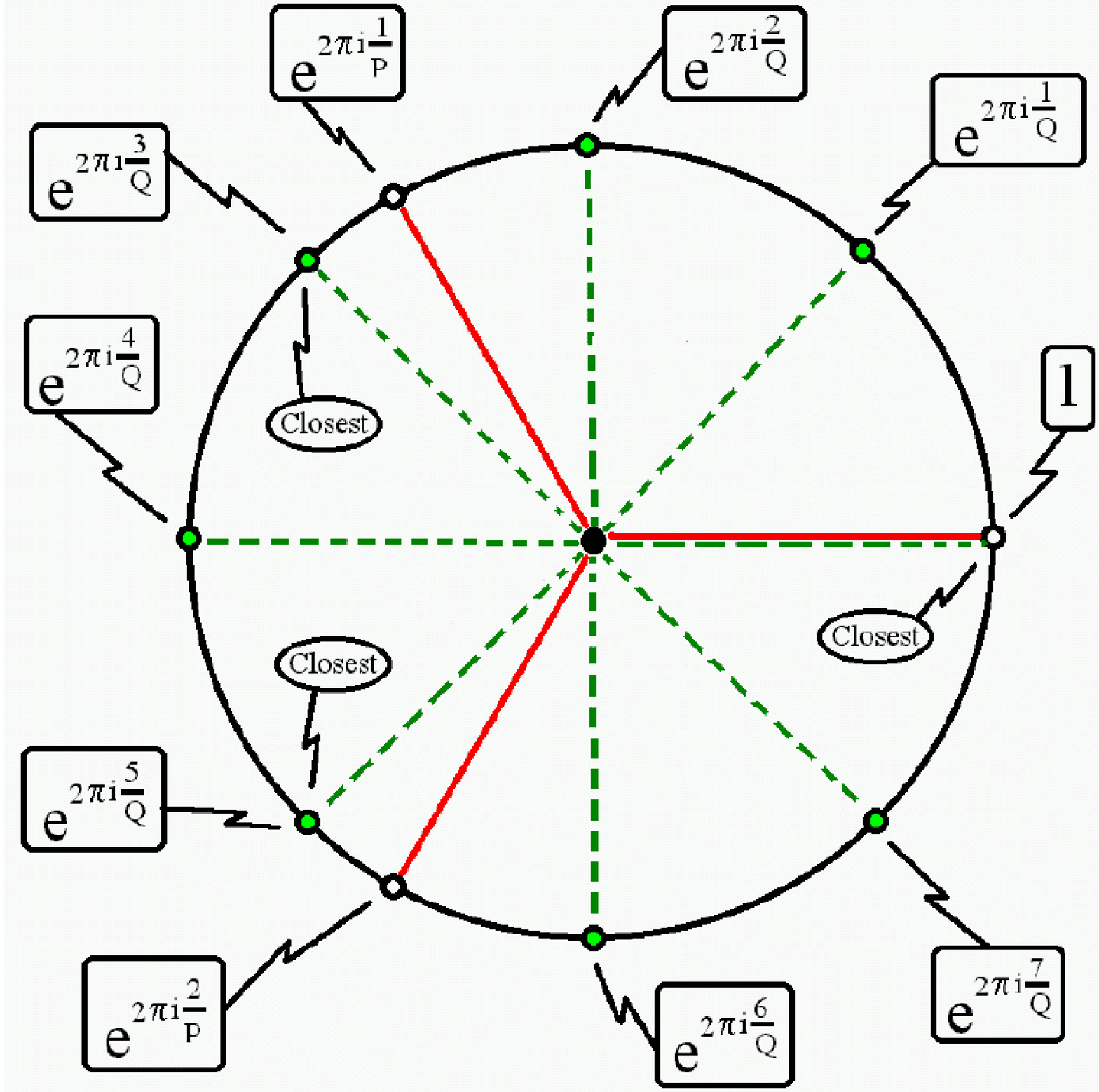}%
}\\
\textbf{Figure 3.} \ The characters of $\mathbb{Z}_{P}$ and $\mathbb{Z}_{Q}$
as points on the circle $\mathbb{S}^{1}$ of radius $1$, with $P=3$ and $Q=8$.
\ The characters $\chi_{1}$, $\chi_{3/8}$, $\chi_{5/8}$ of $\mathbb{Z}_{Q}$
are close respectively to characters $\chi_{1}$, $\chi_{\frac{1}{P}}$,
$\chi_{\frac{2}{P}}$ of $\mathbb{Z}_{P}$. \ They are the characters of
\textsc{Arc}$_{2\pi}$ distance less than $\frac{\pi}{Q}\left(  1-\frac{P}%
{Q}\right)  $ from some character of $\mathbb{Z}_{P}$. \ Also, $\chi
_{\frac{1}{P}}$ and $\chi_{\frac{2}{P}}$ are the primitive characters of
$\mathbb{Z}_{P}$. Unfortunately, since $Q\ngeq P^{2}$, the characters
$\chi_{3/8}$ and $\chi_{5/8}$ of $\mathbb{Z}_{Q}$ are not sufficiently close
respectively to the primitive characters $\chi_{\frac{1}{P}}$ and
$\chi_{\frac{2}{P}}$ of $\mathbb{Z}_{P}$. \ Hence, the continued fraction
algorithm can not be used to find $P$.
\end{center}

\bigskip

\section{When are characters of $\widetilde{A}=\mathbb{Z}_{Q}$ close to some
character of a maximal cyclic subgroup $\mathbb{Z}_{P}$ of $H_{\varphi}?$}

\bigskip\label{When}

\begin{definition}
Let $\mathbb{Z}_{P_{1}}\oplus\mathbb{Z}_{P_{2}}\oplus\cdots\oplus
\mathbb{Z}_{P_{n}}$ be the hidden direct sum decomposition of the hidden
quotient group $H_{\varphi}$, and let $P=\operatorname{lcm}\left(  P_{1}%
,P_{2},\ldots,P_{n}\right)  $. \ A character $\chi_{\frac{y}{Q}}$ of the group
probe $\mathbb{Z}_{Q}$ is said to be \textbf{close} to a character of the
maximal cyclic subgroup $\mathbb{Z}_{P}$ of the hidden quotient group
$H_{\varphi}$ provided either of the following equivalent conditions are satisfied

\begin{itemize}
\item[{%
\begin{tabular}
[c]{l}%
\textbf{Closeness }\\
\textbf{Condition }$\mathbf{1}$%
\end{tabular}
}] There exists \ an integer $d$ such that
\[
\text{\textsc{Arc}}_{2\pi}\left(  \frac{y}{Q}-\frac{d}{P}\right)
\leq\frac{\pi}{Q}\left(  1-\frac{P}{Q}\right)  \text{ ,}%
\]
\end{itemize}

or equivalently,

\begin{itemize}
\item[{%
\begin{tabular}
[c]{l}%
\textbf{Closeness }\\
\textbf{Condition }$\mathbf{1}^{\prime}$%
\end{tabular}
}]
\[
\text{\textsc{Arc}}_{2\pi}\left(  \frac{Py}{Q}\right)  \leq\frac{\pi P}%
{Q}\left(  1-\frac{P}{Q}\right)
\]
If in addition, $Q\geq P^{2}$, then the the character $\chi_{\frac{y}{Q}}$ of
$\mathbb{Z}_{Q}$ is said to be \textbf{sufficiently close} to a character of
the maximal cyclic subgroup $\mathbb{Z}_{P}$.
\end{itemize}
\end{definition}

\bigskip

It immediately follows from the theory of continued fractions \cite{Hardy1,
Lomonaco2} that

\bigskip

\begin{proposition}
If a character $\chi_{\frac{y}{Q}}$ of $\mathbb{Z}_{Q}$ is sufficiently close
to a character $\chi_{\frac{d}{P}}$ of $\mathbb{Z}_{P}$, then $\frac{d}{P}$ is
a convergent of the continued fraction expansion of $\frac{y}{Q}$.
\end{proposition}

\bigskip

However, to determine the sought integer $P$ from the rational $\frac{d}{P}$,
the numerator and denominator of $\frac{d}{P}$ must be relatively prime,
i.e.,
\[
\gcd\left(  d,P\right)  =1\text{ .}%
\]
This leads to the following definition:

\bigskip

\begin{definition}
A character $\chi_{\frac{d}{P}}$ of $\mathbb{Z}_{P}$ is said to be
\textbf{primitive} provided that it is a generator of the dual group
$\widehat{\mathbb{Z}}_{P}$. \ 
\end{definition}

\bigskip

\begin{proposition}
A character $\chi_{\frac{d}{P}}$ of $\mathbb{Z}_{P}$ is a primitive character
if and only $\gcd\left(  d,P\right)  =1$. \ Moreover, the number of primitive
characters of $\mathbb{Z}_{P}$ is $\phi\left(  P\right)  $, where $\phi\left(
P\right)  $ denotes Euler's totient function, i.e., the number of positive
integers less than $P$ which are relatively prime to $P$.
\end{proposition}

\bigskip

\begin{theorem}
Assume that $Q\geq P^{2}$, and that the approximating map $\widetilde{\varphi
}$ is constructed from a maximal Shor transversal. \ Then the probability that
\textsc{QRand}$_{\widetilde{\varphi}}()$ produces a character of the group
probe $\mathbb{Z}_{Q}$ which is sufficiently close to a primitive character of
the maximal cyclic subgroup $\mathbb{Z}_{P}$ of the hidden quotient group
$H_{\varphi}$ satisfies the following bound
\[
Prob_{\widetilde{\varphi}}\left(
\begin{tabular}
[c]{l}%
$\chi$ sufficiently close to some\\
primitive character of $\mathbb{Z}_{P}$%
\end{tabular}
\ \ \right)  \geq\frac{4}{\pi^{2}}\frac{\phi\left(  P\right)  }{P}\left(
1-\frac{P}{Q}\right)  ^{2}%
\]
\end{theorem}

\begin{proof}
Let $\chi_{\frac{y}{Q}}$ be a particular character of the group probe
$\mathbb{Z}_{Q}$ which is sufficiently close to some character of the maximal
cyclic subgroup $\mathbb{Z}_{P}$. \ We now compute the probability that
QRand$_{\widetilde{\varphi}}\left(  {}\right)  $ will produce this particular character.\bigskip

First consider the exceptional case when $Py=0\operatorname{mod}Q$. \ Using
the expression for $\left\|  \widetilde{\varphi}\left(  \chi_{\frac{y}{Q}%
}^{\bullet}\right)  \right\|  ^{2}$ given in \textbf{Corollary
\ref{InequalityCorollary}}, we have
\[
\left\|  \widetilde{\varphi}\left(  \chi_{\frac{y}{Q}}^{\bullet}\right)
\right\|  ^{2}=r\left(  q+1\right)  ^{2}+\left(  P-r\right)  q^{2}\geq
Pq^{2}=P\left(  \frac{Q-r}{P}\right)  ^{2}\geq\frac{1}{P}\left(  Q-P\right)
^{2}\text{ .}%
\]
So
\[
Prob_{\widetilde{\varphi}}\left(  \chi_{\frac{y}{Q}}\right)  =\frac{\left\|
\widetilde{\varphi}\left(  \chi_{\frac{y}{Q}}^{\bullet}\right)  \right\|
^{2}}{Q^{2}}\geq\frac{1}{P}\frac{P\left(  Q-P\right)  ^{2}}{Q^{2}}=\frac{1}%
{P}\left(  1-\frac{P}{Q}\right)  ^{2}\geq\frac{4}{\pi^{2}}\frac{1}{P}\left(
1-\frac{P}{Q}\right)  ^{2}\text{.}%
\]
\bigskip

Next consider the non-exceptional case when $Py\neq0\operatorname{mod}Q$.

In this case, \textbf{Proposition \ref{ArcProposition}} can be applied to both
terms in the numerator of the expression given in \textbf{Corollary
\ref{InequalityCorollary}}. \ Hence,
\begin{align*}
\left\|  \widetilde{\varphi}\left(  \chi_{\frac{y}{Q}}^{\bullet}\right)
\right\|  ^{2}  &  \geq\frac{4}{\pi^{2}}\left(  \frac{r\text{\textsc{Arc}%
}_{2\pi}^{2}\left[  \frac{Py}{Q}y\left(  q+1\right)  \right]  +\left(
P-r\right)  \text{\textsc{Arc}}_{2\pi}^{2}\left[  \frac{Py}{Q}yq\right]
}{\text{\textsc{Arc}}_{2\pi}^{2}\left(  \frac{Py}{Q}y\right)  }\right) \\
& \\
&  \geq\frac{4}{\pi^{2}}\left(  \frac{r\left(  q+1\right)  ^{2}%
\text{\textsc{Arc}}_{2\pi}^{2}\left[  \frac{Py}{Q}y\right]  +\left(
P-r\right)  q^{2}\text{\textsc{Arc}}_{2\pi}^{2}\left[  \frac{Py}{Q}y\right]
}{\text{\textsc{Arc}}_{2\pi}^{2}\left(  \frac{Py}{Q}y\right)  }\right) \\
& \\
&  \geq\frac{4}{\pi^{2}}r\left(  q+1\right)  ^{2}+\frac{4}{\pi^{2}}\left(
P-r\right)  q^{2}\geq\frac{4}{\pi^{2}}rq^{2}+\frac{4}{\pi^{2}}\left(
P-r\right)  q^{2}\\
& \\
&  \geq\frac{4}{\pi^{2}}Pq^{2}=\frac{4}{\pi^{2}}\frac{1}{P}\left(  Q-r\right)
^{2}\geq\frac{4}{\pi^{2}}\frac{1}{P}\left(  Q-P\right)  ^{2}%
\end{align*}
Thus,
\[
Prob_{\widetilde{\varphi}}\left(  \chi_{\frac{y}{Q}}\right)  =\frac{\left\|
\widetilde{\varphi}\left(  \chi_{\frac{y}{Q}}^{\bullet}\right)  \right\|
^{2}}{Q^{2}}\geq\frac{4}{\pi^{2}}\frac{1}{P}\left(  1-\frac{P}{Q}\right)  ^{2}%
\]
\bigskip

So, in either case we have
\[
Prob_{\widetilde{\varphi}}\left(  \chi_{\frac{y}{Q}}\right)  \geq\frac{4}%
{\pi^{2}}\frac{1}{P}\left(  1-\frac{P}{Q}\right)  ^{2}\text{.}%
\]
\bigskip

We now note that there is one-to-one correspondence between the characters of
$\mathbb{Z}_{P}$ and the sufficiently close characters of $\mathbb{Z}_{Q}$.
Hence, there are exactly $\phi\left(  P\right)  $ characters of the group
probe $\mathbb{Z}_{Q}$ which are sufficiently close some primitive character
of the maximal cyclic group $\mathbb{Z}_{P}$. The theorem follows.
\end{proof}

\bigskip

The following theorem can be found in \cite[Theorem 328, Section 18.4]{Hardy1}:

\bigskip

\begin{theorem}%
\[
\lim\inf\frac{\phi(N)}{N/\ln\ln N}=e^{-\gamma}\text{,}%
\]
where $\gamma$ denotes Euler's constant $\gamma=0.57721566490153286061\ldots$
, and where $e^{-\gamma}=0.5614594836\ldots$ . \ 
\end{theorem}

\bigskip

As a corollary, we have:

\bigskip

\begin{corollary}
\label{QRandSuccess}$Prob_{\widetilde{\varphi}}\left(  \chi\text{ sufficiently
close to some primitive character of }\mathbb{Z}_{P}\right)  $ is bounded
below by%
\[
\frac{4}{\pi^{2}\ln2}\cdot\frac{e^{-\gamma}-\epsilon\left(  P\right)  }{\lg\lg
Q}\cdot\left(  1-\frac{P}{Q}\right)  ^{2}\text{ ,}%
\]
where $\epsilon\left(  P\right)  $ is a monotone decreasing sequence
converging to zero. \ In terms of asymptotic notation,
\[
Prob_{\widetilde{\varphi}}\left(  \chi\text{ sufficiently close to some
primitive character of }\mathbb{Z}_{P}\right)  =\Omega\left(  \frac{1}{\lg\lg
Q}\right)  \text{ .}%
\]
\end{corollary}

\bigskip

For a proof of the above, please refer to \cite{Lomonaco2, Shor1}.

\bigskip

\section{Summary of Vintage $\mathbb{Z}_{Q}$ Shor QHSAs}

\bigskip\label{VintageSummary}

Let $\varphi:A\longrightarrow S$ be a map with hidden subgroup structure with
ambient group A free abelian of finite rank $n$, and with image of $\varphi$
finite. \ Then as a culmination of the mathematical developments in sections
11 through 19, we have the following \textbf{vintage} $\mathbb{Z}_{Q}$
\textbf{Shor QHSA} for finding the order $P=\operatorname{lcm}\left(
P_{1},P_{2},\ldots,P_{n}\right)  $ of the maximum cyclic subgroup $Z_{P}$ of
the hidden quotient group $H_{\varphi}=\bigoplus_{j=1}^{n}\mathbb{Z}_{P_{j}}$.
\ A flowchart of this algorithm is given in Figure 4.

\bigskip

\begin{center}
\fbox{\textsc{Vintage}\_\textsc{Shor}$\left(  \varphi,Q,n\right)  $}\bigskip
\end{center}

\begin{itemize}
\item[\qquad] \quad\# \textsc{Input}\textbf{:}$\quad\ \varphi:A\longrightarrow
S$ and $Q$ and rank $n$ of $A$

\item[\qquad] \quad\# \textsc{Output}\textbf{:} $P=\operatorname{lcm}\left(
P_{1},P_{2},\ldots P_{n}\right)  $ if hidden quotient group

\#\qquad\qquad\quad is $H_{\varphi}=\bigoplus_{j=1}^{\overline{n}}%
\mathbb{Z}_{P_{j}}$\bigskip

\item[\fbox{\textbf{Step 1}}] \quad Select a basis $a_{1}^{\prime}%
,a_{2}^{\prime},\ldots,a_{n}^{\prime}$ of $A$ and a generator $\widetilde{a}$
of $\mathbb{Z}_{Q}$\bigskip

\item[\fbox{\textbf{Step 2}}] $\left(  \iota_{\mu}:Z_{Q}\longrightarrow
A\right)  =$ \textsc{Random}\_\textsc{Shor}\_\textsc{transversl}$\left(
\ \left\{  a_{1}^{\prime},a_{2}^{\prime},\ldots,a_{n}^{\prime}\right\}
,\ Q,\ \widetilde{a},\ n\ \right)  $\bigskip

\item[\fbox{\textbf{Step 3}}] \quad Construct $\widetilde{\varphi}%
=\varphi\circ\iota_{\mu}:\mathbb{Z}_{Q}\longrightarrow S$\bigskip

\item[\fbox{\textbf{Step 4}}] $\quad\chi_{\frac{y}{Q}}=\ $\textsc{QRand}%
$_{\widetilde{\varphi}}\left(  {}\right)  $\bigskip

\item[\fbox{\textbf{Step 5}}] $\quad\left(  d^{\prime\prime},P^{\prime\prime
}\right)  =\left(  0,1\right)  $ \ \ \ \ \ \ \ \# 0-th Cont. Frac. Convergent
of $\frac{y}{Q}$

$\left(  d^{\prime},P^{\prime}\right)  =\left(  1,\left\lfloor \frac{Q}%
{y}\right\rfloor \right)  $ \ \ \# 1-th Cont. Frac. Convergent of $\frac{y}{Q}$

\textsc{Inner Loop}

\qquad$\left(  Save\_d^{\prime},Save\_P^{\prime}\right)  =\left(  d^{\prime
},P^{\prime}\right)  $

\qquad$\left(  d^{\prime},P^{\prime}\right)  =\ $\textsc{Next}\_\textsc{Cont}%
\_\textsc{Frac}\_\textsc{ConvergEnt}$\left(  \frac{y}{Q},\left(  d^{\prime
},P^{\prime}\right)  ,\left(  d^{\prime\prime},P^{\prime\prime}\right)
\right)  $

\qquad$\left(  d^{\prime\prime},P^{\prime\prime}\right)  =\left(
Save\_d^{\prime},Save\_P^{\prime}\right)  $

\qquad\textsc{If }$\varphi\left(  P^{\prime}a_{j}^{^{\prime}}\right)
=\varphi\left(  0\right)  $ for all $j=1,2,\ldots,n$\textsc{ Then
Goto}\fbox{\textbf{Step 6}}

\qquad\textsc{If }$\frac{d^{\prime}}{P^{\prime}}=\frac{y}{Q}$\textsc{ Then
Goto} \fbox{\textbf{Step 2}}

\textsc{Inner Loop Boundary}\bigskip

\item[\fbox{\textbf{Step 6}}] \quad\textsc{Output }$P^{\prime}$\textsc{ and Stop}
\end{itemize}

\bigskip%

\begin{center}
\includegraphics[
height=5.0393in,
width=3.039in
]%
{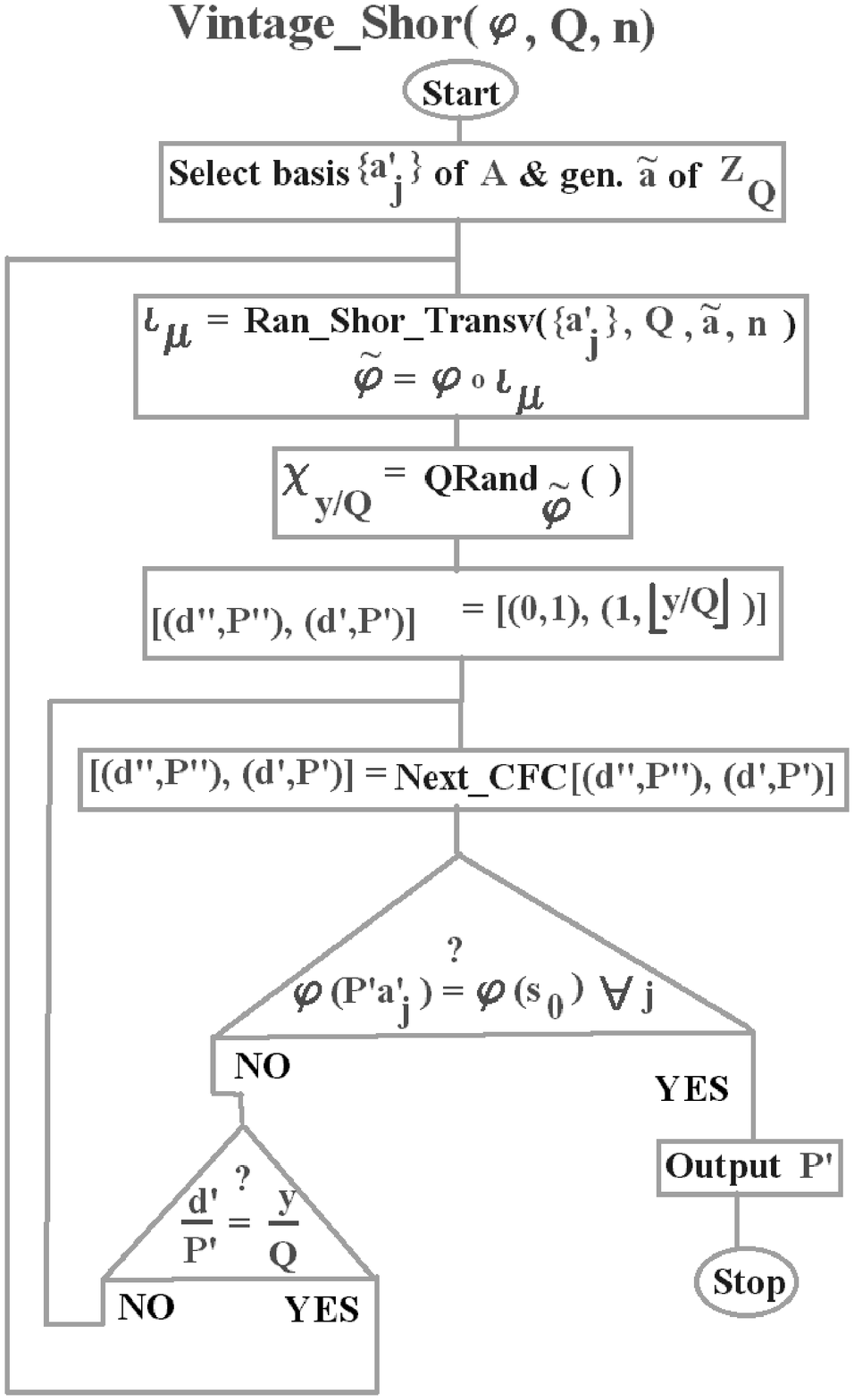}%
\\
\textbf{Figure 4.} \ Flowchart for Vintage $\mathbb{Z}_{Q}$ Shor QHSA. \ This
is a Wandering Shor algorithm.
\end{center}

\vspace{0.5in}

\section{A cursory analysis of complexity}

\bigskip\label{ComplexitySection}

We now make a cursory analysis of the algorithmic complexity of the vintage
$\mathbb{Z}_{Q}$ Shor algorithm. \ By the word ``cursory'' we mean that our
objective is to find an asymptotic bound which is by no means the tightest possible.

\bigskip

Our analysis is based on the following three assumptions:

\begin{itemize}
\item \textbf{Assumption 1.} \ \label{Assumptions}Conjecture \ref{Conjecture1}
(found in Appendix B) is true.\bigskip

\item \textbf{Assumption 2. \ }$U_{\widetilde{\varphi}}$ is of complexity
$O\left(  n^{2}\left(  \lg Q\right)  ^{3}\right)  $.\bigskip

\item \textbf{Assumption 3.} \ The integer $Q$ is chosen so that $Q=2^{L}\geq
P^{2}$, where $P=\operatorname{lcm}\left(  P_{1},P_{2},\ldots,P_{n}\right)  $.
\end{itemize}

\bigskip

The following theorem is an immediate consequence of \textbf{Assumption 2}.

\bigskip

\begin{theorem}
\label{QRand}Let
\[
\widetilde{\varphi}:\mathbb{Z}_{Q}\longrightarrow S
\]
be a map from the cyclic group $\mathbb{Z}_{Q}$ to a set $S$, where $Q=2^{L}$.

If $U_{\widetilde{\varphi}}$ is of algorithmic complexity
\[
O\left(  n^{2}\left(  \lg Q\right)  ^{3}\right)  \text{ ,}%
\]
then the algorithmic complexity of \textsc{QRand}$_{\widetilde{\varphi}%
}\left(  {}\right)  $ is the same, i.e.,
\[
O\left(  n^{2}\left(  \lg Q\right)  ^{3}\right)
\]
\end{theorem}

\begin{proof}
Steps 1 and 3 are each of the same algorithmic complexity as the quantum
Fourier transform\footnote{If instead the Hadamard-Walsh transform is used in
Step 1, then the complexity of Step 1 is $O\left(  \lg Q\right)  $.}, i.e., of
complexity $O\left(  \left(  \lg Q\right)  ^{2}\right)  $. \ (See
\cite[Chapter 5]{Nielsen1}.). \ Thus the dominant step in \textsc{QRand}%
$_{\widetilde{\varphi}}\left(  {}\right)  $\ is Step 2, which is by assumption
of complexity $O\left(  n^{2}\left(  \lg Q\right)  ^{3}\right)  $.
\end{proof}

\bigskip

The complexities of each step of the vintage $Z_{Q}$ Shor algorithm are given
below. \ An accompanying abbreviated flow chart of this algorithm is shown in
Figure 5.

\bigskip

\begin{itemize}
\item[\fbox{\textbf{Step 1}}] Step 1 is of algorithmic complexity is $O\left(
n\right)  $.\bigskip

\item[\fbox{\textbf{Step 2}}] By theorem \ref{ShorTransversal} of section
\ref{FindShorTransversal}, Step 2 is of average case complexity$O\left(
n^{2}\left(  \lg Q\right)  ^{3}\right)  $. \ By corollary
\ref{conditionalprob} of Appendix B, the probability that this step will be
successful, i.e., will produce a maximal Shor transversal, is $\Omega\left(
\left(  \frac{1}{\lg\lg Q}\right)  ^{n}\right)  $.\bigskip

\item[\fbox{\textbf{Step 3}}] Step 3 is of algorithmic complexity $O\left(
n\right)  $.\bigskip

\item[\fbox{\textbf{Step 4}}] By theorem \ref{QRand} given above, Step 4 is of
algorithmic complexity $O\left(  n^{2}\left(  \lg Q\right)  ^{3}\right)  $.
\ By corollary \ref{QRandSuccess} of section \ref{When}, the probability
(given that Step 2 is successful) that this step will be successful, i.e.,
will produce a character sufficiently close to a primitive character of the
maximal cyclic group $\mathbb{Z}_{P}$ is $\Omega\left(  \frac{1}{\lg\lg
Q}\right)  .$ \bigskip

\item[\fbox{\textbf{Step 5}}] This step is of algorithmic complexity $O\left(
n\left(  \lg Q\right)  ^{3}\right)  $. \ (See, for example, \cite{Knuth1}.)\bigskip

\item[\fbox{\textbf{Step 6}}] For Step 5 to branch to this step, both Steps 2
and 4 must be successful. \ Thus the probability of branching to step 6 is
$Prob_{Success}\left(  \text{Step 2}\right)  \cdot Prob_{Success}\left(
\text{Step 2}\right)  =\Omega\left(  \left(  \frac{1}{\lg\lg Q}\right)
^{n+1}\right)  $\bigskip
\end{itemize}

\bigskip

Since the Steps 2 through 5 loop will on average be executed $O\left(  \left(
\lg\lg Q\right)  ^{n+1}\right)  $ times, the average algorithmic complexity of
the Vintage $Z_{Q}$ Shor algorithm is $O\left(  n^{2}\left(  \log Q\right)
^{3}\left(  \lg\lg Q\right)  ^{n+1}\right)  $. \ (This is, of course, not the
tightest possible asymptotic bound.) \ We formalize this analysis as a theorem:\bigskip%

\begin{center}
\includegraphics[
trim=0.001830in 0.021504in 0.000000in -0.021504in,
height=4.5238in,
width=4.0413in
]%
{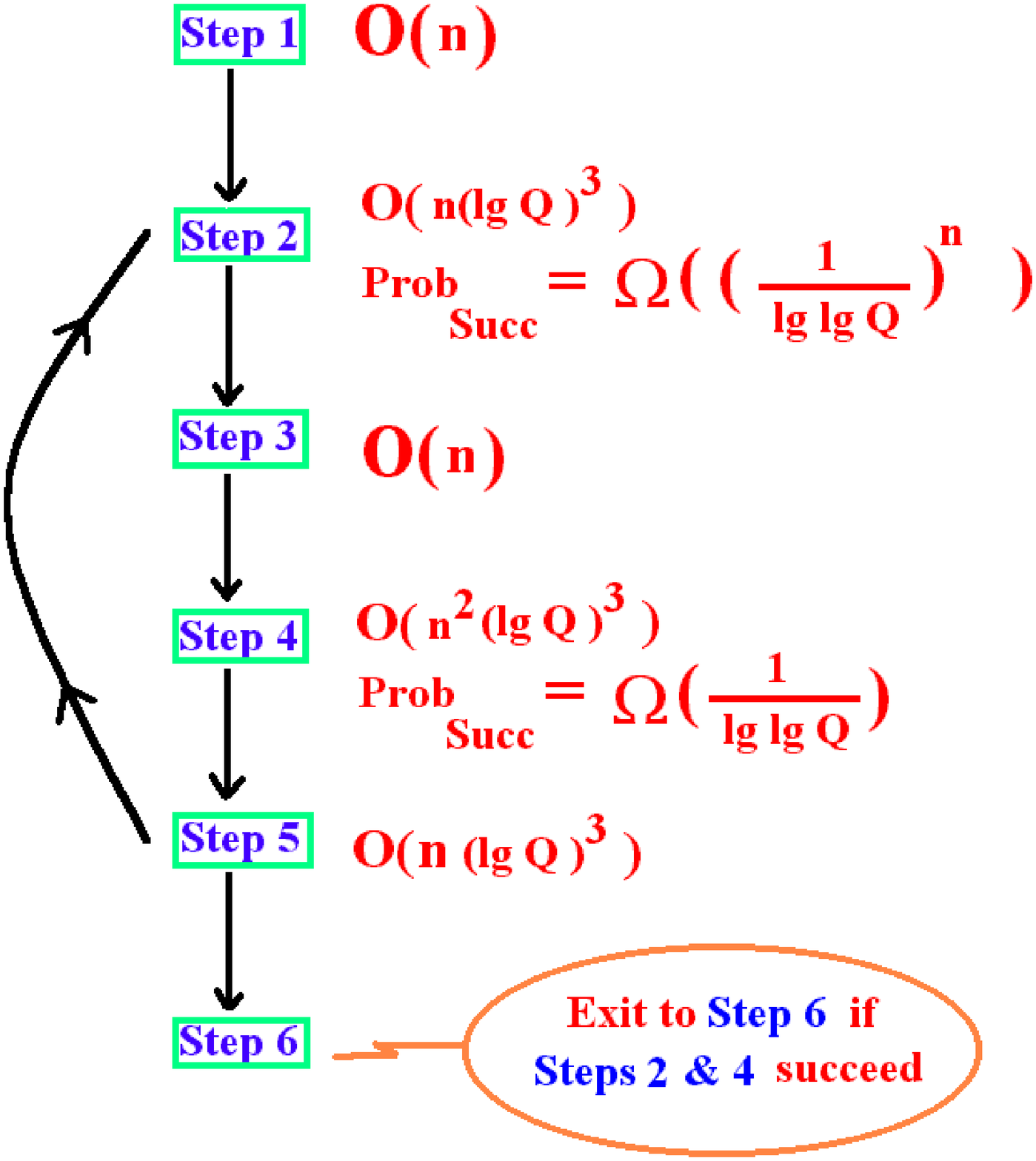}%
\\
\textbf{Figure 5.} \ An abbreviated flowchart of the vintage $Z_{Q}$ Shor
Algorithm. \ The probability of a successful exit to Step 6 is $\Omega\left(
\left(  \frac{1}{\lg\lg Q}\right)  ^{n+1}\right)  $. \ Hence, the average
number of times Steps 2 through 5 are executed is $O\left(  \left(  \lg\lg
Q\right)  ^{n+1}\right)  $.
\end{center}

\bigskip

\begin{theorem}
\label{ZComplexity}Assuming the three assumptions given in section
\ref{ComplexitySection}, the average algorithmic complexity of the Vintage
$\mathbb{Z}_{Q}$ Shor algorithm for finding the maximal cyclic subgroup
$\mathbb{Z}_{P}$ of the hidden quotient group $H_{\varphi}=\bigoplus_{j=1}%
^{n}\mathbb{Z}_{P_{j}}$ is
\[
O\left(  n^{2}\left(  \lg Q\right)  ^{3}\left(  \lg\lg Q\right)
^{n+1}\right)
\]
\end{theorem}

\bigskip

\section{Two alternative vintage $\mathbb{Z}_{Q}$ Shor algorithms}

\bigskip\label{Alternate1}\label{Alternate2}

As two alternatives to the algorithm described in the last two sections, we
give below two other vintage $\mathbb{Z}_{Q}$ Shor algorithms. \ Unlike the
above described algorithm, these two alternative algorithms do not depend on
finding a maximal Shor transversal. \ The first finds the order of the maximal
cyclic subgroup $\mathbb{Z}_{P}$ of the hidden quotient group $H_{\varphi}$.
\ The second finds the entire hidden subgroup $K_{\varphi}$. Flowcharts for
these two quantum algorithms are given in Figures 6 and 7.

\bigskip

An optimal choice for the parameter $K$ of the following algorithm is not
known at this time.

\begin{center}
\fbox{\textsc{Alternative1}\_\textsc{Vintage}\_\textsc{Shor}$\left(
\varphi,Q,n,\ K\right)  $}
\end{center}

\begin{itemize}
\item[\qquad] \quad\# \textsc{Input}\textbf{:}$\quad\ \varphi:A\longrightarrow
S$, $Q$, rank $n$ of $A$, and number of

\item[\qquad] \quad\# \ \ \ \ \ \ \ \ \ \ \ \ \ \ inner loop iterations $K$

\item[\qquad] \quad\# \textsc{Output}\textbf{:} $P=\operatorname{lcm}\left(
P_{1},P_{2},\ldots P_{n}\right)  $ if hidden quotient group

\#\qquad\qquad\quad is $H_{\varphi}=\bigoplus_{j=1}^{\overline{n}}%
\mathbb{Z}_{P_{j}}$\bigskip

\item[\fbox{\textbf{Step 1}}] \quad\textsc{Set }$P=1$\bigskip

\item[\fbox{\textbf{Step 2}}] \quad Select a basis $a_{1}^{\prime}%
,a_{2}^{\prime},\ldots,a_{n}^{\prime}$ of $A$ and a generator $\widetilde{a}$
of $\mathbb{Z}_{Q}$\bigskip

\item[\fbox{\textbf{Step 3}}] \quad\textsc{Outer Loop}\bigskip

\begin{itemize}
\item[\fbox{\textbf{Step 4}}] \quad\textsc{Inner Loop }for $K$ iterations\bigskip

\begin{itemize}
\item[\fbox{\textbf{Step 5}}] \quad$\!$\negthinspace$\!$\negthinspace$\left(
\iota_{\mu}\!\!:Z_{Q}\longrightarrow A\right)  =$ \textsc{Rand}\_\textsc{Shor}%
\_\textsc{transvr}$\left(  \left\{  a_{1}^{\prime},a_{2}^{\prime},\ldots
,a_{n}^{\prime}\right\}  ,Q,\widetilde{a},n\right)  $\bigskip

\item[\fbox{\textbf{Step 6}}] \quad Construct $\widetilde{\varphi}%
=\varphi\circ\iota_{\mu}:\mathbb{Z}_{Q}\longrightarrow S$\bigskip

\item[\fbox{\textbf{Step 7}}] $\quad\chi_{\frac{y}{Q}}=\ $\textsc{QRand}%
$_{\widetilde{\varphi}}\left(  {}\right)  $\bigskip

\item[\fbox{\textbf{Step 8}}] $\quad\left(  d^{\prime\prime},P^{\prime\prime
}\right)  =\left(  0,1\right)  $ \ \ \ \ \# 0-th Cont. Frac. Converg. of
$\frac{y}{Q}$
\end{itemize}
\end{itemize}

\qquad\qquad$\ \left(  d^{\prime},P^{\prime}\right)  =\left(  1,\left\lfloor
\frac{Q}{y}\right\rfloor \right)  $ \# 1-th Cont. Frac. Converg. of
$\frac{y}{Q}$

\qquad\qquad\ \textsc{Innermost Loop}

\qquad\qquad\qquad$\!$\negthinspace$\left(  Save\_d^{\prime},Save\_P^{\prime
}\right)  =\left(  d^{\prime},P^{\prime}\right)  $

\qquad\qquad\qquad$\!$\negthinspace$\left(  d^{\prime},P^{\prime}\right)
=\ $\textsc{Nxt}\_\textsc{Cont}\_\textsc{Frac}\_\textsc{Convrg}$\left(
\frac{y}{Q},\left(  d^{\prime},P^{\prime}\right)  ,\left(  d^{\prime\prime
},P^{\prime\prime}\right)  \right)  $

\qquad\qquad\qquad$\!$\negthinspace$\left(  d^{\prime\prime},P^{\prime\prime
}\right)  =\left(  Save\_d^{\prime},Save\_P^{\prime}\right)  $

\qquad\qquad\qquad$\!$\negthinspace\textsc{If }$\varphi\left(  P^{\prime}%
\iota_{\mu}\left(  \widetilde{a}\right)  \right)  =\varphi\left(  0\right)
$\textsc{ Then Goto} \fbox{\textbf{Step 9}}

\qquad\qquad\qquad$\!$\negthinspace\textsc{If }$\frac{d^{\prime}}{P^{\prime}%
}=\frac{y}{Q}$\textsc{ Then Goto} \fbox{\textbf{Step 4}}

\qquad\qquad\ \textsc{Innermost Loop Boundary}\bigskip

\begin{itemize}
\item
\begin{itemize}
\item[\fbox{\textbf{Step 9}}] $\quad P=\operatorname{lcm}\left(  P,P^{\prime
}\right)  $\bigskip
\end{itemize}

\item[\fbox{\textbf{Step 10}}] $\quad$\textsc{Inner Loop Lower Boundary}\bigskip

\item[\fbox{\textbf{Step 11}}] \quad\textsc{If }$\varphi\left(  Pa_{j}\right)
=\varphi\left(  0\right)  $\textsc{ for }$j=1,2,\ldots,n$\textsc{ Then Goto
}\fbox{\textbf{Step 13}}\bigskip
\end{itemize}

\item[\fbox{\textbf{Step 12}}] \quad\textsc{Outer Loop Lower Boundary}\bigskip

\item[\fbox{\textbf{Step 13}}] \quad\textsc{Output }$P^{\prime}$\textsc{ and Stop}
\end{itemize}%

\begin{center}
\includegraphics[
height=6.0701in,
width=3.5405in
]%
{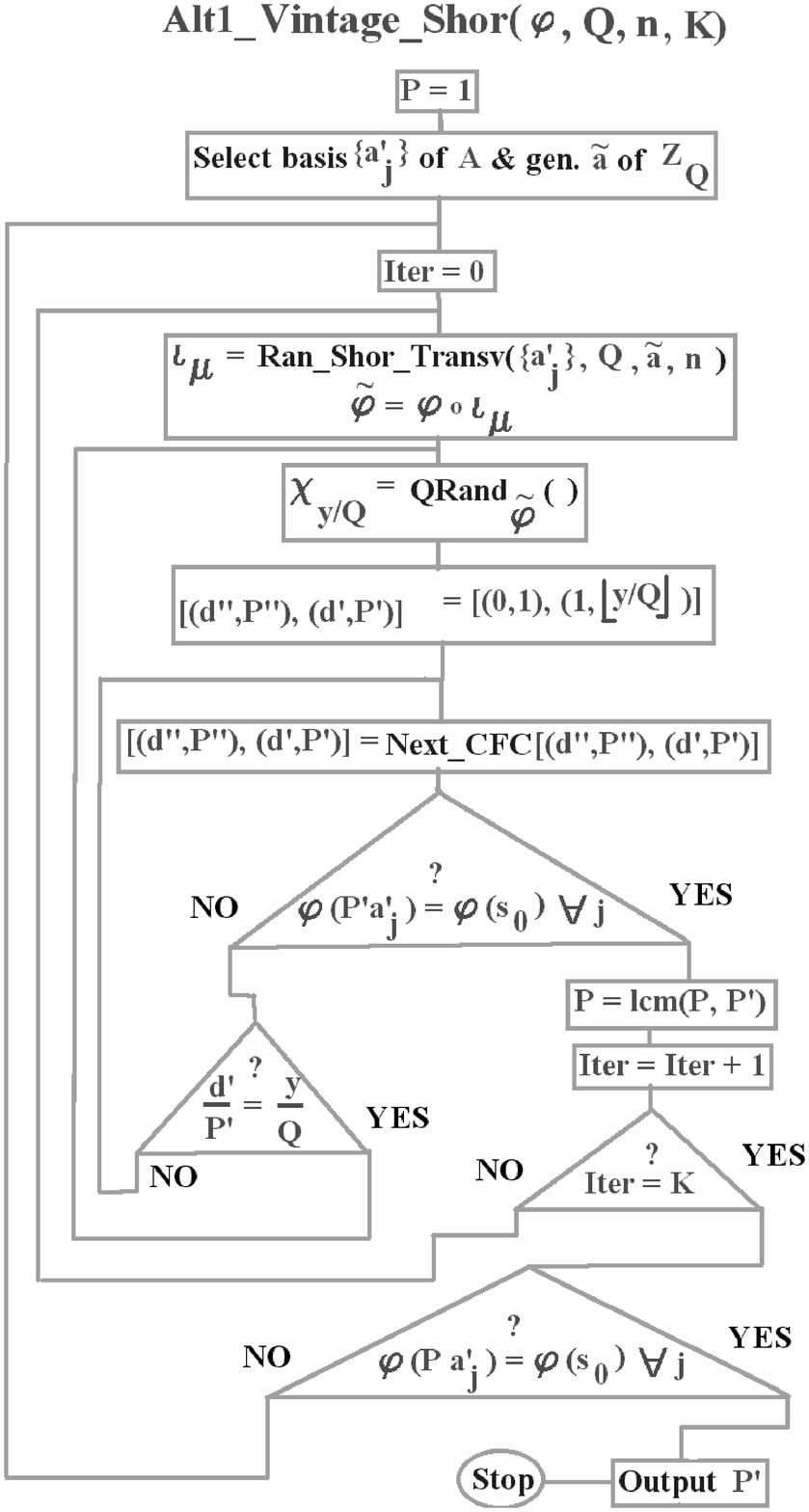}%
\\
\textbf{Figure 6.} \ The First Alternate Vintage $\mathbb{Z}_{Q}$ Shor QHSA.
\ This is a Wandering Shor algorithm.
\end{center}

\bigskip

The following wandering Shor algorithm actually finds the entire hidden
quotient group $H_{\varphi}$, and hence the hidden subgroup $K_{\varphi}$:

\bigskip

\begin{center}
\fbox{\textsc{Alternative2}\_\textsc{Vintage}\_\textsc{Shor}$\left(
\varphi,Q,n\right)  $}\bigskip
\end{center}

\begin{itemize}
\item[\qquad] \quad\# \textsc{Input}\textbf{:}$\quad\ \varphi:A\longrightarrow
S$, $Q$, rank $n$ of $A$

\item[\qquad] \quad\# \textsc{Output}\textbf{:} A matrix $\mathfrak{G}$ with
row span equal to the

\#\qquad\qquad\quad hidden subgroup $K_{\varphi}=\bigoplus_{j=1}^{n}%
P_{j}\mathbb{Z}$\bigskip

\item[\fbox{\textbf{Step 1}}] \quad\textsc{Set }$\mathfrak{G}=\left[
\quad\right]  $\textsc{ and NonZeroRows }$=0$\bigskip

\item[\fbox{\textbf{Step 2}}] \quad Select a basis $a_{1}^{\prime}%
,a_{2}^{\prime},\ldots,a_{n}^{\prime}$ of $A$ and a generator $\widetilde{a}$
of $\mathbb{Z}_{Q}$\bigskip

\item[\fbox{\textbf{Step 3}}] \quad\textsc{Outer Loop Until NonZeroRows }$=n$\bigskip

\begin{itemize}
\item
\begin{itemize}
\item[\fbox{\textbf{Step 4}}] \quad$\!\!\!\!\!\!\left(  \iota_{\mu}%
\!\!:Z_{Q}\longrightarrow A\right)  =$ \textsc{Ran}\_$\!\!$\textsc{Shor}%
\_$\!\!$\textsc{Transvrsl}$\left(  \left\{  a_{1}^{\prime},a_{2}^{\prime
},\ldots,a_{n}^{\prime}\right\}  ,Q,\widetilde{a},n\right)  $\bigskip

\item[\fbox{\textbf{Step 5}}] \quad Construct $\widetilde{\varphi}%
=\varphi\circ\iota_{\mu}:\mathbb{Z}_{Q}\longrightarrow S$\bigskip

\item[\fbox{\textbf{Step 6}}] $\quad\chi_{\frac{y}{Q}}=\ $\textsc{QRand}%
$_{\widetilde{\varphi}}\left(  {}\right)  $\bigskip

\item[\fbox{\textbf{Step 7}}] $\quad\left(  d^{\prime\prime},P^{\prime\prime
}\right)  =\left(  0,1\right)  $ \ \ \ \ \# 0-th Cont. Frac. Converg. of
$\frac{y}{Q}$
\end{itemize}
\end{itemize}

\qquad\qquad$\ \left(  d^{\prime},P^{\prime}\right)  =\left(  1,\left\lfloor
\frac{Q}{y}\right\rfloor \right)  $ \# 1-th Cont. Frac. Converg. of
$\frac{y}{Q}$

\qquad\qquad\ \textsc{Inner Loop}

\qquad\qquad\qquad$\!\!\left(  Save\_d^{\prime},Save\_P^{\prime}\right)
=\left(  d^{\prime},P^{\prime}\right)  $

\qquad\qquad\qquad$\!\!\left(  d^{\prime},P^{\prime}\right)  =\ $%
\textsc{Next}\_$\!\!$\textsc{Cont}\_$\!\!$\textsc{Frac}\_$\!\!$%
\textsc{Converg}$\left(  \frac{y}{Q},\left(  d^{\prime},P^{\prime}\right)
,\left(  d^{\prime\prime},P^{\prime\prime}\right)  \right)  $

\qquad\qquad\qquad$\!\!\left(  d^{\prime\prime},P^{\prime\prime}\right)
=\left(  Save\_d^{\prime},Save\_P^{\prime}\right)  $

\qquad\qquad\qquad$\!\!$\textsc{If }$\varphi\left(  P^{\prime}\iota_{\mu
}\left(  \widetilde{a}\right)  \right)  =\varphi\left(  0\right)  $\textsc{
Then Goto} \fbox{\textbf{Step 8}}

\qquad\qquad\qquad$\!\!$\textsc{If }$\frac{d^{\prime}}{P^{\prime}}=\frac{y}%
{Q}$\textsc{ Then Goto} \fbox{\textbf{Step 11}}

\qquad\qquad\ \textsc{Inner Loop Boundary: Continue}\bigskip

\begin{itemize}
\item
\begin{itemize}
\item[\fbox{\textbf{Step 8}}] $\quad\mathfrak{G}=\left[
\begin{array}
[c]{c}%
\mathfrak{G}\\
------------\\%
\begin{array}
[c]{cccc}%
P^{\prime}\lambda_{1}^{\prime} & P^{\prime}\lambda_{2}^{\prime} & \ldots &
P^{\prime}\lambda_{n}^{\prime}%
\end{array}
\end{array}
\right]  $\bigskip

\item[\fbox{\textbf{Step 9}}] $\quad\mathfrak{G}=$\textsc{ Put}\_\textsc{In}%
\_\textsc{Echelon}\_\textsc{Canonical}\_\textsc{Form}$\left(  \mathfrak{G}%
\right)  $\bigskip

\item[\fbox{\textbf{Step 10}}] \quad\textsc{NonZeroRows }$=$ \textsc{Number}%
\_\textsc{of}\_\textsc{Non}\_\textsc{Zero}\_\textsc{Rows}$\left(
\mathfrak{G}\right)  $\bigskip
\end{itemize}
\end{itemize}

\item[\fbox{\textbf{Step 11}}] $\quad$\textsc{Outer Loop Lower Boundary:
Continue}\bigskip

\item[\fbox{\textbf{Step 12}}] \quad\textsc{Output }matrix $\mathfrak{G}%
$\textsc{ and Stop}\bigskip
\end{itemize}

\bigskip%

\begin{center}
\includegraphics[
height=6.7749in,
width=4.2436in
]%
{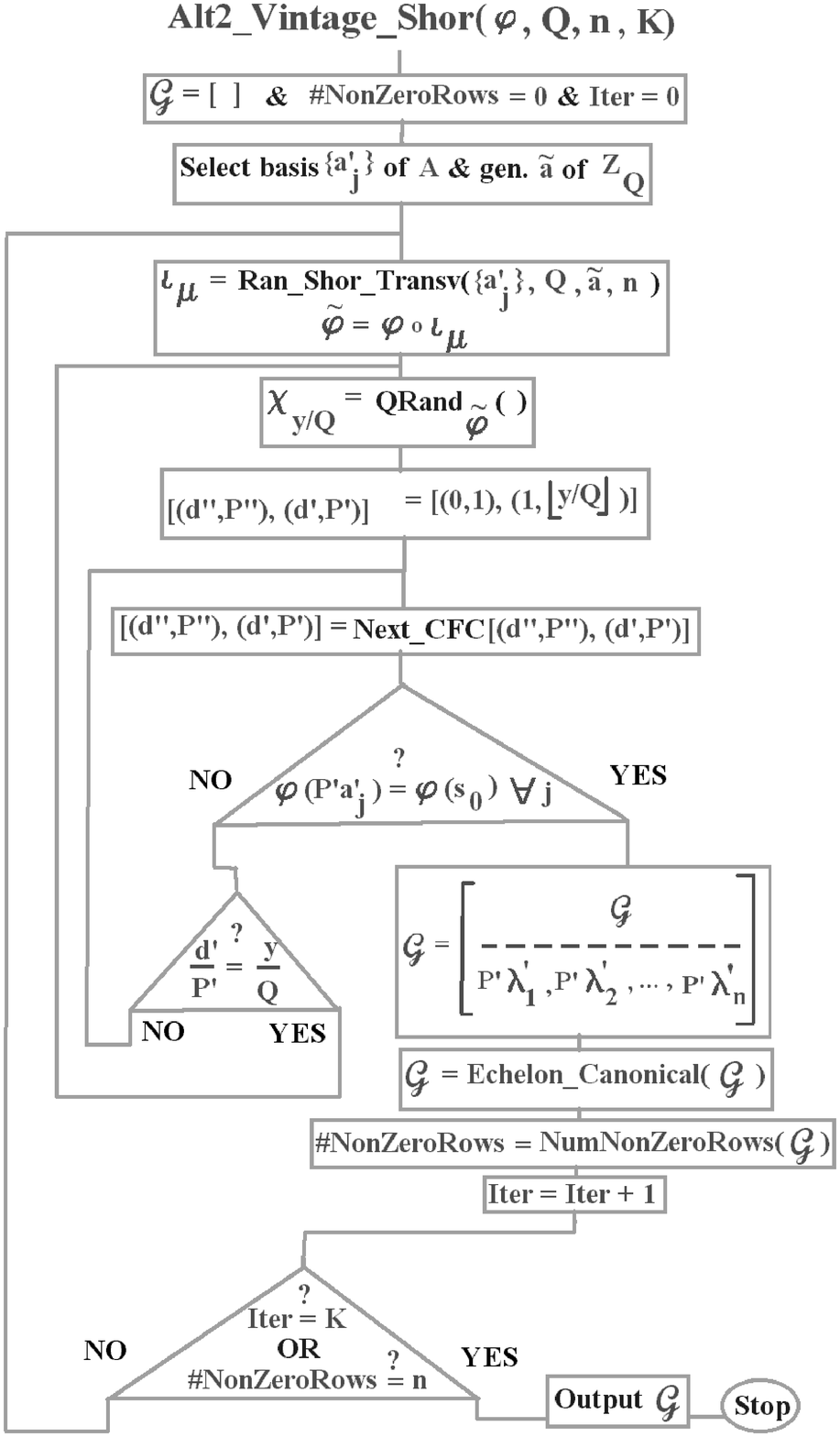}%
\\
\textbf{Figure 7.} \ The Second Alternate Vintage $\mathbb{Z}_{Q}$ Shor QHSA.
\ This is a Wandering Shor algorithm.
\end{center}

\bigskip\pagebreak 

\part{Epilogue\bigskip}

\bigskip

\section{Conclusion}

\bigskip

Each of the three vintage $\mathbb{Z}_{Q}$ Shor QHSAs created in this paper is
a natural generalization of Shor's original quantum factoring algorithm to
free abelian groups $A$ of finite rank $n$ . \ The first two of the three find
a maximal cyclic subgroup $\mathbb{Z}_{P}$ of the hidden quotient group
$H_{\varphi}$. \ The last of the three does more. \ It finds the entire hidden
quotient group $H_{\varphi}$.

\bigskip

We also note that these QHSAs can be viewed from yet another perspective as
\textbf{wandering Shor algorithms} on free abelian groups. \ \ By this we mean
quantum algorithms which, with each iteration, first select a random cyclic
direct summand $\mathbb{Z}$ of the ambient group $A$ and then apply one
iteration of the standard Shor algorithm on $\mathbb{Z}$ to produce a random
character of the ``approximating'' group $\widetilde{A}=\mathbb{Z}_{Q}$. \ 

\bigskip

From this perspective, under the assumptions given in section
\ref{ComplexitySection}, the algorithmic complexity of the first of these
wandering QHSAs is found to be%
\[
O\left(  n^{2}\left(  \lg Q\right)  ^{3}\left(  \lg\lg Q\right)
^{n+1}\right)  \text{ .}%
\]

\bigskip

Obviously, much remains to be accomplished. \ 

\bigskip

It should be possible to extend the vintage $\mathbb{Z}_{Q}$ Shor algorithms
to quantum algorithms with more general group probes of the form
\[
\widetilde{A}=%
{\displaystyle\bigoplus\limits_{j=1}^{m}}
\mathbb{Z}_{Q_{j}}%
\]
for $m>1$. \ This would be a full generalization of Shor's quantum factoring
algorithm to the abelian category. \ 

\bigskip

It is hoped that this paper will provide a useful stepping stone to the
construction of QHSAs on non-abelian groups. \ 

\bigskip

\section{Acknowledgement}

\bigskip

We would like to thank Tom Armstrong, Howard Brandt, Eric Rains, Fernando
Souza, Umesh Vazirani, and Yaacov Yesha for some helpful discussions. \ We
would also like to thank the referee for some helpful suggestions.

\bigskip

\section{Appendix A. Continued fractions}

\qquad\bigskip

We give a brief summary of those aspects of the theory of continued fractions
that are relevant to this paper. (For a more in-depth explanation of the
theory of continued fractions, please refer, for example, to \cite{Hardy1} and
\cite{LeVeque1}.)

\bigskip

Every positive rational number $\xi$ can be written as an expression in the
form
\[
\xi=a_{0}+\frac{1}{a_{1}+\frac{\overset{}{\underset{}{1}}}{a_{2}%
+\frac{\overset{}{\underset{}{1}}}{a_{3}+\frac{\overset{}{\underset{}{1}}%
}{\cdots+\frac{\overset{}{\underset{}{1}}}{\overset{}{a_{N}}}}}}}\text{ ,}%
\]
where $a_{0}$ is a non-negative integer, and where $a_{1},\ldots,a_{N}$ are
positive integers. \ Such an expression is called a (finite, simple)
\textbf{continued fraction}, and is uniquely determined by $\xi$ provided we
impose the condition $a_{N}>1$. \ For typographical simplicity, we denote the
above continued fraction by
\[
\left[  a_{0},a_{1},\ldots,a_{N}\right]  \text{ .}%
\]

The continued fraction expansion of $\xi$ can be computed with the following
recurrence relation, which always terminates if $\xi$ is rational:
\[
\fbox{$\overset{}{\underset{}{%
\begin{array}
[c]{lll}%
\left\{
\begin{array}
[c]{r}%
a_{0}=\left\lfloor \xi\right\rfloor \\
\\
\xi_{0}=\xi-a_{0}%
\end{array}
\right.  \text{ ,} & \text{and if }\xi_{n}\neq0\text{, then} & \left\{
\begin{array}
[c]{l}%
a_{n+1}=\left\lfloor 1/\xi_{n}\right\rfloor \\
\\
\xi_{n+1}=\frac{1}{\xi_{n}}-a_{n+1}%
\end{array}
\right.
\end{array}
}}$}%
\]

\bigskip

The $n$-th \textbf{convergent} ($0\leq n\leq N$) of the above continued
fraction is defined as the rational number $\xi_{n}$ given by
\[
\xi_{n}=\left[  a_{0},a_{1},\ldots,a_{n}\right]  \text{ .}%
\]
Each convergent $\xi_{n}$ can be written in the form, $\xi_{n}=\frac{p_{n}%
}{q_{n}}$, where $p_{n}$ and $q_{n}$ are relatively prime integers (
$\gcd\left(  p_{n},q_{n}\right)  =1$). The integers $p_{n}$ and $q_{n}$ are
determined by the recurrence relation
\[
\fbox{$%
\begin{array}
[c]{lll}%
p_{0}=a_{0}, & p_{1}=a_{1}a_{0}+1, & p_{n}=a_{n}p_{n-1}+p_{n-2},\\
&  & \\
q_{0}=1, & q_{1}=a_{1}, & q_{n}=a_{n}q_{n-1}+q_{n-2}\text{ \ .}%
\end{array}
$}%
\]
The subroutine
\[
\text{\textsc{Next}\_\textsc{Cont}\_\textsc{Frac}\_\textsc{Convergent}}%
\]
found in the vintage $\mathbb{Z}_{Q}$ Shor algorithm given in section
\ref{VintageSummary} is an embodiment of the above recursion.

\bigskip

This recursion is used because of the following theorem which can be found in
\cite[Theorem 184, Secton 10.15]{Hardy1}:

\bigskip

\begin{theorem}
\label{CFConvergent}Let $\xi$ be a real number, and let $d$ and $P$ be
integers with $P>0$. If
\[
\left|  \xi-\frac{d}{P}\right|  \leq\frac{1}{2P^{2}}\text{ ,}%
\]
then the rational number $d/P$ is a convergent of the continued fraction
expansion of $\xi$. \ 
\end{theorem}

\bigskip

\section{Appendix B. Probability Distributions on Integers}

\bigskip

Let
\[
Prob_{Q}:\left\{  1,2,\ldots,Q\right\}  \longrightarrow\left[  0,1\right]
\]
denote the uniform probability distribution on the finite set of integers
$\left\{  1,2,\ldots,Q\right\}  $. \ Thus, the probability that a random
integer $\lambda$ from $\left\{  1,2,\ldots,Q\right\}  $ is divisible by a
given prime $p$ is
\[
Prob_{Q}\left(  \ p\!\mid\!\lambda\ \right)  =\frac{\left\lfloor
Q/p\right\rfloor }{Q}\leq\frac{1}{p}\text{ ,}%
\]
where `$\left\lfloor -\right\rfloor $' denotes the the floor function.

\bigskip

The limit $Prob_{\infty}$ , should it exist, of the probability distribution
$Prob_{Q}$ as $Q$ approaches infinity, i.e.,
\[
Prob_{\infty}=\lim_{Q\longrightarrow\infty}Prob_{\infty}\text{ ,}%
\]
will turn out to be a useful tool. \ Since $Prob_{\infty}$ is not a
probability distribution, we will call it a \textbf{pseudo-probability
distribution} on the integers $\mathbb{Z}$. \ It immediately follows that
\[
Prob_{\infty}\left(  \ p\!\mid\!\lambda\ \right)  =\frac{1}{p}\text{.}%
\]
In this sense, we say that the pseudo-probability of a random integer
$\lambda\in\mathbb{Z}$ being divisible by a given prime $p$ is $1/p$.

\bigskip

\begin{theorem}
Let $n$ be an integer greater than $1$. \ Let $\lambda_{1}^{\prime}%
,\lambda_{2}^{\prime},\ldots,\lambda_{n}^{\prime}$, be $n$ \ integers selected
randomly and independently with replacement from the set $\left\{
1,2,\ldots,Q\right\}  $ according to the uniform probability distribution.
Then the probability that
\[
\gcd\left(  \lambda_{1}^{\prime},\lambda_{2}^{\prime},\ldots,\lambda
_{n}^{\prime}\right)  =1
\]
is
\[
Prob_{Q}\left(  \underset{}{\overset{}{\gcd\left(  \lambda_{1}^{\prime
},\lambda_{2}^{\prime},\ldots,\lambda_{n}^{\prime}\right)  =1}}\right)  =%
{\displaystyle\sum\limits_{k=1}^{Q}}
\mu\left(  k\right)  \left(  \frac{\left\lfloor Q/k\right\rfloor }{Q}\right)
^{n}\text{ ,}%
\]
where `$\ \left\lfloor -\right\rfloor $' and `$\mu\left(  -\right)  $'
respectively denote the floor and M\"{o}bius functions.

Moreover,
\[
Prob_{\infty}\left(  \underset{}{\overset{}{\gcd\left(  \lambda_{1}^{\prime
},\lambda_{2}^{\prime},\ldots,\lambda_{n}^{\prime}\right)  =1}}\right)
=\zeta\left(  n\right)  ^{-1}\text{,}%
\]
where $\zeta\left(  n\right)  $ denotes the Riemann zeta function
$\zeta\left(  n\right)  =\sum_{k=1}^{\infty}\frac{1}{k^{n}}$.
\end{theorem}

\begin{proof}
Let $Primes_{Q}$ denote the set of primes less than or equal to $Q$. \ 

For each prime $p$ and integer $0<j\leq n$, let $A_{pj}$ denote the set
\[
A_{pj}=\left\{  \overrightarrow{\lambda}\in\left\{  1,\ldots,Q\right\}
^{n}:p\mid\lambda_{j}\right\}  \text{ .}%
\]

\bigskip

Since
\[%
{\displaystyle\bigcap\limits_{p\in Primes_{Q}}}
{\displaystyle\bigcup\limits_{j=1}^{n}}
\overline{A}_{pj}=\left\{  \overrightarrow{\lambda}\in\left\{  1,\ldots
,Q\right\}  ^{n}:\forall p\exists j\ \ p\mid\lambda_{j}\right\}  \text{, }%
\]
we have
\[
Prob_{Q}\left(  \gcd\left(  \lambda_{1},\lambda_{2},\ldots,\lambda_{n}\right)
=1\right)  =Prob_{Q}\left(  \bigcap\limits_{p\in Primes_{Q}}\bigcup
\limits_{j=1}^{n}\overline{A}_{pj}\right)  \text{ ,}%
\]
where $\overline{A}_{pj}$ denotes the complement of $A_{pj}$.

\bigskip

We proceed to compute $Prob_{Q}\left(  \bigcap\limits_{p\in Primes_{Q}}%
\bigcup\limits_{j=1}^{n}\overline{A}_{pj}\right)  $ by first noting that:
\[
Prob_{Q}\left(  \bigcap\limits_{p\in Primes_{Q}}\bigcup\limits_{j=1}%
^{n}\overline{A}_{pj}\right)  =1-Prob_{Q}\left(  \bigcup\limits_{p\in
Primes_{Q}}\bigcap\limits_{j=1}^{n}A_{pj}\right)  \text{ .}%
\]
So by the inclusion/exclusion principle, we have
\begin{align*}
Prob_{Q}\left(  \bigcup\limits_{p\in Primes_{Q}}\bigcap\limits_{j=1}^{n}%
A_{pj}\right)   &  =-%
{\displaystyle\sum\limits_{\underset{S\neq\varnothing}{S\subseteq Primes_{Q}}%
}}
\left(  -1\right)  ^{\left|  S\right|  }Prob_{Q}\left(  \bigcap\limits_{p\in
S}\bigcap\limits_{j=1}^{n}A_{pj}\right) \\
& \\
&  =-%
{\displaystyle\sum\limits_{\underset{S\neq\varnothing}{S\subseteq Primes_{Q}}%
}}
\left(  -1\right)  ^{\left|  S\right|  }Prob_{Q}\left(  \bigcap\limits_{j=1}%
^{n}\bigcap\limits_{p\in S}A_{pj}\right)
\end{align*}
Since the $\lambda_{j}$'s are independent random variables, we have
\[
Prob_{Q}\left(  \bigcap\limits_{j=1}^{n}\bigcap\limits_{p\in S}A_{pj}\right)
=%
{\displaystyle\prod\limits_{j=1}^{n}}
Prob_{Q}\left(  \bigcap\limits_{p\in S}A_{pj}\right)
\]
Moreover, it follows from a straight forward counting argument that
\[
Prob_{Q}\left(  \bigcap\limits_{p\in S}A_{pj}\right)  =\left\lfloor
Q/\prod_{p\in S}p\right\rfloor /Q\text{ ,}%
\]
from which we obtain
\[
Prob_{Q}\left(  \bigcap\limits_{j=1}^{n}\bigcap\limits_{p\in S}A_{pj}\right)
=%
{\displaystyle\prod\limits_{j=1}^{n}}
\left\lfloor Q/\prod_{p\in S}p\right\rfloor /Q=\left(  \left\lfloor
Q/\prod_{p\in S}p\right\rfloor /Q\right)  ^{n}\text{ .}%
\]
Thus,
\[
Prob_{Q}\left(  \bigcap\limits_{p\in Primes_{Q}}\bigcup\limits_{j=1}%
^{n}\overline{A}_{pj}\right)  =%
{\displaystyle\sum\limits_{S\subseteq Primes_{Q}}}
\left(  -1\right)  ^{\left|  S\right|  }\left(  \left\lfloor Q/\prod_{p\in
S}p\right\rfloor /Q\right)  ^{n}%
\]
This last expression expands to
\[
1-\sum_{\underset{p\text{ Prime}}{p\leq Q}}\left(  \frac{\left\lfloor
Q/p\right\rfloor }{Q}\right)  ^{n}+\sum_{\underset{p,p^{\prime}\text{ Prime}%
}{p<p^{\prime}\leq Q}}\left(  \frac{\left\lfloor Q/pq\right\rfloor }%
{Q}\right)  ^{n}-\sum_{\underset{p,p^{\prime},p^{\prime\prime}\text{ Prime}%
}{p<p^{\prime}<p^{\prime\prime}\leq Q}}\left(  \frac{\left\lfloor
Q/pp^{\prime}p^{\prime\prime}\right\rfloor }{Q}\right)  ^{n}+\ \ldots\text{ ,}%
\]
which can be rewritten as
\[%
{\displaystyle\sum\limits_{k=1}^{Q}}
\mu\left(  k\right)  \left(  \frac{\left\lfloor Q/k\right\rfloor }{Q}\right)
^{n}%
\]
since $\mu\left(  k\right)  =0$ for all integers $k$ that are not squarefree.

The last part of this theorem follows immediately from the fact that
\[
\lim_{Q\longrightarrow\infty}%
{\displaystyle\sum\limits_{k=1}^{Q}}
\mu\left(  k\right)  \left(  \frac{\left\lfloor Q/k\right\rfloor }{Q}\right)
^{n}=%
{\displaystyle\sum\limits_{k=1}^{\infty}}
\mu\left(  k\right)  \frac{1}{k^{n}}=\zeta\left(  n\right)  ^{-1}\text{ .}%
\]
(See \cite{Rosser1}, \cite{Schoenfeld1}, or \cite{Hardy1}.)
\end{proof}

\bigskip

\begin{corollary}
\label{Omega1Corollary}Let $n$ be an integer greater than $1$, and let
$\lambda_{1}^{\prime},\lambda_{2}^{\prime},\ldots,\lambda_{n}^{\prime}$ be $n$
integers randomly and independently selected with replacement from the set
$\left\{  1,2,\ldots,Q\right\}  $ according to the uniform probability
distribution. \ Let $M$ be a fixed element of the group $SL_{\pm}\left(
n,\mathbb{Z}\right)  $ of invertible $n\times n$ integer matrices. \ Finally,
let $\lambda_{1},\lambda_{2},\ldots,\lambda_{n}$ be $n$ integers given by
\[
\left(  \lambda_{1},\lambda_{2},\ldots,\lambda_{n}\right)  ^{transpose}%
=M\left(  \lambda_{1}^{\prime},\lambda_{2}^{\prime},\ldots,\lambda_{n}%
^{\prime}\right)  ^{transpose}\text{ .}%
\]
Then the probability that
\[
\gcd\left(  \lambda_{1},\lambda_{2},\ldots,\lambda_{n}\right)  =1
\]
is
\[
Prob_{Q}\left(  \underset{}{\overset{}{\gcd\left(  \lambda_{1},\lambda
_{2},\ldots,\lambda_{n}\right)  =1}}\right)  =%
{\displaystyle\sum\limits_{k=1}^{Q}}
\mu\left(  k\right)  \left(  \frac{\left\lfloor Q/k\right\rfloor }{Q}\right)
^{n}\text{ ,}%
\]
where `$\ \left\lfloor -\right\rfloor $' and `$\mu\left(  -\right)  $'
respectively denote the floor and M\"{o}bius functions.

Moreover,
\[
Prob_{\infty}\left(  \underset{}{\overset{}{\gcd\left(  \lambda_{1}%
,\lambda_{2},\ldots,\lambda_{n}\right)  =1}}\right)  =\zeta\left(  n\right)
^{-1}\text{,}%
\]
where $\zeta\left(  n\right)  $ denotes the Riemann zeta function
$\zeta\left(  n\right)  =\sum_{k=1}^{\infty}\frac{1}{k^{n}}$. Hence,
\[
Prob_{Q}\left(  \underset{}{\overset{}{\gcd\left(  \lambda_{1},\lambda
_{2},\ldots,\lambda_{n}\right)  =1}}\right)  =\Omega\left(  \zeta\left(
n\right)  ^{-1}\right)  =\Omega\left(  1\right)
\]
\end{corollary}

\begin{proof}
This corollary immediately follows from the fact that the $\gcd$ is invariant
under the action of $SL_{\pm}\left(  n,\mathbb{Z}\right)  $.
\end{proof}

\bigskip

\begin{remark}
We conjecture that a stronger result holds, namely that \ the function
$\zeta\left(  n\right)  ^{-1}$ is actually a lower bound for $Prob_{Q}\left(
\underset{}{\overset{}{\gcd\left(  \lambda_{1},\lambda_{2},\ldots,\lambda
_{n}\right)  =1}}\right)  $ for $Q\geq n$.
\end{remark}

\bigskip

We need to make the following conjecture to estimate the algorithmic
complexity of Vintage $\mathbb{Z}_{Q}$ algorithms, also called wandering Shor algorithms.

\bigskip

\begin{conjecture}
\label{Conjecture1}Let $n$ be an integer greater than $1$, let $P_{1}%
,P_{2},\ldots,P_{n}$ be $n$ fixed positive integers, and let $\lambda
_{1}^{\prime},\lambda_{2}^{\prime},\ldots,\lambda_{n}^{\prime}$ be $n$
integers randomly and independently selected with replacement from the set
$\left\{  1,2,\ldots,Q\right\}  $ according to the uniform probability
distribution. \ Let $M$ be a fixed element of the group $SL_{\pm}\left(
n,\mathbb{Z}\right)  $ of invertible $n\times n$ integral matrices, and let
\[
\left(  \lambda_{1},\lambda_{2},\ldots,\lambda_{n}\right)  =M\left(
\lambda_{1}^{\prime},\lambda_{2}^{\prime},\ldots,\lambda_{n}^{\prime}\right)
\]

Then the conditional pseudo-probability
\[
Prob_{\infty}\left(  \underset{}{\overset{}{\gcd\left(  \lambda_{j}%
,P_{j}\right)  =1\ \forall j}}\text{ }\left|  \ \overset{}{\underset{}%
{\gcd\left(  \lambda_{1},\lambda_{2},\ldots,\lambda_{n}\right)  =1}}\right.
\right)
\]
is given by
\[
\frac{%
{\displaystyle\prod\limits_{j=1}^{n}}
\frac{\varphi\left(  P_{j}\right)  }{P_{j}}}{\prod\limits_{\underset
{p\mid\operatorname{lcm}\left(  P_{1},\ldots,P_{n}\right)  }{p\text{ Prime}}%
}\left(  1-p^{-n}\right)  }\geq%
{\displaystyle\prod\limits_{j=1}^{n}}
\frac{\varphi\left(  P_{j}\right)  }{P_{j}}\text{ ,}%
\]
where $\zeta\left(  -\right)  $ and $\varphi\left(  -\right)  $ denote
respectively the Riemann zeta and the Euler totient functions.
\end{conjecture}

\bigskip

\begin{proof}
[\textbf{Plausibility Argument}](This is not a proof.)

We treat $Prob_{\infty}$ as if it were a probability distribution on the
integers $\mathbb{Z}^{n}=\left\{  \left(  \lambda_{1},\lambda_{2}%
,\ldots,\lambda_{n}\right)  \right\}  $. \ We assume that $M$ maps this
distribution on itself, and that $Prob_{\infty}\left(  p\mid\lambda
_{j}\right)  $ and $Prob_{\infty}\left(  q\mid\lambda_{j}\right)  $ are
stochastically independent when $p$ and $q$ are distinct primes.

For fixed $j$, the probability $Prob_{\infty}\left(  p\nmid\lambda_{j}\right)
$ that a given prime divisor $p$ of $P_{j}$ does not divide $\lambda
_{j}^{\prime}$ is
\[
1-\frac{1}{p}\text{.}%
\]
Hence, the probability that $P_{j}$ and $\lambda_{j}$ are relatively prime is
\[
Prob_{\infty}\left(  \overset{}{\underset{}{\gcd\left(  P_{j},\lambda
_{j}\right)  =1}}\right)  =%
{\displaystyle\prod\limits_{p\mid P_{j}}}
\left(  1-\frac{1}{p}\right)  \text{ .}%
\]
This can be reexpressed in terms of the Euler totient function as
\[
Prob_{\infty}\left(  \underset{}{\overset{}{\gcd\left(  P_{j},\lambda
_{j}\right)  =}1}\right)  =\frac{\varphi\left(  P_{j}\right)  }{P_{j}}%
\]

Since $\lambda_{1},\lambda_{2},\ldots,\lambda_{n}$ are independent random
variables, we have
\[
Prob_{\infty}\left(  \overset{}{\underset{}{\gcd\left(  P_{j},\lambda
_{j}\right)  =1\forall j}}\right)  =%
{\displaystyle\prod\limits_{j=1}^{n}}
\frac{\varphi\left(  P_{j}\right)  }{P_{j}}\text{ .}%
\]

On the other hand, the probability that a given prime $p$ does not divide all
the integers $\lambda_{1},\lambda_{2},\ldots,\lambda_{n}$ $\ is$%
\[
1-\frac{1}{p^{n}}\text{.}%
\]
Thus,%
\[
Prob_{\infty}\left(  \underset{}{\overset{}{p\nmid\gcd\left(  \lambda
_{1},\lambda_{2},\ldots,\lambda_{n}\right)  }}\forall p\text{ s.t. }%
p\nmid\operatorname{lcm}\left(  P_{1},P_{2},\ldots,P_{n}\right)  \right)
\]
is given by the expression
\[%
{\displaystyle\prod\limits_{p\nmid\operatorname{lcm}\left(  P_{1},P_{2}%
,\ldots,P_{n}\right)  }}
\left(  1-p^{-n}\right)  =\frac{\zeta\left(  n\right)  ^{-1}}{%
{\displaystyle\prod\limits_{p\mid\operatorname{lcm}\left(  P_{1},P_{2}%
,\ldots,P_{n}\right)  }}
\left(  1-p^{-n}\right)  }\text{ ,}%
\]
where we have used the fact \cite{Hardy1} that
\[
\zeta\left(  n\right)  ^{-1}=\prod_{p\text{ Prime}}\left(  1-\frac{1}{p^{n}%
}\right)  \text{.}%
\]

We next note that the events $\forall j\ \gcd\left(  P_{j},\lambda_{j}\right)
=1$ and $\forall p\ p\nmid\operatorname{lcm}\left(  P_{1,}P_{2},\ldots
,P_{n}\right)  \longrightarrow p\nmid\gcd\left(  P_{1,}P_{2},\ldots
,P_{n}\right)  $ are stochastically independent since they respectively refer
to the disjoint sets of primes $\left\{  p:p\mid\operatorname{lcm}\left(
P_{1},P_{2},\ldots,P_{n}\right)  \right\}  $ and $\left\{  p:p\nmid
\operatorname{lcm}\left(  P_{1},P_{2},\ldots,P_{n}\right)  \right\}  $.
\ Hence, the probability of the joint event
\[
Prob_{\infty}\left(  \underset{}{\overset{}{\gcd\left(  P_{j},\lambda
_{j}\right)  =1\ \forall j\text{ AND }\gcd\left(  \lambda_{1},\lambda
_{2},\ldots,\lambda_{n}\right)  =1}}\right)
\]
is given by the expression
\[
\frac{\zeta\left(  n\right)  ^{-1}%
{\displaystyle\prod\limits_{j=1}^{n}}
\frac{\varphi\left(  P_{j}\right)  }{P_{j}}}{%
{\displaystyle\prod\limits_{p\mid\operatorname{lcm}\left(  P_{1},P_{2}%
,\ldots,P_{n}\right)  }}
\left(  1-p^{-n}\right)  }\text{ .}%
\]

Using exactly the same argument as that used to find an expression for
\[
Prob_{\infty}\left(  \underset{}{\overset{}{p\nmid\gcd\left(  \lambda
_{1},\lambda_{2},\ldots,\lambda_{n}\right)  }}\forall p\text{ s.t. }%
p\nmid\operatorname{lcm}\left(  P_{1},P_{2},\ldots,P_{n}\right)  \right)
\text{ ,}%
\]
we have
\[
Prob_{\infty}\left(  \underset{}{\overset{}{\gcd\left(  \lambda_{1}%
,\lambda_{2},\ldots,\lambda_{n}\right)  =1}}\right)  =\zeta\left(  n\right)
^{-1}\text{.}%
\]
Hence the conditional probability
\[
Prob_{\infty}\left(  \overset{}{\underset{}{\gcd\left(  P_{j},\lambda
_{j}\right)  =1\ \forall j\ \left|  \overset{}{\ \gcd}\left(  \lambda
_{1},\lambda_{2},\ldots,\lambda_{n}\right)  =1\right.  }}\right)
\]
is given by the expression
\[
\frac{%
{\displaystyle\prod\limits_{j=1}^{n}}
\frac{\varphi\left(  P_{j}\right)  }{P_{j}}}{%
{\displaystyle\prod\limits_{p\mid\operatorname{lcm}\left(  P_{1},P_{2}%
,\ldots,P_{n}\right)  }}
\left(  1-p^{-n}\right)  }%
\]

Finally, since
\[%
{\displaystyle\prod\limits_{p\mid\operatorname{lcm}\left(  P_{1},P_{2}%
,\ldots,P_{n}\right)  }}
\left(  1-p^{-n}\right)  \leq1\text{ ,}%
\]
it follows that the conditional probability
\[
Prob_{\infty}\left(  \overset{}{\underset{}{\gcd\left(  P_{j},\lambda
_{j}\right)  =1\ \forall j\ \left|  \overset{}{\ \gcd}\left(  \lambda
_{1},\lambda_{2},\ldots,\lambda_{n}\right)  =1\right.  }}\right)
\]
is bounded below by the expression
\[%
{\displaystyle\prod\limits_{j=1}^{n}}
\frac{\varphi\left(  P_{j}\right)  }{P_{j}}\text{ .}%
\]
\end{proof}

\bigskip

The following is an immediate corollary of the above conjecture.\bigskip

\begin{corollary}
\label{conditionalprob}Let $n$ be an integer greater than $1$, and let
$P_{1},P_{2},\ldots,P_{n}$ be $n$ fixed positive integers. \ Let $\lambda
_{1}^{\prime},\lambda_{2}^{\prime},\ldots,\lambda_{n}^{\prime}$ be $n$
integers randomly and independently selected with replacement from the set all
integers $\mathbb{Z}$ according to the uniform probability distribution. \ Let
$M$ be a fixed element of the group $SL_{\pm}\left(  n,\mathbb{Z}\right)  $ of
invertible $n\times n$ integer matrices. \ Finally, let $\lambda_{1}%
,\lambda_{2},\ldots,\lambda_{n}$ be $n$ integers given by
\[
\left(  \lambda_{1},\lambda_{2},\ldots,\lambda_{n}\right)  ^{transpose}%
=M\left(  \lambda_{1}^{\prime},\lambda_{2}^{\prime},\ldots,\lambda_{n}%
^{\prime}\right)  ^{transpose}\text{ .}%
\]

Then, assuming conjecture \ref{Conjecture1}, we have
\[
Prob_{\infty}\left(  \underset{}{\overset{}{\gcd\left(  \lambda_{j}%
,P_{j}\right)  =1\ \forall j}}\text{ }\left|  \ \overset{}{\underset{}%
{\gcd\left(  \lambda_{1},\lambda_{2},\ldots,\lambda_{n}\right)  =1}}\right.
\right)  =\Omega\left(
{\displaystyle\prod\limits_{j=1}^{n}}
\frac{1}{\lg\lg P_{j}}\right)  \text{ ,}%
\]
where $\Omega\left(  -\right)  $ denotes the asymptotic lower bound `big-omega.'

Thus, if $Q$ is greater than each $P_{j}$, we have
\[
Prob_{Q}\left(  \underset{}{\overset{}{\gcd\left(  \lambda_{j},P_{j}\right)
=1\ \forall j}}\text{ }\left|  \ \overset{}{\underset{}{\gcd\left(
\lambda_{1},\lambda_{2},\ldots,\lambda_{n}\right)  =1}}\right.  \right)
=\Omega\left(  \left(  \frac{1}{\lg\lg Q}\right)  ^{n}\right)
\]
\end{corollary}

\begin{proof}
Since\footnote{See \cite[Theorem 328, Section 18.4]{Hardy1}.}
\[
\underline{\lim}\frac{\varphi\left(  n\right)  \ln\ln n}{n}=e^{-\gamma}\text{
,}%
\]
where $\gamma$ denotes Euler's constant, we have that
\[
\frac{\varphi\left(  P_{j}\right)  }{P_{j}}=\Omega\left(  \frac{1}{\lg\lg
P_{j}}\right)
\]
Thus, an asymptotic lower bound for the above conditional probability is given
by the expression
\[
\Omega\left(
{\displaystyle\prod\limits_{j=1}^{n}}
\frac{1}{\lg\lg P_{j}}\right)  \text{ .}%
\]
\end{proof}


\bigskip

\begin{thebibliography}{9}                                                                                                %

\bibitem {Alber1}Alber, G., T. Beth, M. Horodecki, P. Horodecki, R. Horodecki,
M. Rotteler, H. Weinfurther, R. Werner, and A. Zeilinger, \textbf{``Quantum
Information: An Introduction to Basic Theoretical Concepts and Experiments,''}
Springer, (2001).

\bibitem {Bach1}Bach, Eric, and Jeffrey Shallit, \textbf{``Algorithmic Number
Theory: Volume I: Efficient Algorithms,''} MIT Press, (1997).

\bibitem {Bernstein1}Bernstein, Ethan, and Umesh Vazirani, \textbf{Quantum
Complexity Theory}, SIAM\ J. of Computing, Vol. 26, No. 5, (1997), pp 1411-1473.

\bibitem {Boneh1}Boneh, Dan, and Richard J. Lipton, \textbf{Quantum
cryptanalysis of hidden linear functions}, in \textbf{``Lecture Notes in
Computer Science -- Advances in Cryptology -- CRYPTO'95,''} D. Coppersmith
(ed.), Springer-Verlag, Berlin, (1995), pp 424-437.

\bibitem {Brassard1}Brassard, Gilles, and Paul Bratley,
``\textbf{Algorithmics: Theory and Practice,''} Printice-Hall, (1988).

\bibitem {Brassard2}Brassard, Gilles and Peter Hoyer, \textbf{An exact quantum
polynomial-time algorithm for Simon's problem}, Proceedings of Fifth Israeli
Symposium on Theory of Computing and Systems -- ISTCS, IEEE Computer Society
Press, (1997), pp. 12-23.

\bibitem {Cartan1}Cartan, Henri, and Samuel Eilenberg, \textbf{``Homological
Algebra,''} Princeton University Press, (1956).

\bibitem {Cheung1}Cheung, Kevin K.H., and Michele Mosca, \textbf{Decomposing
finite abelian groups}, http://xxx.lanl.gov/abs/cs.DS/0101004.

\bibitem {Cleve1}Cleve, Richard, Artur Ekert, Chiara Macchiavello, and Michele
Mosca, \textbf{Quantum Algorithms Revisited}, Phil. Trans. Roy. Soc. Lond., A,
(1997). http://xxx.lanl.gov/abs/quant-ph/9708016

\bibitem {Coppersmith1}Coppersmith, D., \textbf{An approximate quantum Fourier
transform used in quantum factoring,} IBM Research Report RC 19642, (1994).

\bibitem {Cormen1}Cormen, Thomas H., Charles E. Leiserson, and Ronald L.
Rivest, ``\textbf{Introduction to Algorithms},'' McGraw-Hill, (1990).

\bibitem {Cox1}Cox, David, John Little, and Donal O'Shea, \textbf{``Ideals,
Varieties, and Algorithms: An Introduction to Computational Algebraic Geometry
and Commutative Algebra''} (second edition), Springer-Verlag, (1996).

\bibitem {Cox2}Cox, David, John Little, and Donal O'Shea, \textbf{``Using
Algebraic Geometry,''} Springer, (1998).

\bibitem {Ekert1}Ekert, Artur K.and Richard Jozsa, \textbf{Quantum computation
and Shor's factoring algorithm}, Rev. Mod. Phys., 68,(1996), pp 733-753.

\bibitem {Ettinger2}Ettinger, Mark, and Peter Hoyer, \textbf{On Quantum
Algorithms for Noncommutative Hidden Subgroups}, (1998). \ http://xxx.lanl.gov/abs/quant-ph/9807029

\bibitem {Ettinger1}Ettinger, Mark, Peter Hoyer, Emanuel Knill, \textbf{Hidden
Subgroup States are Almost Orthogonal}, http://xxx.lanl.gov/abs/quant-ph/9901034.

\bibitem {Fulton1}Fulton, William, and Joe Harris, \textbf{``Representation
Theory,''} Springer-Verlag, (1991).

\bibitem {Gathen1}Gathen, Joachim von zur, and Jurgen Gerhard, ``Modern
Computer Algebra,'' Cambridge University Press, (1999).

\bibitem {Geddes1}Geddes, Keith O., Stephen R. Czapor, and George Labahn,
\textbf{``Algorithms for Computer Algebra,''} Kluwer Academic Publishers,
Boston, (1992).

\bibitem {Hall1}Hall, Marshall, Jr., \textbf{``The Theory of Groups,''}
Macmillan, New York, (1959).

\bibitem {Hardy1}Hardy, G.H., and E.M. Wright, ``\textbf{An Introduction to
the Theory of Numbers},'' Oxford Press, (1965).

\bibitem {Hirvensalo1}Hirvensalo, Mika, \textbf{``Quantum Computing,''}
Springer, (2001).

\bibitem {Hoyer1}Hoyer, Peter, \textbf{Efficient quantum transforms}, http://xxx.lanl.gov/abs/quant-ph/9702028.

\bibitem {Ivanyos1}Ivanyos, Gabor, Frederic Magniez, and Miklos Santha,
\textbf{Efficient quantum algorithms for some instances of the non-Abelian
hidden subgroup problem}, (2001). http://xxx.lanl.gov/abs/quant-ph/0102014

\bibitem {Jozsa2}Jozsa, Richard, \textbf{Quantum algorithms and the Fourier
transform}, quant-ph preprint archive 9707033 17 Jul 1997.

\bibitem {Jozsa3}Jozsa, Richard, Proc. Roy. Soc. London Soc., Ser. A, 454,
(1998), 323 - 337.

\bibitem {Jozsa4}Jozsa, Richard, Quantum factoring, discrete logarithms and
the hidden subgroup problem, IEEE Computing in Science and Engineering, (to
appear). \ http://xxx.lanl.gov/abs/quant-ph/0012084

\bibitem {Kemeny1}Kemeny, John G., and J. Laurie Snell, ``\textbf{Finite
Markov Chains},'' Van Nostrand, (1960).

\bibitem {Kitaev1}Kitaev, A., \textbf{Quantum measurement and the abelian
stabiliser problem,} (1995), quant-ph preprint archive 9511026.

\bibitem {Knuth1}Knuth, Donald E., ``\textbf{The Art of Computer
Programming,''} (second edition), Addison-Wesley, Reading, Massachusetts, (1981).

\bibitem {LeVeque1}LeVeque, , William Judson, \textbf{``Topics in Number
Theory,''} Addison-Wesley, (1956).

\bibitem {Lomonaco1}Lomonaco, Samuel J., Jr., \textbf{A Rosetta Stone for
quantum mechanics with an introduction to quantum computation,} in
\textbf{``Quantum Computation: A Grand Mathematical Challenge for the
Twenty-First Century and the Millennium''} PSAPM/58, American Mathematical
Society, Providence, RI, (2002). (http://xxx.lanl.gov/abs/quant-ph/0007045)

\bibitem {Lomonaco2}Lomonaco, Samuel J., Jr., \textbf{Shor's quantum factoring
algorithm,} in \textbf{``Quantum Computation: A Grand Mathematical Challenge
for the Twenty-First Century and the Millennium,''} PSAPM/58, American
Mathematical Society, Providence, RI, (2002). (http://xxx.lanl.gov/abs/quant-ph/0010034)

\bibitem {Lomonaco3}Lomonaco, Samuel J., Jr., \textbf{The non-abelian Fourier
transform and quantum computation}, MSRI Streaming Video, (2000), http://www.msri.org/publications/ln/msri/2000/qcomputing/lomonaco/1/index.html

\bibitem {Mosca1}Mosca, Michelle, and Artur Ekert, \textbf{The Hidden Subgroup
Problem and Eigenvalue Estimation on a Quantum Computer}, Proceedings of the
1st NASA International Conference on Quantum Computing and Quantum
Communication, Springer-Verlag, (to appear). (http://xxx.lanl.gov/abs/quant-ph/9903071)

\bibitem {Nielsen1}Nielsen, Michael A., and Isaac L. Chuang, \textbf{``Quantum
Computation and Quantum Information,''} Cambridge University Press, (2000).

\bibitem {Preskill1}Preskill, John, \textbf{``Quantum Computation,''} Lecture
Notes, http://www.theory.caltech.edu/people/preskill/ph229/\#lecture

\bibitem {Pueschel1}Pueschel, Markus, Martin Roetteler, and Thomas Beth,
\textbf{Fast Quantum Fourier Transforms for a Class of Non-abelian Groups},
(1998). \ http://xxx.lanl.gov/abs/quant-ph/9807064

\bibitem {Roetteler1}Roetteler, Martin, and Thomas Beth,
\textbf{Polynomial-Time Solution to the Hidden Subgroup Problem for a Class of
non-abelian Groups}, (1998). \ http://xxx.lanl.gov/abs/quant-ph/9812070

\bibitem {Rosser1}Rosser, J. Barkley, and Lowell Schoenfeld,
\textbf{Approximate formulas for some functions of prime numbers,} Illinois J.
Math., v.6, (1962), pp. 64-94.

\bibitem {Russell1}Russell, Alexander, and Amnon Ta-Shma, \textbf{Normal
Subgroup Reconstruction and Quantum Computation Using Group Representations},
STOC, (2000).

\bibitem {Schoenfeld1}Schoenfeld, Lowell, \textbf{Sharper bounds for the
Chebyshev functions }$\mathbf{\theta}\left(  \mathbf{x}\right)  $\textbf{ and
}$\mathbf{\psi}\left(  \mathbf{x}\right)  $\textbf{. \ II,} Math. Comp., vol.
30, No. 134, (1976), pp. 337-360.

\bibitem {Shor1}Shor, Peter W., \textbf{Polynomial time algorithms for prime
factorization and discrete logarithms on a quantum computer}, SIAM\ J. on
Computing, 26(5) (1997), pp 1484 - 1509. (http://xxx.lanl.gov/abs/quant-ph/9508027)

\bibitem {Shor2}Shor, Peter W., \textbf{Introduction to quantum algorithms,}
in \textbf{``Quantum Computation: A Grand Mathematical Challenge for the
Twenty-First Century and the Millennium,''} PSAPM/58, American Mathematical
Society, Providence, RI, (2002). (http://xxx.lanl.gov/abs/quant-ph/0005003)

\bibitem {Simon1}Simon, Daniel R., \textbf{On the power of quantum
computation}, SIAM J. Comput., Vol. 26, No. 5, (1997), pp 1474-1483.

\bibitem {van Dam}van Dam, Wim, and Sean Hallgren, \textbf{Efficient Quantum
Algorithms for Shifted Quadratic Character Problems}, http://xxx.lanl.gov/abs/quant-ph/0011067.

\bibitem {Vazirani1}Vazirani, Umesh, \textbf{On the power of quantum
computation}, Philosophical Tranactions of the Royal Society of London, Series
A, 354:1759-1768, August 1998.

\bibitem {vanDam1}van Dam, Wim, and Lawrence Ip, \textbf{Quantum Algorithms,
for Hidden Coset Problems}, manuscript, http://www.cs.caltech.edu/\symbol{126}hallgren/hcp.pdf
\end{thebibliography}
\end{document}